\documentclass[12pt, a4paper]{article}

\usepackage{tikz-cd}
\usetikzlibrary{arrows,snakes,shapes.arrows,decorations.markings}
     \tikzset{>=triangle 90}
     \tikzstyle{bbc}=[draw,circle,fill=black,scale=.75]
     \tikzstyle{rc}=[circle,fill=red,scale=.6]
     \tikzstyle{wc}=[draw,circle,scale=.75]

\usepackage{color}

\usepackage{latexsym}
\usepackage{amsmath, mathdots, amscd}
\usepackage{multirow}
\usepackage{amssymb}
\usepackage{graphicx}
\usepackage{euscript}
\usepackage{amsfonts}
\usepackage{verbatim}
\usepackage{epsfig}
\usepackage{hyperref}
\usepackage[margin=2cm,bottom=2.0cm]{geometry}
\usepackage{float}
\usepackage{cite}

\def\be{\begin{eqnarray}}
\def\ee{\end{eqnarray}}
\def\0{\nonumber}

\providecommand{\Tr}{\textnormal{Tr}}

\providecommand{\ZZ}{\mathbb{Z}}

\def\0{\nonumber}

\def\c{{\bf c}}

\def\bZ{\mathbb{Z}}

\begin{document}
\begin{titlepage}
 
\vskip -0.5cm
 
\begin{flushright}

\end{flushright}
 
\vskip 1cm
\begin{center}
 
{\LARGE\bf \boldmath $\mathcal{N}=1$ SCFTs from F-theory on Orbifolds} 
 
 \vskip 2cm
 
{\large Simone Giacomelli$^1$ and Raffaele Savelli$^{2}$}

 \vskip 0.9cm
 
 {\it  $^1$Dipartimento di Fisica, Universit\`a di Milano-Bicocca,\\ Piazza della Scienza 3, I-20126 Milano, Italy \\[2mm]
 
 $^2$Dipartimento di Fisica \& INFN, Universit\`a di Roma ``Tor Vergata'', \\ Via della Ricerca
Scientifica 1, I-00133 Roma, Italy
 }
 \vskip 2cm
 
\abstract{\noindent We study four-dimensional superconformal field theories living on the worldvolume of $D3$ branes probing minimally-supersymmetric F-theory backgrounds, focusing on the case of orbi-orientifold setups with and without 7-branes. We observe that these theories are closely related to compactifications of six-dimensional $\mathcal{N}=(1,0)$ theories on a torus with flux, where the flux quanta is mapped in Type IIB to the defining data of the orbifold group. We analyze the cases of class $\mathcal{S}_k$ theories as well as of compactifications of the E-string and of orbi-instanton theories. We also classify $\mathcal{S}$-fold configurations in F-theory preserving minimal supersymmetry in four dimensions and their mass deformations.} 

\end{center}

\end{titlepage}

\tableofcontents

\section{Introduction}

Supersymmetric field theories constitute a very important theoretical laboratory for the exploration of the non-perturbative dynamics of quantum field theory. This is exemplified by the discovery of infrared (IR) dualities for supersymmetric gauge theories, such as Seiberg duality \cite{Seiberg:1994pq} and mirror symmetry \cite{Intriligator:1996ex}, and the discovery of the Seiberg-Witten solution \cite{Seiberg:1994rs, Seiberg:1994aj} for $\mathcal{N}=2$ gauge theories in four dimensions (4d). In particular the Seiberg-Witten soution opened the way to the discovery of many strongly-coupled theories in 4d, starting with \cite{Argyres:1995jj, Argyres:1995xn}. The landscape of interacting $\mathcal{N}=2$ theories (without any obvious lagrangian description) was then considerably enlarged by exploiting their geometric realization in string theory, either via compactification as pioneered in \cite{Katz:1996fh, Shapere:1999xr}, or by introducing branes in string/M-theory whose worldvolume hosts the desired field theory as in \cite{Hanany:1996ie, Witten:1997sc} (and many follow-up works), and more recently in \cite{Gaiotto:2009we}. 

The above results allowed us to make a huge progress in our understanding of the landscape of interacting conformal field theories (at least in the supersymmetric case) and paved the way for a classification program of those with eight supercharges (see e.g.~the recent review \cite{Argyres:2022mnu} and references therein). In this respect, a remarkable result is the discovery of $\mathcal{N}=3$ theories \cite{Garcia-Etxebarria:2015wns, Garcia-Etxebarria:2016erx, Aharony:2016kai}, which were previously believed not to exist. Their realization exploits a new geometric construction in Type IIB, called $\mathcal{S}$-fold, which involves a twist by the S-duality group of the string theory, and can be thought of as a non-perturbative generalization of an orientifold $O3$ plane. This setup was later generalized by combining $\mathcal{S}$-folds and 7-branes with constant axio-dilaton, leading to a new class of $\mathcal{N}=2$ superconformal field theories (SCFTs) \cite{Apruzzi:2020pmv, Giacomelli:2020jel, Giacomelli:2020gee, Bourget:2020mez, Heckman:2020svr}. An interesting feature of this class of theories is that they admit an alternative construction in terms of six-dimensional (6d) SCFTs with minimal supersymmetry (see e.g.~\cite{Apruzzi:2013yva, Heckman:2013pva, DelZotto:2014hpa, Gaiotto:2014lca, Bhardwaj:2015xxa, Heckman:2015bfa, Bhardwaj:2015oru}) compactified on a torus with almost commuting holonomies (Stiefel-Whitney twist) \cite{Ohmori:2018ona, Heckman:2022suy}. The relevant 6d $\mathcal{N}=(1,0)$ theories can be realized on the worldvolume of $M5$ branes in M-theory probing orbifolds and/or a $M9$ wall \cite{Horava:1996ma}. This fact provides a M-theory/Type IIB correspondence which turned out to be a key ingredient for the exploration of $\mathcal{N}=2$ $\mathcal{S}$-fold SCFTs. 

The purpose of this work is to initiate the exploration of  $\mathcal{N}=1$ $\mathcal{S}$-fold theories, which are supposed to lead to a vast new class of 4d theories with minimal supersymmetry. We start by classifying possible $\mathcal{S}$-fold backgrounds preserving four supercharges, resulting in a large class of $K3$-fibered Calabi-Yau (CY) fourfolds, and discuss their deformations. Since at present little is known about non-perturbative F-theory backgrounds preserving four supercharges, and our understanding of the landscape of $\mathcal{N}=1$ SCFTs is still quite limited, a systematic analysis of  $\mathcal{N}=1$ $\mathcal{S}$-folds is currently very challenging. Therefore we limit ourselves to making a preliminary step in this direction. In particular, we aim at understanding how the M-theory/Type IIB correspondence observed in the $\mathcal{N}=2$ case generalizes to theories with four supercharges. Interestingly, we find that the correspondence does indeed generalize: We obtain a connection between $D3$ branes probing Type IIB $\mathcal{N}=1$ orbi-orientifold backgrounds and torus compactifications of 6d $\mathcal{N}=(1,0)$ theories with flux halving the amount of supersymmetry. These constitute an extremely vast family of $\mathcal{N}=1$ SCFTs which has not been fully  explored yet. However, several results are already available in the literature, and we take them as the starting point of our analysis. A (partial) list of references on the subject is \cite{Gaiotto:2015usa, Coman:2015bqq, Razamat:2016dpl, Kim:2017toz, Bah:2017gph, Bourton:2017pee, Kim:2018bpg, Apruzzi:2018oge, Razamat:2018gro, Razamat:2018gbu, Razamat:2018zel, Zafrir:2018hkr, Razamat:2019vfd, Razamat:2019mdt, Razamat:2019ukg, Pasquetti:2019hxf, Bah:2019vmq, Bah:2020jas, Bah:2020uev, Bah:2021brs, Bourton:2021das, Razamat:2020bix, Hwang:2021xyw, Nazzal:2021tiu, Razamat:2022gpm, Bah:2021iaa, Bah:2022wot, Kim:2023qbx}. 

One of the characteristic features of compactifications of 6d theories to 4d preserving minimal supersymmetry is that, unlike the $\mathcal{N}=2$ class $\mathcal{S}$ case, the 4d theory is not just specified by the topology of the Riemann surface, but we can also turn on fluxes (and holonomies) for the global symmetry of the 6d theory. This had already been noticed for 6d $\mathcal{N}=(2,0)$ theories (see e.g.~\cite{Benini:2009mz, Bah:2012dg, Bonelli:2013pva, Xie:2013gma, Giacomelli:2014rna}) and is a key ingredient of the compactification of $\mathcal{N}=(1,0)$ theories, as discussed in the above references. Accordingly, we find that in the $\mathcal{N}=1$ version of the M-theory/Type IIB correspondence mentioned above the flux plays a prominent role: The torus compactification with flux of the 6d $\mathcal{N}=(1,0)$ theories living on the worldvolume of $M5$ branes corresponds to 4d theories living on the worldvolume of $D3$ branes probing an F-theory background, and the properties of the background depend on the amount of flux. We test this proposal for $M5$ branes probing an Abelian orbifold ($A$-type conformal matter), $M5$s probing a smooth $M9$ wall (E-string), and $M5$s probing a $M9$ wrapping a $\mathbb{Z}_2$-orbifolded complex plane; on the Type IIB side we limit ourselves to backgrounds involving Abelian orbifolds and orientifolds whose lagrangian description can be determined with known technology. We find that for all the above 6d theories the information about the flux is encoded in the corresponding orbifold group $\Gamma$ in Type IIB: The order of $\Gamma$ depends on the amount of flux, and only a proper subgroup of $\Gamma$ acting on a $\mathbb{C}^2$ appears in the M-theory background. In the special situation when the resulting 4d theory has enhanced supersymmetry (happening for a particular value of the flux), $\Gamma$ reduces to a subgroup of SU$(2)$ acting on $\mathbb{C}^2$ and coincides with the orbifold group specifying the 6d SCFT in M-theory (see \cite{Giacomelli:2022drw}). 

We would like to stress that the torus compactifications with flux of 6d theories map to Type IIB backgrounds involving orbifolds and orientifolds only for specific choices of flux. We check that for all the above 6d theories there are choices of flux with this property. For generic flux this type of correspondence may still hold provided we enlarge our setup and consider Type IIB backgrounds involving more general CY geometries. The study of $A_1$ conformal matter presented in \cite{Bah:2021iaa} suggests that this is indeed the case. In that context the relevant CY geometries are the cones over $Y^{p,q}$ Sasaki-Einstein manifolds. It would be important to explore this problem further, hoping that this will provide a new perspective both on the compactification of 6d theories and on $\mathcal{N}=1$ $\mathcal{S}$-fold theories. We also need to point out that the Type IIB theories we propose do not coincide exactly with the 6d theories on $T^2$: They are obtained by removing from the 6d theories certain chiral gauge singlets and/or activating certain relevant mass deformations. The role of mass deformations was also noticed in the more supersymmetric case in \cite{Giacomelli:2022drw}. 

The rest of the paper is organized as follows: In Section \ref{sec2} we discuss F-theory orbifolds and classify possible $\mathcal{N}=1$ - preserving $\mathcal{S}$-folds. In Section \ref{Pert:Sec} we review the perturbative orbifold and orientifold technology we need for our analysis and discuss all possible conformal theories we can get in Type IIB for orbifold order two. An exhaustive scan for orbifold order three and four is presented in Appendix \ref{ExampleLoworder1}. In Section \ref{sec4} we compare $T^2$ compactifications with flux of 6d theories with the perturbative Type IIB theories. We identify for each 6d theory the relevant orbi-orientifold background and determine the corresponding choice of flux. Finally, we present our concluding remarks in Section \ref{sec5}.

\section{Probing F-theory on orbifolds}\label{sec2}

In this section we describe the type of string geometry we are interested to probe by means of $D3$ branes. After recalling in Section \ref{SUSYProbe} the conditions to preserve minimal supersymmetry on the probe, we introduce what we dubbed $\mathcal{N}=1$ $\mathcal{S}$-fold in Section \ref{N=1Sfold} and generalize it in Section \ref{GenDisc}. Finally, in Section \ref{MassDef}, we present some general features of mass deformations for the above $\mathcal{N}=1$ theories.

\subsection{Supersymmetries on the probe}\label{SUSYProbe}

Consider a large diffeomorphism of a two-torus, i.e.~a matrix
\be\label{Sduality}
\left(\begin{array}{ll}a&b\\c&d\end{array}\right)\in{\rm SL}(2,\mathbb{Z})\,,
\ee
acting on its complex structure $\tau$ by the usual fractional-linear transformation
\begin{equation}
\tau\longrightarrow\frac{a\tau+b}{c\tau+d}\,.
\end{equation}
The kernel of the above representation is the center $\ZZ_2\simeq\{I_2,-I_2\}\subset{\rm SL}(2,\mathbb{Z})$, where $I_2$ is the two-by-two identity matrix. This means that any value of $\tau$ is fixed by this order-$2$ subgroup. There are moreover special values of $\tau$ which are fixed by other cyclic subgroups of SL$(2,\mathbb{Z})$. We choose a frame such that all of the orbifold operations on the torus are described as follows\footnote{For the generators, we have chosen the specific representatives (of the respective PSL$(2,\ZZ)$ classes) displayed in the last column, in order to have an action of the same order on the holomorphic $(1,0)$-form of the torus, as will be clear momentarily.}
\be\label{ZkOptions}
\begin{tabular}{|c|c|c|}
 \hline  $\tau$&$\ZZ_k$&generator \\ \hline any &$\ZZ_2$ & $\left(\begin{array}{cc}-1&0\\ 0&-1 \end{array}\right)$ \\ \hline $e^{\pi {\rm i}/3}$ & $\ZZ_3$ & $\left(\begin{array}{cc}0&-1\\ 1&-1 \end{array}\right)$ \\   \hline
${\rm i}$&$\ZZ_4$&$\left(\begin{array}{cc}0&-1\\ 1&0 \end{array}\right)$\\ \hline $e^{\pi {\rm i}/3}$ & $\ZZ_6$ & $\left(\begin{array}{cc}1&-1\\ 1&0 \end{array}\right)$\\ \hline
\end{tabular}
\ee

Parametrizing the torus locally by two real coordinates $(x_1,x_2)$ forming a doublet under \eqref{Sduality}, we have that the local complex coordinate $t\equiv x_1-\tau x_2$ transforms under SL$(2,\mathbb{Z})$ as
\be\label{T2CoordTransf}
t\stackrel{S}{\longrightarrow}\frac{t}{c\tau+d}\,.
\ee
It is easy to see that, for all of the SL$(2,\ZZ)$ transformations described above, \eqref{T2CoordTransf} amounts to the phase rotation
\be\label{T2CoordRot}
t\stackrel{S}{\longrightarrow} e^{-2\pi {\rm i}/k}\,t\,,\qquad k=2,3,4,6\,,
\ee
where $k$ labels the four different options listed in \eqref{ZkOptions}.

If one considers a stack of $D3$ branes probing flat space ($\mathbb{C}^3$ in Type IIB or $\mathbb{C}^3\times T^2$ in F-theory), $\tau$ plays the role of the complexified gauge coupling of the $\mathcal{N}=4$ worldvolume theory, and \eqref{Sduality} the role of S-duality transformations. Associated to any given SL$(2,\ZZ)$ transformation \eqref{Sduality}, there is a U$(1)$ rotation which acts on objects of charge $q$ by multiplying them with the phase $e^{-{\rm i}q\,{\rm arg}(c\tau+d)}$. In particular, the supercharges preserved by the field theory, which can be packaged into $4$ Weyl fermions $\{Q^j\}_{j=1,\ldots,4}$, transform non-trivially under S-duality, and more precisely have charge $1/2$ under the aforementioned U$(1)$. This means that, for the orbifold transformations in \eqref{ZkOptions} we have
\be\label{SdualityOnQ}
Q^j\stackrel{S}{\longrightarrow} e^{-\pi {\rm i}/k}\,Q^j\,,\qquad k=2,3,4,6\,.
\ee

The SU$(4)$ $R$-symmetry of the field theory is identified with the isometry group of the space $\mathbb{C}^3$ transverse to the $D3$ branes. Parametrizing such a space with complex coordinates $\{z_1,z_2,z_3\}$, the Cartan torus of the $R$-symmetry acts on them simply as
\be\label{SO(2)3}
(z_1,z_2,z_3)\stackrel{R}{\longrightarrow}(e^{{\rm i}\Psi_1}z_1,e^{{\rm i}\Psi_2}z_2,e^{{\rm i}\Psi_3}z_3)\,,
\ee
where $\{\Psi_a\}_{a=1,2,3}$ are three angles defined modulo $2\pi$.
The supercharges transform instead in the fundamental representation of the $R$-symmetry, which means that, under the Cartan torus corresponding to the U$(1)^3$ rotation \eqref{SO(2)3}, they behave as follows
\be\label{RsymmetryOnQ}
(Q^1,Q^2,Q^3,Q^4)\stackrel{R}{\longrightarrow}\left(e^{\tfrac{{\rm i}}{2}(\Psi_1+\Psi_2+\Psi_3)}Q^1,e^{\tfrac{{\rm i}}{2}(\Psi_1-\Psi_2-\Psi_3)}Q^2,e^{\tfrac{{\rm i}}{2}(-\Psi_1+\Psi_2-\Psi_3)}Q^3,e^{\tfrac{{\rm i}}{2}(-\Psi_1-\Psi_2+\Psi_3)}Q^4\right).
\ee

The subgroups of SL$(2,\ZZ)$ listed in \eqref{ZkOptions} as well as (cyclic subgroups of) the $R$-symmetry are honest global symmetries of the field theory and, as such, can be combined and quotiented to obtain less supersymmetric theories. This amounts to introducing $O3$ planes and non-perturbative generalizations thereof, at the origin of the internal $\mathbb{C}^3$, thus creating orbifold singularities now commonly known under the name of ``$\mathcal{S}$-folds'' \cite{Garcia-Etxebarria:2015wns,Aharony:2016kai}.

Certain special bound states of mutually non-perturbative 7-branes can also be accommodated in this picture. Their associated monodromy matrices are the ones listed in \eqref{ZkOptions} as well as the negatives thereof (which are the other representatives in each PSL$(2,\ZZ)$ class). The background value of the axio-dilaton can therefore be held constant in the presence of such bound states. This allows us to write the metric in the vicinity of any of these 7-brane stacks as
\be
{\rm d} s^2 \sim |z_3|^{-\frac{N_7}{6}}{\rm d}z_3\, {\rm d}\bar{z}_3 \equiv {\rm d}\left(z_3^{\frac{1}{\Delta_7}}\right)\, {\rm d}\left(\bar{z}_3^{\frac{1}{\Delta_7}}\right)\,,
\ee
where we have the 7-branes extended along $\mathbb{C}^2_{\{z_1,z_2\}}$, and we see that the metric on their transverse plane has a conical singularity at the location of the stack $z_3=0$. In the above formula, $N_7$ denotes the total number of 7-branes constituting the given bound state, and we have defined the quantity $\Delta_7$ ($2\pi$ divided by the absolute value of the deficit angle of the singularity) which determines the periodicity of the phase transverse to it. In other words, encircling any such bound state counterclockwise, the transverse coordinate undergoes the phase shift
\be\label{RotationStack}
z_3\longrightarrow e^{2\pi {\rm i}/\Delta_7}z_3\,.
\ee
Clearly $\Delta_7=1$ corresponds to no 7-branes. The other allowed values are\footnote{They can be easily obtained using the fact that the quantity $2\cos(\frac{\pi N_7}{6})$ is the trace of the SL$(2,\ZZ)$ monodromy matrix, and thus it is an integer.} $\Delta_7=\tfrac65,\tfrac43,\tfrac32,2,3,4,6$. Recalling Eq.~\eqref{T2CoordRot}, the last four options correspond to orbifold limits of the local elliptic $K3$ of F-theory $(\mathbb{C}_{z_3}\times T^2_t)/\ZZ_k$.

One can combine 7-branes and $\mathcal{S}$-folds \cite{Apruzzi:2020pmv}, recalling that the latter does act as a quotient on the transverse directions (as well as on the longitudinal ones), as in Eq.~\eqref{SO(2)3}. Denoting by $\ell$ the order of the $\mathcal{S}$-fold, the combination of the latter with a given stack of 7-branes results in the relation $k=\ell \Delta_7=2,3,4,6$. In this way, also the first three bound states, when suitably combined with $\mathcal{S}$-folds, may be geometrized by elliptic fibrations of orbifold type. The resulting fourfold is a quotient of a CY fourfold, where the quotient acts simultaneously over fiber and base.

\subsection{$\mathcal{N}=1$ $\mathcal{S}$-folds}\label{N=1Sfold}

Prompted by a standard construction in F-theory \cite{Beasley:2008dc}, we can obtain a pretty straightforward $\mathcal{N}=1$ generalization of the $\mathcal{S}$-folds. Consider the following identification on the four local complex coordinates of the F-theory fourfold
\be\label{K3fiberedCY4}
(z_1,z_2,z,t)\;\sim\;\left(e^{{\rm i}(\Psi+2\pi/\ell)}\,z_1\,,\,e^{{\rm i}(\Psi-2\pi/\ell)}\,z_2\,,\,e^{{\rm i}(2\pi/k-2\Psi)}\,z_3\,,\,e^{-2\pi {\rm i}/k}\,t\right)\,,
\ee
which, at fixed $\ell,k$, depends on a single parameter $\Psi=\frac{\Psi_1+\Psi_2}{2}$,\footnote{A priori this parameter is an angle, and thus can be taken to be continuous. However, for the orbifold quotient \eqref{K3fiberedCY4} to make sense, $\Psi$ should be a rational multiple of $\pi$.} because we have set $\Psi_3=\frac{2\pi}{k}-2\Psi$ by imposing the CY condition. As we will see, this guarantees that the quotient preserves at least four supercharges. Except for certain specific values of $\Psi$, however, the fourfold structure is not that of a direct product of two $K3$s (or quotients thereof), but rather the local elliptic $K3$ with coordinates $\{z_3,t\}$ is non-trivially fibered over the K\"ahler (non-CY) surface, call it $S$, with local coordinates $\{z_1,z_2\}$, which we regard as the internal part of the 7-brane worldvolume. This can be seen by noting that $S$ has the following orbifold structure
\be\label{FibrationPicture}
S=\mathbb{C}^2/\Gamma\,,\qquad\Gamma\subset {\rm U}(2)\,,
\ee
i.e.~it is the quotient by a discrete subgroup of U$(2)$, and not of SU$(2)$ which would instead characterize orbifold limits of $K3$ surfaces. The embedding of $\Gamma$ into the central U$(1)$ of U$(2)$ is parametrized by $2\Psi$, which therefore determines the topology of the canonical bundle of $S$, $K_S$.\footnote{In a global context, the non-triviality of $K_S$ is typically accompanied by the non-triviality of the normal bundle of $S$ inside the base of the elliptic fibration. Here, instead, we stick to trivial normal bundles because we focus in the vicinity of a single 7-brane stack.} The CY condition of the fibration (and thus supersymmetry) is then guaranteed by the fact that the holomorphic $(2,0)$-form of the $K3$ fiber, which can be locally written as ${\rm d}z_3\wedge{\rm d}t$, transforms as a section of $K_S$, as can be seen from the orbifold-action assignments in Eq.~\eqref{K3fiberedCY4}. That supersymmetry is safe can also be realized by noting that the 7-brane stack can be made to wrap a local $K3$ surface by a topological twist on its tangent bundle. Since the normal coordinate $z_3$ is charged under $K_S$, one can kill the central-U$(1)$ part of the spin connection of $S$ by simply shifting it by (a suitable multiple of) the U$(1)$-$R$-symmetry connection. This gives us
\be
S_{\{z_1,z_2\}}\;\stackrel{\rm twist}{\Longrightarrow} \;{K3}_{\{\tilde{z}_1,\tilde{z}_2\}}\,\qquad{\rm with}\qquad (\tilde{z}_1,\tilde{z}_2)\;\sim\;(e^{2\pi {\rm i}/\ell}\,\tilde{z}_1,e^{-2\pi {\rm i}/\ell}\,\tilde{z}_2)\,.
\ee
The situation after the twist leaves us just with the $\mathcal{S}$-fold quotient, acting on the 7-brane worldvolume coordinates as a cyclic subgroup of SU$(2)$.

Using Eqs.~\eqref{SdualityOnQ} and \eqref{RsymmetryOnQ}, we can see that the quotient \eqref{K3fiberedCY4} acts on the supercharges of the parent $\mathcal{N}=4$ $D3$-brane theory as
\be\label{QuotientOnQ}
(Q^1,Q^2,Q^3,Q^4)\sim (Q^1\,,\,e^{{\rm i}\left(\Psi-\frac{2\pi}{k}(1-\Delta_7)\right)}Q^2\,,\,e^{{\rm i}\left(\Psi-\frac{2\pi}{k}(1+\Delta_7)\right)}Q^3\,,\,e^{-2{\rm i}\Psi}Q^4)\,.
\ee
As anticipated, the four supercharges $Q^1$ are always preserved for quotients of the form \eqref{K3fiberedCY4}, thanks to the CY condition. At fixed $\Psi$, all the options are labeled by the integer $\ell\geq1$ satisfying $\ell\Delta_7=2,3,4,6$. When $\Psi=0$ (mod $2\pi$), we recover the class of $\mathcal{N}=2$ $\mathcal{S}$-folds studied in \cite{Apruzzi:2020pmv}, preserving both $Q^1$ and $Q^4$. If one removes the 7-branes too (setting $\Delta_7=1$), $Q^2$ is also preserved and one goes back to the original $\mathcal{N}=3$ - preserving $\mathcal{S}$-folds (or $\mathcal{N}=4$ for $k=2$) \cite{Garcia-Etxebarria:2015wns}. 

When $\Psi=\pi$, again both $Q^1$ and $Q^4$ are preserved, but the ensuing $\mathcal{N}=2$ four-dimensional theory does not originate from the $\mathcal{N}=2$ $\mathcal{S}$-folds considered in \cite{Apruzzi:2020pmv}. Indeed, Eq.~\eqref{K3fiberedCY4} becomes
\be
(z_1,z_2,z_3,t)\;\sim\;\left\{\begin{array}{l}\left(e^{2\pi {\rm i}/\ell}\,z_1\,,\,e^{-2\pi {\rm i}/\ell}\,z_2\,,\,e^{2\pi {\rm i}/k}\,z_3\,,\,e^{-2\pi {\rm i}/k}\,t\right)\\ \\ (-z_1,-z_2,z_3,t) \,,\end{array}\right.
\ee
i.e.~there is an extra $\ZZ_2$ quotient on the 7-brane worldvolume coordinates. If we switch off the $\mathcal{S}$-fold, setting $\ell=1$, we land on the low-energy theories of $D3$ branes probing $\ZZ_2$ quotients of D$_4$ and E$_n$ stacks. See \cite{Giacomelli:2022drw} for an in-depth analysis of the probe theory of such $\mathcal{N}=2$ orbifolds.

Looking at Eq.~\eqref{QuotientOnQ}, we would be tempted to conclude that there are more $\mathcal{N}=2$ - preserving theories, besides those at $\Psi=0,\pi$: They would be at $\Psi=\frac{2\pi}{k}(1\pm\Delta_7)$. However, this is not quite correct. Geometrically, for these values of $\Psi$, the fourfold becomes (a quotient of) the direct product of two local $K3$s, like for $\Psi=0,\pi$, except that none of the two $K3$s is elliptically fibered over the plane transverse to the 7-branes.\footnote{For $\Psi=\frac{2\pi}{k}(1+\Delta_7)$ we have ${K3}_{\{z_1,z_3\}}\times{K3}_{\{z_2,t\}}$, whereas for $\Psi=\frac{2\pi}{k}(1-\Delta_7)$ we have ${K3}_{\{z_1,t\}}\times{K3}_{\{z_2,z_3\}}$. Note that for $\ell=1$ and $\ell=2$ these choices are equivalent.} As a consequence, the $\mathcal{N}=2$ superalgebra preserved by the background is not fully aligned with that preserved by the 3/7-branes. However, they still intersect on the $\mathcal{N}=1$ superalgebra generated by $Q^1$, thus preserving minimal supersymmetry. Full $\mathcal{N}=2$ alignment is instead recovered at $\Psi=\pi$.
\\

\subsection{General discussion}\label{GenDisc}

Let us combine all the details discussed so far in a uniform framework. Consider a stack of 7-branes transverse to $\mathbb{C}_{z_3}$, identified by the number $\Delta_7$, and call $\ell$ the order of the $\mathcal{S}$-fold, which must be such that $\ell\Delta_7=k=2,3,4,6$.

First focus on $\mathcal{N}=2$ theories. In this realm, we can envisage two somewhat ``extremal'' cases. On the one hand, we have the theories on $D3$ branes probing the $\mathcal{N}=2$ $\mathcal{S}$-folds considered in \cite{Apruzzi:2020pmv}, namely quotients (by cyclic groups) of the flat space, i.e.
\be
\left[\left(\mathbb{C}^2_{\{z_1,z_2\}}\right)\times\left(\mathbb{C}_{z_3}\times T^2_t\right)\right]\Big/\,\mathbb{Z}_h\,,
\ee
where $h={\rm LCM}(\ell,k)$. This space is defined by the following identification on its four local complex coordinates
\be\label{StandardN2S}
(z_1,z_2,z_3,t)\;\sim\;\left(e^{2\pi {\rm i}/\ell}\,z_1\,,\,e^{-2\pi {\rm i}/\ell}\,z_2\,,\,e^{2\pi {\rm i}/k}\,z_3\,,\,e^{-2\pi {\rm i}/k}\,t\right)\,.
\ee
On the other hand, we have the theories on $D3$ branes probing a 7-brane stack wrapping an $ADE$ orbifold of $\mathbb{C}^2$ \cite{Giacomelli:2022drw}. This instead corresponds to fourfolds with the geometry of a product. For simplicity, throughout the paper we restrict our attention to $A$-type orbifolds, and write such fourfolds as
\be\label{ProductN=2}
\left(\mathbb{C}^2_{\{z_1,z_2\}}\right)\Big/\,\mathbb{Z}_m\quad\times\;\;\,\left(\mathbb{C}_{z_3}\times T^2_t\right)\Big/\,\mathbb{Z}_k\,,
\ee
where $m$ is the order of the orbifold wrapped by the 7-brane stack. In terms of local coordinates, we have two independent identifications
\be
(z_1,z_2,z_3,t)\;\sim\;\left\{\begin{array}{l}\left(e^{2\pi {\rm i}/m}\,z_1\,,\,e^{-2\pi {\rm i}/m}\,z_2\,,\,z_3\,,\,t\right)\\ \\ \left(z_1\,,\,z_2\,,\,e^{2\pi {\rm i}/k}\,z_3\,,\,e^{-2\pi {\rm i}/k}\,t\right)\,. \end{array}\right.
\ee

Clearly, we can combine these two extremal cases into a single interpolating geometry, which therefore (modulo considering $D$ and $E$ quotients of $\mathbb{C}^2$) represents the most general orbifold geometry compatible with $\mathcal{N}=2$ supersymmetry:
\be
\left[\left(\mathbb{C}^2_{\{z_1,z_2\}}\right)\big/\,\mathbb{Z}_m\;\;\;\times\;\;\,\left(\mathbb{C}_{z_3}\times T^2_t\right)\right]\;\Big/\;\mathbb{Z}_h\,.
\ee
In terms of local coordinates, we have
\be\label{GeneralN=2}
(z_1,z_2,z_3,t)\;\sim\;\left\{\begin{array}{l}\left(e^{2\pi {\rm i}/m}\,z_1\,,\,e^{-2\pi {\rm i}/m}\,z_2\,,\,z_3\,,\,t\right)\\ \\ \left(e^{2\pi {\rm i}/\ell}\,z_1\,,\,e^{-2\pi {\rm i}/\ell}\,z_2\,,\,e^{2\pi {\rm i}/k}\,z_3\,,\,e^{-2\pi {\rm i}/k}\,t\right)\,. \end{array}\right.
\ee
Choosing $m=1$ we get the standard $\mathcal{N}=2$ $\mathcal{S}$-folds \eqref{StandardN2S}, whereas setting $\ell=1$ (no $\mathcal{S}$-fold) we land on the product geometries \eqref{ProductN=2}. Note, moreover, that the family \eqref{GeneralN=2} intersects the one (compactly) defined by \eqref{K3fiberedCY4} only at $m=1,\Psi=0$ and at $m=2,\Psi=\pi$.

Let us now move on to $\mathcal{N}=1$ theories. There are (at least) two ways to halve the supersymmetry preserved by the above construction. On the one hand, we can add extra 7-brane stacks, intersecting the one already present (i.e.~transverse to $\mathbb{C}_{z_1}$ or to $\mathbb{C}_{z_2}$). Eq.~\eqref{RsymmetryOnQ}, together with Eq.~\eqref{RotationStack}, tells us that a 7-brane stack transverse to $\mathbb{C}_{z_i}$ preserves the eight supercharges $\{Q^1,Q^{i+1}\}$, for $i=1,2,3$. Therefore, having at least two of them simultaneously breaks supersymmetry to the $\mathcal{N}=1$ superalgebra generated by $Q^1$. We will consider such situations in Section \ref{E-string}, to investigate the connection between flux compactifications the E-string theory to certain orbi-orientifold theories.

On the other hand, we can remain with just a single 7-brane stack (say, transverse to $\mathbb{C}_{z_3}$), but twist it in such a way as to make the canonical bundle of its internal worldvolume non-trivial. This adds an extra parameter, $\Psi$ (a rational multiple of $\pi$), controlling the embedding of the orbifold group inside the central U(1) of the U(2) holonomy group of the spin connection on the worldvolume. As discussed at length above, and summarized in \eqref{FibrationPicture}, this operation amounts to fibering the local $K3$ with coordinates $\{z_3,t\}$ over the surface $S$ locally parametrized by the complex coordinates $\{z_1,z_2\}$. This gives us a way of determining the most general Abelian orbifold geometry compatible with $\mathcal{N}=1$ supersymmetry:
\be\label{GeneralN=1}
\left[\begin{array}{lll}\left(\mathbb{C}_{z_3}\times T^2_t\right)&\longrightarrow&{\rm CY}_4\\ &&\;\;\downarrow\\ &&\;\;\mathbb{C}^2_{\{z_1,z_2\}}/\,\mathcal{A}_{m,\Psi}\end{array}\right]\;\;\Big/\;\mathbb{Z}_h\,,
\ee
where $\mathcal{A}_{m,\Psi}$ is an Abelian subgroup of U($2$), and the $\mathbb{Z}_h$ quotient acts on the total space of the fibration. In terms of local coordinates, we have three independent identifications:
\be\label{GeneralN=1bis}
(z_1,z_2,z_3,t)\;\sim\;\left\{\begin{array}{l}\left(e^{2\pi {\rm i}/m}\,z_1\,,\,e^{-2\pi {\rm i}/m}\,z_2\,,\,z_3\,,\,t\right)\\ \\ \left(e^{2\pi {\rm i}/\ell}\,z_1\,,\,e^{-2\pi {\rm i}/\ell}\,z_2\,,\,e^{2\pi {\rm i}/k}\,z_3\,,\,e^{-2\pi {\rm i}/k}\,t\right) \\ \\ \left(e^{{\rm i}\Psi}\,z_1\,,\,e^{{\rm i}\Psi}\,z_2\,,\,e^{-2{\rm i}\Psi}\,z_3\,,\,t\right)\,,\end{array}\right.
\ee
where the last relation is associated to the non-trivial fibration, making the base a non-CY orbifold of $\mathbb{C}^2$. If one wants to go beyond Abelian orbifolds, the base of the fibration \eqref{GeneralN=1} should be substituted by $\mathbb{C}^2_{\{z_1,z_2\}}/\Gamma$, with $\Gamma$ any non-Abelian finite subgroup of U(2). 

In what follows, we stick to the Abelian case, and check that indeed the parametrization \eqref{GeneralN=1bis} is general enough to accommodate any $\mathcal{N}=1$ - preserving Abelian orbifold. Indeed, any finite Abelian group $\mathcal{A}\subset$ U(2) is such that
\be\label{FinAbSub}
\mathcal{A}\;\subseteq\;\mathbb{Z}_m\times\mathbb{Z}_m\,,\qquad\quad m={\rm max}_{\gamma\in\mathcal{A}}\,\rm{ord}(\gamma)\,,
\ee 
which trivially follows from the fact that the most general element of $\mathcal{A}$ is a diagonal matrix of the form $\gamma={\rm diag}(e^{2\pi {\rm i}v_1/m},e^{2\pi {\rm i}v_2/m})$, with $v_1,v_2$ two integers. Hence, we can safely choose $\Psi=2\pi/m$, and be sure that any finite Abelian subgroup of U(2) we may want to quotient by will be a subgroup of the $\mathbb{Z}_m\times\mathbb{Z}_m$ used for the quotient \eqref{GeneralN=1bis}.

If $\mathcal{A}$ has two generators, it is of the form $\mathcal{A}\simeq\mathbb{Z}_{m_1}\times\mathbb{Z}_{m_2}$, and therefore \eqref{FinAbSub} is fulfilled by setting $m={\rm LCM}(m_1,m_2)$. If instead $\mathcal{A}\simeq\mathbb{Z}_n$, acting as $(z_1,z_2)\to(e^{2\pi {\rm i}v_1/n}z_1,e^{2\pi {\rm i}v_2/n}z_2)$, we can find its embedding \eqref{FinAbSub} simply by solving the integral system
\be\label{IntegrSyst}
\left\{\begin{array}{l}n(v_++v_-)=mv_1 \\ n(v_+-v_-)=mv_2 \end{array}\right.
\ee
for the unknowns $v_+,v_-$, which specify the action of the first and the second $\mathbb{Z}_m$ respectively.\footnote{As is clear from \eqref{GeneralN=1bis}, the first $\mathbb{Z}_m$ is inside the Cartan of U(2), whereas the second $\mathbb{Z}_m$ is inside the center of U(2).} At least one among $v_1$ and $v_2$ must share no divisor with $n$ (otherwise the order of $\mathcal{A}$ would be lower than $n$). Without loss of generality, let us say that $v_1$ has such a property, and thus set it to $1$. Calling $v_2\equiv n'$, the system \eqref{IntegrSyst} is then solved by
\be
v_\pm=(1+\theta)\,\frac{1\pm n'}{2}\,,
\ee
where
\be\label{solm}
m=(1+\theta)\,n\qquad {\rm and}\qquad \theta=\,\left\{\begin{array}{l}0\;\;n'\,{\rm odd}\\ 1\;\;n'\,{\rm even}\,.\end{array} \right.
\ee
Clearly, if $n$ is odd, $\theta$ can be set to zero, regardless of the parity of $n'$. Famously, when $n$ and $n'$ are coprime, the orbifolds $\mathbb{C}^2/\mathcal{A}$ are resolved by a linear array of $r$ rational curves with self-intersections $\{-c_i\}_{i=1,\ldots,r}$, satisfying
\be
\frac{n}{n'}\,=\,c_1-\frac{1}{c_2-\frac{1}{c_3-\cdots}}\,.
\ee

\subsection{Mass deformations}\label{MassDef}
In this section we briefly discuss mass deformations of the SCFTs engineered by probing the F-theory backgrounds described above, focusing in particular on the new features arising at $\mathcal{N}=1$, as opposed to $\mathcal{N}=2$ theories. For simplicity, consider the spaces defined by \eqref{GeneralN=1bis} with $m=\ell=1$. Generilizing to more complicated Abelian orbifolds of the base is straightforward.

The Casimir invariants of the flavor symmetry appear in this context as versal deformations of the Weierstrass model defining the F-theory elliptic fibration
\be\label{WeierstrassFth}
y^2=x^3+f(z)x+g(z)\,,
\ee
where $f,g$ are two polynomials depending on the base coordinate $z$ transverse to the 7-branes. The above equation describes a local elliptic $K3$, which for our purposes we will take to be homeomorphic to the orbifold $(\mathbb{C}\times T^2)/\ZZ_k$, with $k=2,3,4,6$, depending on the choice of $f$ and $g$ (see the table below)
\begin{center}\begin{tabular}{|c|c|c|}\hline
$k$ & $f$ & $g$  \\ \hline\hline $2$ & $z^2$ & $z^3$ \\ \hline $3$ & $-$ & $z^4$ \\ \hline $4$ & $z^3$ & $-$ \\ \hline $6$ & $-$ & $z^5$\\ \hline
\end{tabular}\end{center}

The holomorphic ($2,0$)-form of the space \eqref{WeierstrassFth} is given by
\be\label{Holo20}
\Omega^{(2,0)}=\frac{{\rm d}z\wedge{\rm d}x}{y}\,,
\ee
which is an invariant quantity under the $\ZZ_k$-orbifold action. To each coordinate is assigned a weight, which physically corresponds to the $\mathcal{N}=1$ R-charge,\footnote{As is manifest from Eq.~\eqref{RsymmetryOnQ} the R-charge is the diagonal combination of the three Cartan generators in \eqref{SO(2)3}.}, in such a way that \eqref{WeierstrassFth} is homogeneous and \eqref{Holo20} has weight $1$. All three coordinates $x,y,z$ are invariant under the action of the $\ZZ_k$ orbifold. Recalling Eq.~\eqref{RotationStack}, this fact implies that
\be
z=z_3^k\,,
\ee
which in turn fixes the weights as follows: $[z]=k$, $[x]=2(k-1)$, $[y]=3(k-1)$, $[f]=4(k-1)$, $[g]=6(k-1)$. Versal deformations appear as additional monomials in Eq.~\eqref{WeierstrassFth}, whose coefficients are Casimir invariants of the flavor symmetry. They are as many as the rank of this symmetry and are summarized in the table below, with the subscript indicating their weight
\begin{center}\begin{tabular}{|c|c|c|}\hline
$k$ & flavor & masses  \\ \hline\hline $2$ & $SO(8)$ & $M_2,M_4,M_4',M_6$  \\ \hline $3$ & $E_6$ & $M_2,M_5,M_6,M_8,M_9,M_{12}$  \\ \hline $4$ &  $E_7$ & $M_2,M_6,M_8,M_{10},M_{12},M_{14},M_{18}$ \\ \hline $6$ & $E_8$ & $M_2,M_8,M_{12},M_{14},M_{18},M_{20},M_{24},M_{30}$ \\ \hline
\end{tabular}\end{center}
Now, in the $\mathcal{N}=2$ case, the $M_d$'s must be constants. However, this is generically not the case for $\mathcal{N}=1$ theories, in the sense that they may get a dependence on $z_1,z_2$, the 7-brane internal-worldvolume coordinates. From the point of view of the field theory on the probe this couples the chiral fields corresponding to $z_1,z_2$ to the rest of the theory.

Let us see how the $K3$-fibering procedure previously described to halve the supersymmetry of the probe theory constrains the behavior of the flavor invariants $M_d$'s. To this end, it is convenient to rewrite \eqref{WeierstrassFth} and \eqref{Holo20} in the coordinates used in the above discussion, which means writing the elliptic fibration in an R-symmetry-invariant manner
\be\begin{array}{ll}
\tilde{y}^2=\tilde{x}^3+(1-\eta)\tilde{x}+\eta\,\\ \\ 
\Omega^{(2,0)}={\rm d}z_3\wedge{\rm d}t\,,
\end{array}\qquad\qquad\eta=\left\{\begin{array}{cl}{\rm any}& \quad SO(8) \\ 0 &\quad E_7\\ 1&\quad E_6, E_8  \end{array}\right.\ee
The new fiber coordinates have vanishing R-charge and are defined as follows
\be
\tilde{x}\equiv z^{-2\frac{k-1}{k}}x\,, \qquad \tilde{y}\equiv z^{-3\frac{k-1}{k}}y\,,\qquad{\rm d}t=z^{\frac{k-1}{k}}\frac{{\rm d}x}{y}\,.
\ee
The advantage of this different formulation is that it makes it easy to extract the behavior of the elliptic $K3$ and of its versal deformations once we fiber it over $S$. As we already remarked, supersymmetry demands $\Omega^{(2,0)}$ to transform as a section of $K_S$, meaning $\Omega^{(2,0)}\to e^{-2{\rm i}\Psi} \Omega^{(2,0)}$, which can be easily verified by looking at the last relation in \eqref{GeneralN=1bis}. Moreover, deformations now all appear in terms of the invariant combinations\footnote{Some of them will be multiplied by $\tilde{x}$, depending on their behavior under the $\ZZ_k$-orbifold action.}
\be
\frac{M_d(z_1,z_2)}{z_3^d}\,,\qquad\quad M_d(z_1,z_2)\to e^{-2d{\rm i}\Psi} M_d(z_1,z_2)\,.
\ee
In other words, if the $K3$-fibration determines a $\ZZ_n$ orbifold of $\mathbb{C}^3$, i.e.~$\Psi=2\pi/n$, then $M_d(z_1,z_2)$ is going to be a polynomial of degree $-2d$ mod $n$ in its variables. Therefore, only when $2d$ is a multiple of $n$ is it consistent to choose a constant $M_d$.

In the only perturbative situation, i.e.~$k=2$, we can directly reason in terms of the mass matrix $\mathcal{M}$ for the fundamental degrees of freedom, which transforms in the adjoint of the (unbroken) flavor group and behaves as
\be\label{MassMatrix}
\mathcal{M}(z_1,z_2)\, \longrightarrow \, e^{-4\pi {\rm i}/n}\mathcal{M}(z_1,z_2)
\ee
under the $\ZZ_n$-orbifold action, namely it is a polynomial of degree $n-2$ in its variables. This is a quantity that will manifest itself in the quiver description of these lagrangian theories, which we are now going to discuss.

\section{Perturbative focus}\label{Pert:Sec}

Some of the theories defined by the quotient \eqref{GeneralN=1bis} admit a description within perturbative string theory: They are those with $k=2$ and all involve orientifolds of some kind. They (together with other similar constructions) have been the subject of recent studies \cite{Bianchi:2020fuk,Antinucci:2020yki,Antinucci:2021edv} (see also \cite{Franco:2007ii,Garcia-Etxebarria:2012ypj,Bianchi:2013gka,Garcia-Etxebarria:2013tba,Garcia-Etxebarria:2015hua,Garcia-Etxebarria:2016bpb} for earlier works). In this section we review such orbi-orientifold setups focusing on the case of a single orbifold generator, and emphasize a number of key aspects, with the purpose of laying the ground to their connection with compactifications of 6d theories and possibly to their non-perturbative generalizations. We can distinguish two basic cases, which we discuss in turn in Section \ref{orbi-orient}: Either one has a stack of $D7$/$O7$ embedded in a CY orbifold of $\mathbb{C}^3$, i.e.~$\ell=1,\Delta_7=2$, or the CY orbifold of $\mathbb{C}^3$ is orientifolded by $O3$ planes i.e.~$\ell=2,\Delta_7=1$.\footnote{Combining these two cases, we obtain a $\ZZ_4$ quotient of the elliptic fiber, which locks us at strong coupling. This, however, does not contradict the fact that a weakly-coupled coexistence of $O3$ and $O7$ planes is possible, as shown e.g.~by taking two T-dualities of Type I with $D5$/$O5$. The reason is that in the latter case one has a $\ZZ_2\times\ZZ_2$ quotient of the F-theory fourfold or, in other words, that in the former case the $O3$ involution acts on the base of the elliptic fibration (as opposed to its double cover).} In Section \ref{AnomConf} we discuss the conditions for having a non-anomalous SCFT on the probe, and then describe in Section \ref{ExampleLoworder} \emph{all} the existing ones for orbifold order two. The cases with orbifold order three and four are detailed in Appendix \ref{ExampleLoworder1}.

\subsection{Orbi-orientifold backgrounds}\label{orbi-orient}

Let us begin by describing the orbifold quotient. Before starting, note that the finite subgroups of SU(3) by which we want to quotient the internal $\mathbb{C}^3$ are many more than those of U(2), needed for the quotient \eqref{FibrationPicture}. However, since we admit a 7-brane stack parallel to $z_1,z_2$, then we can only allow quotients by subgroups that do not mix $z_1,z_2$ with $z_3$. This leads us to consider only finite subgroups of U(2). In particular, finite Abelian subgroups, since they have a diagonalizable action, are the same for U(2) and SU(3), and we will only focus on them. Since here we restrict to a single generator, we are going to consider theories living on $D3$ branes probing a D$_4$ stack embedded in an orbifold $\mathbb{C}^3/\mathbb{Z}_n$, with $\mathbb{Z}_n\subset\;$SU(3). We now describe this family of theories in some detail, leaving the number of $D3$/$D7$ branes, and the orbifold action generic.\footnote{To retrieve this family from \eqref{GeneralN=1bis}, pick $\ell=1$, $k=2$, and $m$ as in \eqref{solm}.}

Consider the following $\mathbb{Z}_n$ quotient of the internal space $\mathbb{C}^3_{\{z_1,z_2,z_3\}}$
\be\label{orbifoldZnSu3}
z_a\;\sim\;e^{\frac{2\pi {\rm i}v_a}{n}}\,z_a\qquad\quad a=1,2,3\,,
\ee
where $\{v_a\}_a$ are integers modulo $n$ such that $\sum_av_a=0$. We also consider $M$ $D7$ branes, all stretched along the complex plane $z_3=0$. We probe this background with $N$ $D3$ branes,\footnote{When we will introduce the orientifold, we will adopt the convention such that $M$ and $N$ do not count orientifold images.} which, for generic values of the $v_a$'s, give rise at low energy to a $\mathcal{N}=1$ quiver gauge theory on their worldvolume. To begin with, we have to choose an embedding of $\mathbb{Z}_n$ into the Chan-Paton factors of both $D3$ and $D7$ branes, which essentially means splitting the stack into $n$ sub-stacks, each in correspondence with an irreducible representation of $\mathbb{Z}_n$. The easiest choice of phases is as follows \cite{Aldazabal:2000sa}
\be\label{CPembOrb}
\Gamma_3=\bigoplus_{p=0}^{n-1} e^{\frac{2\pi {\rm i} p}{n}} I_{N_p}\,,\qquad\qquad\Gamma_7=\bigoplus_{p=0}^{n-1} e^{\frac{2\pi {\rm i} p}{n}} I_{M_p}\,,
\ee
where $\{N_p\}_p$ and $\{M_p\}_p$ denote partitions of $N$ and $M$ respectively, and $p$ (defined modulo $n$) labels the irreducible representation. Let us discuss the massless spectrum of 3-3 and 3-7 strings after this orbifold projection, before introducing the orientifold one. Gauge bosons come from the Neveu-Schwarz sector of 3-3 strings, and invariance under the orbifold action requires
\be
A=\Gamma_3 A \Gamma_3^{-1}\,,
\ee
where the sum over the Chan-Paton indices is implicit. This clearly breaks the original $U(N)$ gauge symmetry as follows
\be
U(N)\,\longrightarrow\,\bigotimes_{p=0}^{n-1}\,U(N_p)\,.
\ee
Analogous conclusion holds for the flavor symmetry provided by the 7-7 sector:
\be
A^{D7}=\Gamma_7 A^{D7} \Gamma_7^{-1}\,.
\ee
Due to the identification \eqref{orbifoldZnSu3}, the orbifold projection on the complex scalars of the three chiral multiplets is instead
\be
e^{\frac{2\pi {\rm i} v_a}{n}}\Phi^a= \Gamma_3 \Phi^a\Gamma_3^{-1}\,,\qquad\quad {\rm no\, sum \,on\;\;} a=1,2,3\,,
\ee
which cuts the original $N^2$ chiral multiplets down to
\be\label{SpectrumAfterOrbifold3-3}
\bigoplus_{a=1}^3\bigoplus_{p=0}^{n-1}\left(N_p\,,\bar{N}_{p-v_a}\right)\,.
\ee
Analogous conclusion holds for the mass terms provided by the 7-7 sector:
\be\label{OrbifoldMass}
e^{\frac{2\pi {\rm i} v_3}{n}}\Phi^{D7}= \Gamma_7 \Phi^{D7}\Gamma_7^{-1}\,,
\ee
which is indeed compatible with Eq.~\eqref{MassMatrix}. In Eq.~\eqref{SpectrumAfterOrbifold3-3} $N_p$ denotes the fundamental representation of $U(N_p)$ and $\bar{N}_q$ the antifundamental of $U(N_q)$, and we call $\Phi^a_{p,p-v_a}$ the scalar field in the chiral multiplet transforming under the above summand. We conventionally choose strings stretching from node $p$ to node $q$ to transform under the fundamental of node $p$ times the antifundamental of node $q$. Open strings in the 3-7 sector contribute additional chiral multiplets in the $D3$-brane worldvolume theory. In particular, scalars arise again from the Neveu-Schwarz sector, which contains fermion zero-modes in the Dirichlet-Neumann directions (i.e.~the plane $\mathbb{C}^2_{\{z_1,z_2\}}$). The orbifold projections are
\be\label{ProjQtQ}
e^{\frac{\pi {\rm i}(v_1+v_2)}{n}}Q = \Gamma_3 Q \Gamma_7^{-1}\,,\qquad\quad e^{\frac{\pi {\rm i}(v_1+v_2)}{n}}\tilde{Q} = \Gamma_7 \tilde{Q} \Gamma_3^{-1}\,,
\ee
where $Q$ and $\tilde{Q}$ denote the scalars in the chiral multiplets of 3-7 and 7-3 strings, represented by $N\times M$ and $M\times N$ matrices respectively. The surviving chiral multiplets therefore are
\be\label{QtildeQ}
\bigoplus_{p=0}^{n-1}\left[\left(N_p\,,\bar{M}_{p+\tfrac12v_{3}}\right)\quad\oplus\quad\left(M_p\,,\bar{N}_{p+\tfrac12v_{3}}\right)\right]\,,
\ee
and we denote by $Q_{p,p+v_3/2}$ and $\tilde{Q}_{p,p+v_3/2}$ the respective scalar fields. Clearly this works only when $v_3$ is an even number. Of course, this is an actual restriction for even $n$ only.\footnote{In order to accommodate the cases with even $n$ and odd $v_3$, one simply chooses $e^{\pi {\rm i}/n}\Gamma_7$ for the $D7$ Chan-Paton embedding \cite{Aldazabal:2000sa}. We will encounter such cases too.} For the sake of simplicity, let us for the moment assume that $v_3$ is even. The field theory with the above spectrum has the following superpotential
\be\label{OrbW}
W_{\rm orb}=\sum_p\left[\epsilon_{abc}\Tr\left(\Phi^a_{p,p-v_a}\Phi^b_{p-v_a,p-v_a-v_b}\Phi^c_{p-v_a-v_b,p}\right)+\Tr\left(Q_{p,p+v_3/2}\Phi^3_{p+v_3/2,p-v_3/2}\tilde{Q}_{p-v_3/2,p}\right)\right]\hspace{-.1cm}.\,
\ee

\subsubsection*{$\bf O7$}

Now, let us introduce the orientifold $O7$$^-$ projection,\footnote{We will never consider the $O7$$^+$ plane, since it necessarily leads to an uncanceled 7-brane tadpole, and hence to a varying axio-dilaton, resulting in non-conformal field theories.} by fixing its embedding into the Chan-Paton factors. If we picture the stacks of fractional branes as the vertices of a regular $n$-polygon in $\mathbb{R}^2$ centered at the origin, we can visualize inequivalent orientifold involutions of the system as reflections in the dihedral group $\mathbb{D}_n$ modulo conjugation \cite{Giacomelli:2022drw}. Hence, for odd $n$ there is only a single orientifold involution up to equivalences, whereas for even $n$ there are two, according to whether one reflects with respect to an axis passing through vertices or edges. If we choose the former involution, we pick the representative that identifies the gauge/flavor node $p$ with the gauge/flavor node $n-p$ (or equivalently that identifies each representation of the orbifold group with its complex conjugate). This leads us to constrain the partitions of $N$ and $M$, imposing that $N_p=N_{n-p}$, $M_p=M_{n-p}$. If we choose, instead, the latter involution we pick the representative that identifies the gauge/flavor node $p$ with the gauge/flavor node $1-p$ and impose the corresponding constraints on the partitions of $N$ and $M$.

We focus for simplicity of exposition on the former involution, the one admitting fixed nodes. It is straightforward to carry over the following discussion for the other type of involution. The matrices acting on the Chan-Paton spaces are
\begin{eqnarray}
\Omega_3=\left(\begin{array}{cccccc}{\rm i}J_{2N_0}&0&\ldots&\ldots&0&0\\ 0&0&\ldots&\ldots&0&{\rm i}I_{N_1}\\ \vdots&\vdots&&&{\rm i}I_{N_2}&0\\  \vdots&\vdots&&\iddots&&\vdots\\ 0&0&-{\rm i}I_{N_2}&&&0\\0&-{\rm i}I_{N_1}&0&\ldots&\ldots&0 \end{array}\right)\qquad\Omega_7=\left(\begin{array}{cccccc}I_{M_0}&0&\ldots&\ldots&0&0\\ 0&0&\ldots&\ldots&0&I_{M_1}\\ \vdots&\vdots&&&I_{M_2}&0\\  \vdots&\vdots&&\iddots&&\vdots\\ 0&0&I_{M_2}&&&0\\0&I_{M_1}&0&\ldots&\ldots&0 \end{array}\right)\nonumber
\end{eqnarray}
where $J_{2N_0}$ is the symplectic matrix of size $2N_0$. The orientifold projection on the spectrum is as follows
\begin{eqnarray}\label{OmegaProjAPhiQ}
A=-\Omega_3A^T\Omega_3\,, && A^{D7}=-\Omega_7(A^{D7})^T\Omega_7\,,\\
\Phi^{a}=(-1)^{\delta_{a,3}}\,
\Omega_3\left(\Phi^a\right)^T\Omega_3\,, && \Phi^{D7}=\,-
\Omega_7\left(\Phi^{D7}\right)^T\Omega_7\,, \label{PhiOmega}
\\ &Q=\Omega_3\tilde{Q}^T\Omega_7\,,& 
\end{eqnarray}
where the transposition is there because the orientifold involution reverses the orientation of the string. This breaks the gauge and flavor groups respectively to
\begin{eqnarray}
{\rm Gauge}:&& \bigotimes_{p=1}^{\left[\frac{n-1}{2}\right]}U(N_p)\,\otimes \,USp\left(2N_0\right)\quad  \left[\otimes \;USp\left(2N_{n/2}\right)\;{\rm if\;}n \;{\rm even}\right]\,,\\
{\rm Flavor}:&&\bigotimes_{p=1}^{\left[\frac{n-1}{2}\right]}U(M_p)\,\otimes \,SO(2M_0)\quad  \left[\otimes \;SO(2M_{n/2})\;{\rm if\;}n \;{\rm even}\right]\,.
\end{eqnarray}

To determine the surviving spectrum of chiral fields, we reason in the following way. Let us first discuss the fields of type $\Phi$. Some of them are special because they connect gauge nodes which are the orientifold image of one another: At fixed value of $a$, this happens exactly when $2p=v_a$ (modulo $n$). Due to the first of Eq.~\eqref{PhiOmega}, in the quotient theory, the fields $\Phi^1$ and $\Phi^2$ transform in the antisymmetric representation of the node resulting from the identification of the two nodes (which is necessarily a node with a unitary group), whereas $\Phi^3$ transforms in the symmetric representation of that same node. Of course, when such a node is a $U(1)$, $\Phi^1$ and $\Phi^2$ are projected out, while $\Phi^3$ survives and has vanishing charge. To count how many $\Phi$ fields are left for each value of $a$, we need to know how many values of $p$ solve the equation $2p=v_a$. If $n$ is odd, there is obviously a single solution, whereas when $n$ is even there are zero or two solutions depending on whether $v_a$ is odd or even respectively. At any rate, at fixed $a$, there are several \emph{pairs} of $\Phi$ fields connecting nodes which are not orientifold images of one another, and the orientifold exchanges the two constituents of each such pair: The quotient theory contains only one combination of these two constituents. All in all, the surviving flavor-singlet chiral multiplets fill up the following representations
\begin{equation}\label{PhiFields}
\bigoplus_{a=1}^3\;\left\{\begin{array}{cc}\frac{1}{2}\;\bigoplus_{p=0}^{n-1}\left(N_p\,,\bar{N}_{p-v_a}\right)&\quad 2p\neq v_a\\ \\ \bigoplus_{p=0}^{n-1} \frac{N_p\left(N_p-(-1)^{\delta_{a,3}}\right)}{2}&\quad 2p=v_a\,,
\end{array}
\right.
\end{equation}
where the $\tfrac12$ in the first row just means that we need to retain only the invariant combination of a field/image-field pair. For the other type of orientifold involution, the one admitting no fixed nodes, the condition distinguishing the two lines in \eqref{PhiFields} becomes $2p=v_a+1$.

The treatment of the $Q,\tilde{Q}$ fields is easier. Again we need to single out the pairs appearing in Eq.~\eqref{QtildeQ} that are their own orientifold image: This happens exactly for $2p=-v_3/2$. For all the other values of $p$, the pair in \eqref{QtildeQ} is sent to a different pair by the orientifold involution, and the projection consists in retaining only the invariant combination of the two pairs. All in all, the surviving flavor-charged chiral multiplets fill up the following representations\footnote{In the special cases of $\mathcal{N}=2$ - preserving orbifolds ($v_3=0$), the second line in \eqref{QtildeQFields} includes the orientifold-image chiral fields $\left(M_p\,,\bar{N}_{p}\right)$ too, which make up a self-image hypermultiplet for every value of $p$.}
\begin{eqnarray}\label{QtildeQFields}
\cfrac{1}{2}\;\;\;\bigoplus_{p=0}^{n-1}\;\left[\left(N_p\,,\bar{M}_{p+\tfrac12v_{3}}\right)\;\oplus\;\left(M_p\,,\bar{N}_{p+\tfrac12v_{3}}\right)\right]&&\quad 2p\neq -\frac{v_3}{2}\nonumber \\ \\ \bigoplus_{p=0}^{n-1}\,\;\left(N_p\,,\bar{M}_{p+\tfrac12v_{3}}\right)&&\quad 2p= -\frac{v_3}{2}\,.\nonumber
\end{eqnarray}
For the other type of orientifold involution, the one admitting no fixed nodes, the condition distinguishing the two lines in \eqref{QtildeQFields} becomes $2p=-v_3/2+1$.

Finally, the superpotential \eqref{OrbW} will contain terms that are exchanged by the orientifold involution and terms which are instead their own image.

\subsubsection*{$\bf O3$}

Let us briefly discuss the other perturbative subclass of theories obtained by probing the background \eqref{GeneralN=1bis}. Here we consider no $D7$ branes, and the Chan-Paton embedding of the orbifold group $\Gamma_3$ is the same as in \eqref{CPembOrb}.  There are moreover two inequivalent actions of the orientifold involution on the Chan-Paton space, which, for the involution with fixed nodes, look precisely like $\Omega_7$ and $\Omega_3$ above, except that they both have gauge indices: Let us call them $\Omega_3^\pm$ respectively, indicating that they are associated to $O3$$^\pm$ planes. Gauge bosons and scalars behave as follows
\begin{eqnarray}
A\Omega_3^\pm&=&\pm (A\Omega_3^\pm)^T\,,\\
\Phi^a\Omega_3^\pm&=&\pm (\Phi^a\Omega_3^\pm)^T\,.
\end{eqnarray}
This means that fixed nodes have $USp,SO$ gauge groups respectively, and those bifundamental chiral fields that are their own image transform in the symmetric, antisymmetric representation of unitary nodes respectively.

Also in this case, for even values of $n$, involutions without fixed nodes are possible. The general rules dictating what are the allowed orbi-orientifold configurations can be derived using the brane-dimer techniques \cite{Franco:2005rj}. While referring the reader to the literature for a general treatment, we will explicitly discuss in Section \ref{ExampleLoworder} and in Appendix \ref{ExampleLoworder1} the case of Abelian orbifolds of low order.

\subsection{Anomalies and conformality}\label{AnomConf}

For generic choices of $\{N_p\}_p$ and $\{M_p\}_p$, the four-dimensional field theory will suffer from cubic non-Abelian triangle anomalies, as a consequence of the fact that the corresponding string construction has uncanceled tadpoles in the \emph{twisted} Ramond-Ramond sector \cite{Uranga:2000xp,Bianchi:2000de}.
While $USp$ and $SO$ gauge groups are free of anomalies, since all perturbatively-realized representations of theirs are self-conjugate, we must ensure cancelation of such anomalies for every $SU$ gauge group. Fixing the value of $p$ corresponding to a unitary gauge node, every chiral field transforming in the fundamental (antifundamental) representation contributes $+1$ ($-1$) to the anomaly, and every chiral field transforming in the symmetric (antisymmetric) representation contributes $N_p+4$ ($N_p-4$). The contribution are opposite if the fields are antichiral. 

On the contrary, mixed $U(1)$/non-Abelian anomalies cancel by a four-dimensional Green-Schwarz mechanism, mediated by the exchange of twisted Ramond-Ramond fields \cite{Ibanez:1998qp}. Moreover, anomalous $U(1)$'s are spontaneously broken and hence get a tree-level mass, disappearing from the low-energy spectrum, whereas non-anomalous $U(1)$ combinations are generally projected out by the orientifold. For these reasons, in what follows we will ignore all $U(1)$ factors in the gauge nodes.

We would like to look for superconformal quiver gauge theories with an exactly marginal coupling for each gauge node. This amounts to imposing the vanishing of the one-loop beta function corresponding to every gauge node.\footnote{We do not consider the possibility of having ``empty'' nodes, i.e.~nodes associated to the trivial gauge group.} The coefficient of the beta function reads\footnote{Since we are not seeking fixed points of the RG flow, we set all the anomalous dimensions to zero.}
\be
\beta \;\propto\; \sum_C\mathcal{I}_{\mathcal{R}_C} - 3 \mathcal{I}_{\rm Adj}\,,
\ee
where $\mathcal{I}_\mathcal{R}$ is the index of the representation $\mathcal{R}$ of the gauge group, and $C$ labels the chiral fields charged under the given gauge node. Let us list here the relevant indices for the set of theories at hand:
\begin{equation}\label{SCTcond}
SU(N): \left\{\begin{array}{l}\mathcal{I}_{\rm F}=\tfrac12\\ \\ \mathcal{I}_{A}=\tfrac N2-1\\ \\ \mathcal{I}_{S}=\tfrac N2+1\\ \\ \mathcal{I}_{\rm Adj}=N\end{array}\right. \quad \;\;USp(2N): \left\{\begin{array}{l}\mathcal{I}_{\rm F}=\tfrac12\\ \\ \mathcal{I}_{A}=N-1\\ \\ \mathcal{I}_{S\equiv{\rm Adj}}=N+1\end{array}\right.\quad \;\;SO(2N): \left\{\begin{array}{l}\mathcal{I}_{\rm F}=1\\ \\ \mathcal{I}_{S}=2N+2\\ \\ \mathcal{I}_{A\equiv{\rm Adj}}=2N-2\end{array}\right.
\end{equation}
\\
As expected, the sum of all the (suitably normalized) beta functions, whose associated gauge coupling coincide with the vev of the axio-dilaton, is always controlled by the net number of $D7$-brane charge present in the background, which vanishes for both the perturbative scenarios discussed above. However, the request that all beta functions vanish independently, when combined with the anomaly cancellation conditions, turns out to be generically too restrictive to be satisfied. There are indeed only  sporadic $\mathcal{N}=1$ SCFTs with such perturbative realizations in string theory.\footnote{This is in contrast to the $\mathcal{N}=2$ case, where both perturbative and non-perturbative solutions exist for every order of the orbifold, and can be classified \cite{Giacomelli:2022drw}.} We ignore whether there is a general criterion to decide when solutions to all the above conditions exist, or a physically-grounded rule to classify them. In the following we limit ourselves to list all of the existing ones for orbifold orders two, three, and four, with the aim of studying their interactions and mass deformations.

\subsection{Examples of conformal theories}\label{ExampleLoworder}

In this section we will apply the techniques reviewed before to construct all of the conformal theories arising from orientifolds and orbifolds of order two. We will continue with the systematic analysis of orbifolds of order three and four in Appendix \ref{ExampleLoworder1}.

First of all, let us remark that in the smooth case, there are only three conformal situations:
\begin{enumerate}
\item The $O3$$^{-}$ projection: $\mathcal{N}=4$ $SO(2N)$ SYM;
\item The $O3$$^{+}$ projection: $\mathcal{N}=4$ $USp(2N)$ SYM;
\item The $O7$$^{-}$ projection with $4$ $D7$s: $\mathcal{N}=2$ $USp(2N)$ SQCD with $1$ antisymmetric and $4$ fundamental flavors.
\end{enumerate}
The  $O7$$^{+}$ as well as all mixed projections yield non-conformal theories.

In the case of $\mathbb{C}^3/\ZZ_2$, the orbifold obviously involves only two of the three internal coordinates. First, let us discuss the cases without $D7$/$O7$. Here it is easy to see that there are two conformal configurations, both with $\mathcal{N}=2$ supersymmetry, which arise by taking a \emph{mixed} $O3$$^\pm$ projection: They are distinguished according to whether the two nodes are fixed or exchanged by the involution. In the first case, the mixed projection manifests itself at the level of the gauge groups, and the ensuing family of field theories is described by the following quiver
\begin{center}
\begin{tikzpicture}[thick, scale=0.8]
  \node[circle, draw, inner sep= 1pt](L1) at (6,-4){$USp(2N)$};
  \node[circle, draw, inner sep= 1pt](L2) at (12,-4){$SO(2N+2)$};
 \path[every node/.style={font=\sffamily\small,
  		fill=white,inner sep=1pt}]
(L1) edge  (L2);
\end{tikzpicture}\vspace{-.9cm}
\end{center}\be\label{ZZ2O3} \ee\\
where the link represents a hypermultiplet in the bifundamental representation. In the second case, the mixed projection manifests itself at the level of the matter fields, and we get the following quiver\footnote{Strictly speaking the gauge group would be $U(N)$, but since we are interested in the interacting SCFT in the IR, we will always disregard the IR-free $U(1)$.}
\begin{center}
\begin{tikzpicture}[thick, scale=0.8]
  \node[circle, draw, inner sep= 2pt](L1) at (10,-2){$SU(N)$};
   \path[every node/.style={font=\sffamily\small,
  		fill=white,inner sep=1pt}]
(L1) edge [loop, in=20, out=60, looseness=7] node[above=2mm] {$A$}  (L1)
(L1) edge [loop, in=-30, out=-70, looseness=6] node[below=1.5mm] {$S$}  (L1);
\end{tikzpicture}\vspace{-1cm}\be\label{Z2MixedQuiver2}\ee
\end{center}
where the two loops represent two hypermultiplets, one in the antisymmetric and one in the symmetric representation of the gauge group.\\

If instead we consider the $O7$ plane, and consequently place on top of it the 4 $D7$ branes needed to keep the axio-dilaton constant, we have to decide whether the complex line not affected by the orbifold projection is trasverse or longitudinal to the $D7$/$O7$ stack. The first scenario leads to a number of families of $\mathcal{N}=2$ SCFTs, which have recently been analyzed in depth in \cite{Giacomelli:2022drw}. The second scenario, instead, breaks supersymmetry to $\mathcal{N}=1$. Let us discuss this one here in detail.

Consider the following orbifold action on the three internal complex coordinates
\be\label{Z2storto}
(z_1,z_2,z_3)\;\sim\;\left(z_1,-z_2,-z_3\right)\,,
\ee
and place a $D7$/$O7$$^-$ stack at $z_3=0$. Recalling the general discussion of Section \ref{orbi-orient}, the treatment of the 3-7 sector is a bit trickier: We need to choose the orbifold action on the $D3$ and $D7$ Chan-Paton spaces respectively as $\Gamma_3={\rm diag}(I_{N_0},-I_{N_1})$ and $\Gamma_7={\rm diag}({\rm i}\,I_{M_0},-{\rm i}\,I_{M_1})$. This leads to the following 3-7 and 7-3 fields surviving the orbifold projection \eqref{ProjQtQ}
\be
Q=\left(\begin{array}{cc}0&Q_{01}\\ Q_{10}&0\end{array}\right)\,,\qquad \tilde{Q}=\left(\begin{array}{cc}\tilde{Q}_{00}&0\\ 0&\tilde{Q}_{11}\end{array}\right)\,.
\ee
At this point we need to specify the orientifold projection. It turns out that, in order for the orbifold-projected spectrum to be orientifold invariant, we are forced to choose a \emph{mixed} orientifold involution, which treats differently gauge and flavor nodes. Like in the $\mathcal{N}=2$ case, there are again two inequivalent choices. Let us discuss them in turn.

One choice is that the two gauge nodes are fixed, while the flavor nodes are the orientifold image of one another, i.e.
\be
\Omega_3=\left(\begin{array}{cc}{\rm i}J_{2N_0}&0\\ 0&{\rm i}J_{2N_1}\end{array}\right)\,,\qquad \hat{\Omega}_7=\left(\begin{array}{cc}0&I_{M_0}\\ I_{M_0}&0\end{array}\right)\,.
\ee
Using Eq.~\eqref{OmegaProjAPhiQ}, we deduce that the gauge group is broken to $USp(2N_0)\times USp(2N_1)$, while the flavor group to $U(M_0)$; moreover the surviving spectrum comprises two chiral multiplets $A^0,A^1$ transforming both in the antisymmetric representation of $USp(2N_0),USp(2N_1)$ respectively, two bifundamental chiral multiplets $\Phi_{01}^2,\Phi_{10}^3$ transforming in the $(2N_0,2N_1),(2N_1,2N_0)$ respectively,\footnote{We have disregarded $\Phi_{10}^2,\Phi_{01}^3$, because the orientifold connects them to $\Phi_{01}^2,\Phi_{10}^3$, as $\Phi_{10}^2=-J_{N_1}(\Phi_{01}^2)^TJ_{N_0}$ and $\Phi_{01}^3=J_{N_0}(\Phi_{10}^3)^TJ_{N_1}$.} and finally $M_0$ pairs of fundamental chiral multiplets $Q_{01},\tilde{Q}_{11}$ transforming in the $2N_0,2N_1$ respectively.\footnote{We have disregarded $Q_{10},\tilde{Q}_{00}$, because the orientifold connects them to $\tilde{Q}_{11},Q_{01}$, as $Q_{10}={\rm i}J_{N_1}(\tilde{Q}_{11})^T$ and $\tilde{Q}_{00}=-{\rm i}(Q_{01})^TJ_{N_0}$.}

Since there appear no complex gauge representations, there are no anomalies in this theory. Using \eqref{SCTcond}, it is immediate to see that the vanishing of the sum of the two beta functions imposes $M_0=8$,\footnote{This seems in contradiction with $D7$-brane tadpole cancelation, which would require the presence of $4$ $D7$ branes (and $4$ image $D7$ branes). The reason we get twice this amount will be clear momentarily.} whereas the vanishing of the difference of the two beta functions sets $N_0=N_1$ (call it $N$). Therefore we get a one-parameter family of $\mathcal{N}=1$ superconformal field theories, described by the following quiver
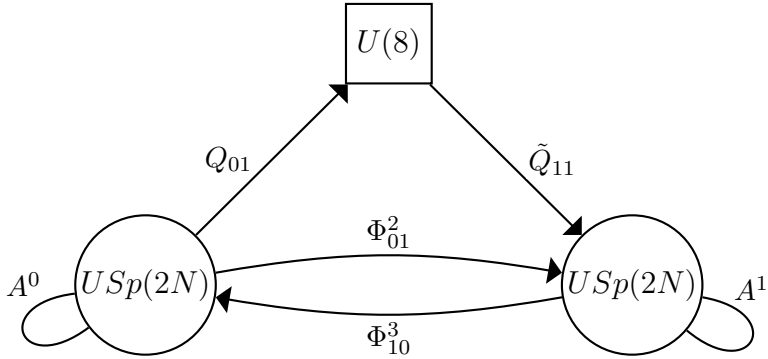
\begin{center}
\begin{tikzpicture}[thick, scale=0.8]
  \node[circle, draw, inner sep= 1pt](L1) at (6,-4){$USp(2N)$};
  \node[circle, draw, inner sep= 1pt](L2) at (14,-4){$USp(2N)$};
\node[draw, rectangle, minimum width=30pt, minimum height=30pt](L5) at (10,0){$U(8)$};
 \path[every node/.style={font=\sffamily\small,
  		fill=white,inner sep=1pt}]
(L1) edge [bend left=10, ->] node[above=0.5mm] {$\Phi^{2}_{01}$} (L2)
(L2) edge [bend left=10, ->] node[below=0.5mm] {$\Phi^{3}_{10}$} (L1)
(L5) edge [->] node[right=2.5mm] {$\tilde{Q}_{11}$} (L2)
(L1) edge [->] node[left=2.1mm] {$Q_{01}$} (L5)
(L1) edge [loop, in=-143, out=-173, looseness=6] node[above=4.5mm] {$A^0$} (L1)
(L2) edge [loop, in=-10, out=-40, looseness=6] node[above=4.8mm] {$A^1$}  (L2);
\end{tikzpicture}\vspace{-.5cm}
\end{center}\be\label{Z2MixedQuiver}\ee
There is also a superpotential containing, as usual, as many cubic terms as there are closed paths in the above quiver. Hence we have
\be\label{WZ2MixedQuiver}
W=\Tr_{\rm L}(A^0\Phi^2_{01}\Phi^3_{10})-\Tr_{\rm R}(A^1\Phi^3_{10}\Phi^2_{01})+\Tr_{\rm F}(\tilde{Q}_{11}\Phi^3_{10}Q_{01})\,,
\ee
where Tr$_{\rm L,R,F}$ means trace over the indices of the left gauge group, right gauge group, and flavor group respectively. Note that $\Phi^2_{01}$ is \emph{not} coupled to the matter in the fundamental because there is only one $D7$-brane stack, transverse to the direction $z_3$.

A very similar class of theories was discussed in Section 3 of \cite{Kim:2017toz} and named ``the basic theory'', in the context of flux torus compactifications of the E-string theory. However, contrary to the present case, in that context one has a superpotential term for \emph{each} triangle of the quiver. Actually from a field-theoretic standpoint there is a very good reason to include this extra term as well, since it leads to a one-dimensional conformal manifold. 
We will have more to say about these configurations and their connection to the E-string in Section \ref{E-string}, where we show that we can recover the desired superpotential term simply by tilting the orientifold plane and the 7-brane stack.

The second option we have is to choose an orientifold involution that does the opposite of what we just described, namely it interchanges the two gauge nodes and fixes the two flavor ones. This is implemented on the respective Chan-Paton spaces by the maps
\be
\hat{\Omega}_3=\left(\begin{array}{cc}0&{\rm i}I_{N_0}\\ -{\rm i}I_{N_0}&0\end{array}\right)\,,\qquad \Omega_7=\left(\begin{array}{cc}I_{2M_0}&0\\ 0&I_{2M_1}\end{array}\right)\,.
\ee
Again, using Eq.~\eqref{OmegaProjAPhiQ}, we deduce that the gauge group is broken to $SU(N)$, while the flavor group to $SO(2M_0)\times SO(2M_1)$; moreover the surviving spectrum comprises a chiral field in the adjoint of the gauge group, a pair of chiral/antichiral fields in the antisymmetric, a pair of chiral/antichiral fields in the symmetric, and finally $2M_1$ fundamentals and $2M_0$ antifundamentals. The last two sets of fields generate a gauge anomaly, unless they balance each other, i.e.~$M_0=M_1$. Moreover, using  \eqref{SCTcond}, we see that insisting on conformality requires to set $M_0=M_1=0$, i.e.~no $D7$ branes! Therefore we get back the one-parameter family of $\mathcal{N}=2$ superconformal field theories, described by the quiver \eqref{Z2MixedQuiver2}.

The absence of $D7$ branes in this second configuration as opposed to the presence of twice the expected number of $D7$ branes in the first situation \eqref{Z2MixedQuiver} leads us to argue that these two inequivalent orientifold involutions actually have a common origin in a T-dual Type IIA setting, being associated to the two $O7$$^{-}$ planes arising at the fixed points of the T-duality circle from an $O8$$^{-}$ plane wrapping the circle. The presence of the orbifold in Type IIB is traced back to the presence of $NS5$ branes in Type IIA, which separate into intervals the $D4$ branes where the field theory lives. All of this has a nice M-theory lift, with $M5$ branes wrapping a smooth co-dimension $4$ manifold inside an $M9$ wall. The following table summarizes the various dual pictures:
\begin{center}
\begin{tabular}{c|ccccccccccc}
&0&1&2&3&4&5&6&7&8&9&10\\ \hline  $M5$ &$\times$&$\times$&$\times$&$\times$&$\circ$&$\circ$&$ $&$ $&&$\circ$&$\circ$\\ $M9$ &$\times$&$\times$&$\times$&$\times$&$\times$&$\times$&$\times$&$\times$&&$\times$&$\times$\\ \hline $D4$ &$\times$&$\times$&$\times$&$\times$& & &&&& $\times$& \\ $NS5$ &$\times$&$\times$&$\times$&$\times$&$\times$&$\times$ &&&& & \\ $D8$/$O8$ &$\times$&$\times$&$\times$&$\times$&$\times$ & $\times$&$\times$&$\times$&& $\times$&\\ \hline $D3$ &$\times$&$\times$&$\times$&$\times$& & &&&& & \\ Orb. &&&&&& &$\times$&$\times$&$\times$& $\times$& \\ $D7$/$O7$ &$\times$&$\times$&$\times$&$\times$&$\times$ & $\times$&$\times$&$\times$&& $\bullet$--$\bullet$&\\ \hline $D4$ &$\times$&$\times$&$\times$&$\times$& & &&$\times$&& & \\ $NS5$ &$\times$&$\times$&$\times$&$\times$&$\times$&$\times$ &&&& & \\ $D6$/$O6$ &$\times$&$\times$&$\times$&$\times$&$\times$ & $\times$&$\times$&$\bullet$--$\bullet$&& $\bullet$--$\bullet$&
\end{tabular}
\end{center}
Directions $4,5$ correspond to the coordinate $z_1$, $6,7$ to $z_2$, and $8,9$ to $z_3$. Crosses mean that the object fills the corresponding direction completely. Circles mean that the $M5$ branes wrap a smooth holomorphic curve inside $\mathbb{R}^4_{4\,5\,9\,10}$. Bullets separated by lines mean that there are $Op$ planes at the extrema of an interval in the corresponding direction. We pass from the first to the second block by reducing along the M-theory circle $S^1_{10}$, and from the second to the third block by T-dualizing along $S^1_{9}$. In the Type IIA setting we have a tadpole-free stack composed of an $O8$$^{-}$ plane and $8$ pairs of $D8$/image-$D8$ branes. After T-duality, in the absence of Wilson lines for the gauge field on the $D8$ branes, the $O8$$^{-}$ plane gives rise to two $O7$$^{-}$ planes at the end points of an interval, whereas the $8$ $D8$/image-$D8$ pairs dualize to $8$ $D7$/image-$D7$ pairs stack on top of one of the two $O7$ planes. The $D3$-brane probes can only ``see'' one of these two end points at a time. It is then natural to associate the two inequivalent families of quiver gauge theories seen earlier as the field theories arising at low energies when the probe approaches either one of the two end points. The last block in the above table represents yet another Type IIA frame, which can be obtained from the Type IIB picture by T-dualizing along $S^1_{7}$. This time we have $D6$ branes and $O6$ planes, and the two inequivalent orientifold projections are seen as associated to the two halves in which each $O6$ plane splits when crossing an $NS5$ brane along the direction $6$. Such configurations, named ``forks'', were analyzed in detail in \cite{Park:1999eb}.

\section{From 6d theories on a torus to {\bf $D3$} branes in Type IIB}\label{sec4}

In \cite{Giacomelli:2020gee, Giacomelli:2022drw} it was argued that there is a close connection between $\mathcal{N}=2$ SCFTs engineered by probing with $D3$ branes a background involving orbifolds/$\mathcal{S}$-folds and 7-branes on one side and the $T^2$ compactification of certain 6d $\mathcal{N}=(1,0)$ theories on the other side. More precisely, in \cite{Giacomelli:2020gee} it was found that $D3$ branes probing $\mathcal{N}=2$ $\mathcal{S}$-folds lead to the same 4d theories one gets by compactifying 6d orbi-instanton theories on $T^2$, provided that one introduces certain holonomies along the cycles of the torus to account for the $\mathcal{S}$-fold. If one instead considers as in \cite{Giacomelli:2022drw} $D3$ branes probing an orbifold (with or without 7-branes), the SCFT one gets is a relevant deformation of the 6d theory compactified on $T^2$, where the RG flow is triggered by a mass deformation. In this section we would like to understand how this correspondence generalizes to theories with four supercharges. The answer we find is that 6d $\mathcal{N}=(1,0)$ theories on a torus with flux (introduced to break to minimal supersymmetry in 4d) are related to $D3$ branes probing a background preserving four supercharges in Type IIB. In general the Type IIB theory is obtained via an RG flow (triggered by a relevant deformation) starting from the 6d theory on $T^2$. For specific choices of flux for the 6d theory, the corresponding $D3$-brane theory probes a system of orbifolds and 7-branes as opposed to a more general CY space. We will focus on these examples, since we are able to provide quantitative checks for our claim, although recent results in the literature \cite{Bah:2021iaa} suggest that the correspondence is more general. We will discuss compactifications of various types of SCFTs engineered in M-theory by probing orbifolds and an $M9$ wall with $M5$ branes. We start with class $\mathcal{S}_k$ theories ($M5$ branes probing a $\mathbb{C}^2/\mathbb{Z}_k$) and then consider E-string theories ($M5$s inside an $M9$ wall) and the simplest example of orbi-instanton theories, in which the $M5$ branes are probing the $M9$ wall wrapped on a $\mathbb{C}^2/\mathbb{Z}_2$ orbifold.

\subsection{Class $\mathcal{S}_k$ on a torus with flux and {\bf $D3$} branes on orbifolds} 

In this section we discuss compactifications of the 6d SCFT living on the worldvolume of $N$ $M5$ branes probing a orbifold singularity $\mathbb{C}^2/\mathbb{Z}_k$ in M-theory. We will refer to this theory as $\mathcal{T}_{N,k}$. It has $SU(k)\times SU(k)\times U(1)$ global symmetry. Here we will consider compactifications to 4d leading to the so-called class $\mathcal{S}_k$ introduced in \cite{Gaiotto:2015usa}, in which only the Cartan subgroup of the 6d global symmetry is preserved.
Let us start from a class $\mathcal{S}_k$ trinion (with two full and one minimal punctures). This is a free theory of the form 
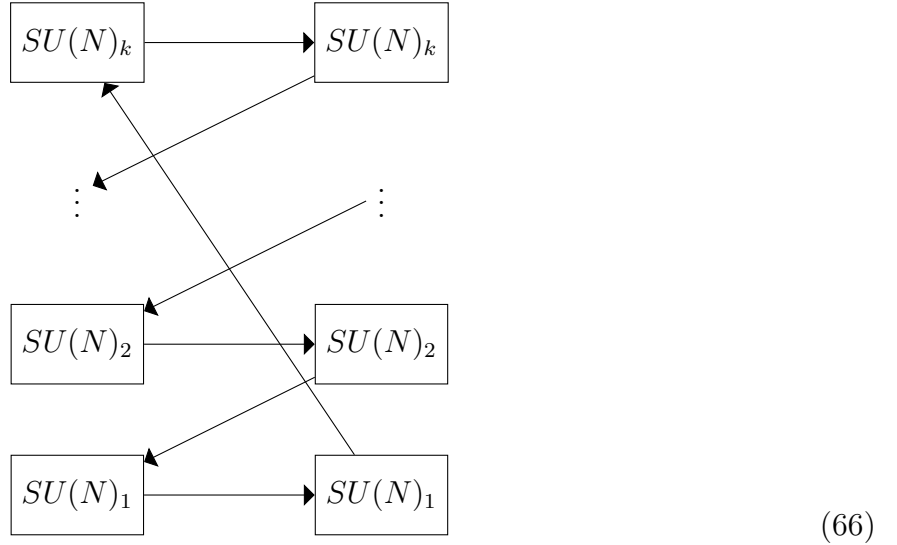
\begin{equation}\label{trinion1}
\begin{tikzpicture} 
\node[draw, rectangle, minimum width=30pt, minimum height=30pt](A1) at (0,0){$SU(N)_k$};
\node[draw, rectangle, minimum width=30pt, minimum height=30pt](B1) at (4,0){$SU(N)_k$}; 
\node[](A2) at (0,-2){$\vdots$};
\node[](B2) at (4,-2){$\vdots$};
\node[draw, rectangle, minimum width=30pt, minimum height=30pt](A3) at (0,-4){$SU(N)_{2}$};
\node[draw, rectangle, minimum width=30pt, minimum height=30pt](B3) at (4,-4){$SU(N)_{2}$};
\node[draw, rectangle, minimum width=30pt, minimum height=30pt](A4) at (0,-6){$SU(N)_1$};
\node[draw, rectangle, minimum width=30pt, minimum height=30pt](B4) at (4,-6){$SU(N)_1$};

\draw[->] (A1)--(B1); 
\draw[->] (B1)--(A2); 
\draw[->] (A3)--(B3); 
\draw[->] (B2)--(A3); 
\draw[->] (B3)--(A4);
\draw[->] (A4)--(B4);
\draw[->] (B4)--(A1);   
\end{tikzpicture}
\end{equation}
where we have $2k$ $SU(N)$ flavor symmetry groups coming from the two full punctures. The minimal puncture instead contributes a $U(1)$ global symmetry. Actually in \eqref{trinion1} there are $2k$ $U(1)$ symmetries under which the bifundamental chirals are charged: One comes from the minimal puncture, as we have just mentioned, and under it all horizontal arrows in \eqref{trinion1} have charge 1, whereas all diagonal arrows have charge $-1$. There is a second one, which we call $U(1)_t$, inherited from the Abelian factor appearing in the 6d global symmetry and all the bifundamentals have charge 1 under it. Finally, we have the remaining $2k-2$ Abelian factors, that come from the non-Abelian part of the global symmetry of the 6d SCFT, which we name $SU(k)_\beta\times SU(k)_\gamma$.  Our convention is that the $j$-th horizontal chiral has charge one under $U(1)_{\beta_j}$ whereas the $j$-th diagonal chiral is charged under $U(1)_{\gamma_j}$. The $k$-th horizontal chiral has charge $-1$ under all $\beta$ symmetries while the $k$-th diagonal bifundamental has charge $-1$ under all $\gamma$ symmetries. We can connect these trinions together and form spheres with multiple minimal punctures via gluing, which is performed by gauging the diagonal $SU(N)^k$ symmetry (denoted by circles in \eqref{gluing}) and adding extra chiral multiplets (denoted in red in \eqref{gluing}) and cubic superpotential terms\footnote{Here we are considering the $\Phi$-gluing of \cite{Gaiotto:2015usa, Razamat:2016dpl}.} corresponding to all closed triangles in \eqref{gluing} (see \cite{Gaiotto:2015usa} for the details): 
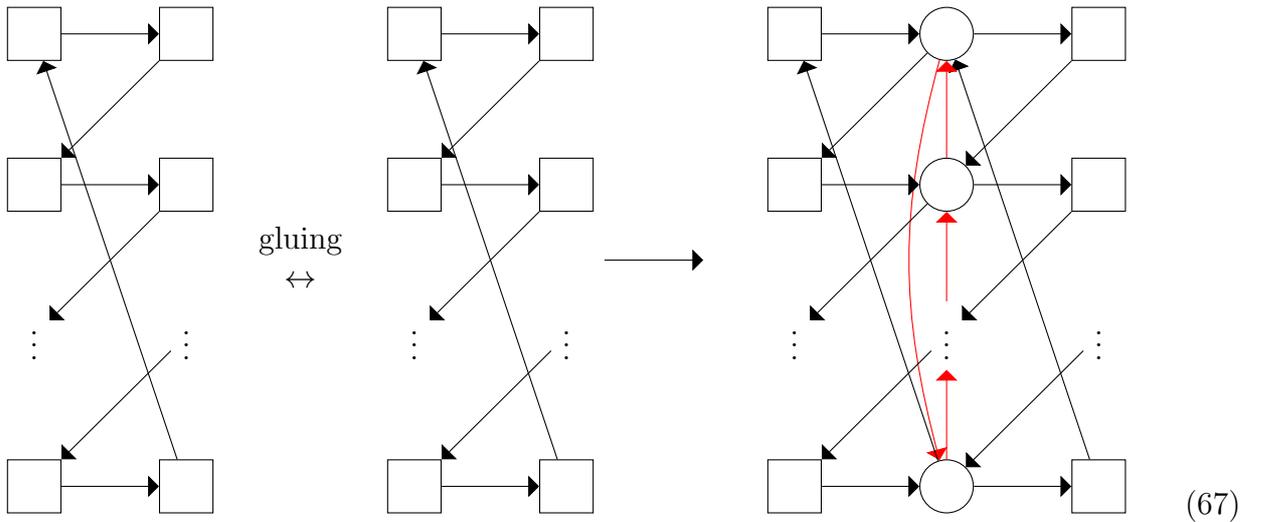
\begin{equation}\label{gluing}
\begin{tikzpicture} 
\node[draw, rectangle, minimum width=20pt, minimum height=20pt](A1) at (0,0){};
\node[draw, rectangle, minimum width=20pt, minimum height=20pt](B1) at (2,0){}; 
\node[draw, rectangle, minimum width=20pt, minimum height=20pt](A2) at (0,-2){};
\node[draw, rectangle, minimum width=20pt, minimum height=20pt](B2) at (2,-2){};
\node[](A3) at (0,-4){$\vdots$};
\node[](B3) at (2,-4){$\vdots$};
\node[draw, rectangle, minimum width=20pt, minimum height=20pt](A4) at (0,-6){};
\node[draw, rectangle, minimum width=20pt, minimum height=20pt](B4) at (2,-6){};

\draw[->] (A1)--(B1); 
\draw[->] (B1)--(A2); 
\draw[->] (A2)--(B2); 
\draw[->] (B2)--(A3); 
\draw[->] (B3)--(A4);
\draw[->] (A4)--(B4);
\draw[->] (B4)--(A1);  

\node[] at (3.5,-3) {$\begin{array}{c}\text{gluing}\\ \leftrightarrow\\\end{array}$}; 

\node[draw, rectangle, minimum width=20pt, minimum height=20pt](C1) at (5,0){};
\node[draw, rectangle, minimum width=20pt, minimum height=20pt](D1) at (7,0){}; 
\node[draw, rectangle, minimum width=20pt, minimum height=20pt](C2) at (5,-2){};
\node[draw, rectangle, minimum width=20pt, minimum height=20pt](D2) at (7,-2){};
\node[](C3) at (5,-4){$\vdots$};
\node[](D3) at (7,-4){$\vdots$};
\node[draw, rectangle, minimum width=20pt, minimum height=20pt](C4) at (5,-6){};
\node[draw, rectangle, minimum width=20pt, minimum height=20pt](D4) at (7,-6){};

\draw[->] (C1)--(D1); 
\draw[->] (D1)--(C2); 
\draw[->] (C2)--(D2); 
\draw[->] (D2)--(C3); 
\draw[->] (D3)--(C4);
\draw[->] (C4)--(D4);
\draw[->] (D4)--(C1);

\draw[->] (7.5,-3)--(8.8,-3);  

\node[draw, rectangle, minimum width=20pt, minimum height=20pt](a1) at (10,0){};
\node[draw, circle, minimum width=20pt](b1) at (12,0){};
\node[draw, rectangle, minimum width=20pt, minimum height=20pt](c1) at (14,0){}; 
\node[draw, rectangle, minimum width=20pt, minimum height=20pt](a2) at (10,-2){};
\node[draw, circle, minimum width=20pt](b2) at (12,-2){};
\node[draw, rectangle, minimum width=20pt, minimum height=20pt](c2) at (14,-2){};
\node[](a3) at (10,-4){$\vdots$};
\node[](b3) at (12,-4){$\vdots$};
\node[](c3) at (14,-4){$\vdots$};
\node[draw, rectangle, minimum width=20pt, minimum height=20pt](a4) at (10,-6){};
\node[draw, circle, minimum width=20pt](b4) at (12,-6){};
\node[draw, rectangle, minimum width=20pt, minimum height=20pt](c4) at (14,-6){};
\path[every node/.style={font=\sffamily\small,
  		fill=white,inner sep=1pt}]
(b1) edge [bend right=15,->, color=red]  (b4);

\draw[->] (a1)--(b1); 
\draw[->] (b1)--(c1);
\draw[->] (b1)--(a2); 
\draw[->] (c1)--(b2); 
\draw[->] (a2)--(b2); 
\draw[->] (b2)--(c2); 
\draw[->] (b2)--(a3); 
\draw[->] (c2)--(b3);
\draw[->] (b3)--(a4);
\draw[->] (c3)--(b4);
\draw[->] (a4)--(b4);
\draw[->] (b4)--(c4);
\draw[->] (b4)--(a1);
\draw[->] (c4)--(b1); 
\draw[->, color=red] (b4)--(b3); 
\draw[->, color=red] (b3)--(b2);
\draw[->, color=red] (b2)--(b1);  

\end{tikzpicture}
\end{equation}
If we glue together $nk$ of these building blocks forming a circular quiver,\footnote{The fact that the number of trinions should be a multiple of $k$ has to do with the concept of color of a full puncture as discussed in \cite{Gaiotto:2015usa}. This constraint is necessary if we want to preserve all the $2k$ Abelian symmetries. See \cite{Bah:2017gph} for a discussion about gluings violating this constraint.} the resulting 4d theory describes the 6d $\mathcal{N}=(1,0)$ SCFT $\mathcal{T}_{N,k}$ compactified on a torus with $nk$ minimal punctures. 
We can now notice that the same 4d gauge theory can be realized by probing with $N$ $D3$ branes a orbifold $\mathbb{C}^3/\Gamma$ with $\Gamma=\mathbb{Z}_k\times \mathbb{Z}_{nk}$. The orbifold action on the $\mathbb{C}^3$ coordinates is 
$$\mathbb{Z}_k:\; (\omega_k,\omega_k^{-1},1)\,;\qquad \mathbb{Z}_{nk}:\; (1,\omega_{nk},\omega_{nk}^{-1})\,,$$ 
where $\omega_k$ denotes indeed the $k$-th root of unity. 

Now we want to remove the punctures on the torus and trade them for a flux (meaning we change the Chern class of the corresponding line bundle on the Riemann surface) for the Abelian symmetries $\beta_i$ and $\gamma_j$. This is done by activating an expectation value (proportional to the identity matrix) for the bifundamentals charged under the symmetry and adding chiral singlets which flip baryonic operators (see \cite{Gaiotto:2015usa}) built out of the other $k-1$ bifundamentals charged under the minimal-puncture symmetry: If we give vev to a horizontal chiral we flip all the other horizontal chirals in the same group and similarly for diagonal fields.  In this way we obtain a 4d gauge theory corresponding to the 6d SCFT compactified on a torus with flux. Since we are free to choose which bifundamentals to give a vev, with this procedure of closing punctures we can get many different theories, labelled by the value of the flux for the various Abelian symmetries. 

The question we are interested in is whether this operation has a Type IIB counterpart or not. In \cite{Bah:2021iaa} it was noticed that in the case of class $\mathcal{S}_{2}$ theories, in which we have just two $\beta$ and $\gamma$ symmetries, the models corresponding to a torus with flux are closely related to the worldvolume theories on $D3$ branes probing the CY cone on $Y^{p,q}$ Sasaki-Einstein manifolds. More precisely, they noticed that if we remove all the flipping fields from the class $\mathcal{S}_{2}$ models one is left precisely with the $Y^{p,q}$ theories. In particular, if we turn on $2p$ units of flux for one of the two $U(1)$ symmetries only, we land on $Y^{p,p}$ models, whose corresponding CY in Type IIB is a $\mathbb{Z}_{2p}$ orbifold of $\mathbb{C}^3$ acting as $$\mathbb{Z}_{2p}:\; (\omega_{2p},\omega_{2p},\omega_{2p}^{-2})\,.$$ 

At this stage we can observe that the above statement has a clear counterpart for $k$ generic. If we start from a torus with $nk$ minimal punctures and close them in such a way that we turn on $nk$ units of flux for a single $U(1)$ symmetry we obtain, modulo flipping fields, a Type IIB orbifold model. The resulting orbifold group is $\mathbb{Z}_{nk(k-1)}$ and acts on $\mathbb{C}^3$ as follows: \be\label{orbi1}  \mathbb{Z}_{nk(k-1)}:\; (\omega,\omega^{k-1},\omega^{-k})\,,\ee 
where $\omega$ denotes a $nk(k-1)$-th root of unity. We therefore see that the Type IIB theory corresponds to a relevant deformation of the class $\mathcal{S}_{k}$ theory on a torus. The information about the amount of flux, at least for this specific choice, is encoded in the orbifold order on the Type IIB side and the number of $M5$ branes in M-theory coincides with the number of $D3$ branes in Type IIB. 

In order to derive our claim, let us exemplify the calculation in the case $k=4$ and $n=2$. The quiver corresponding to the torus with eight minimal punctures is 
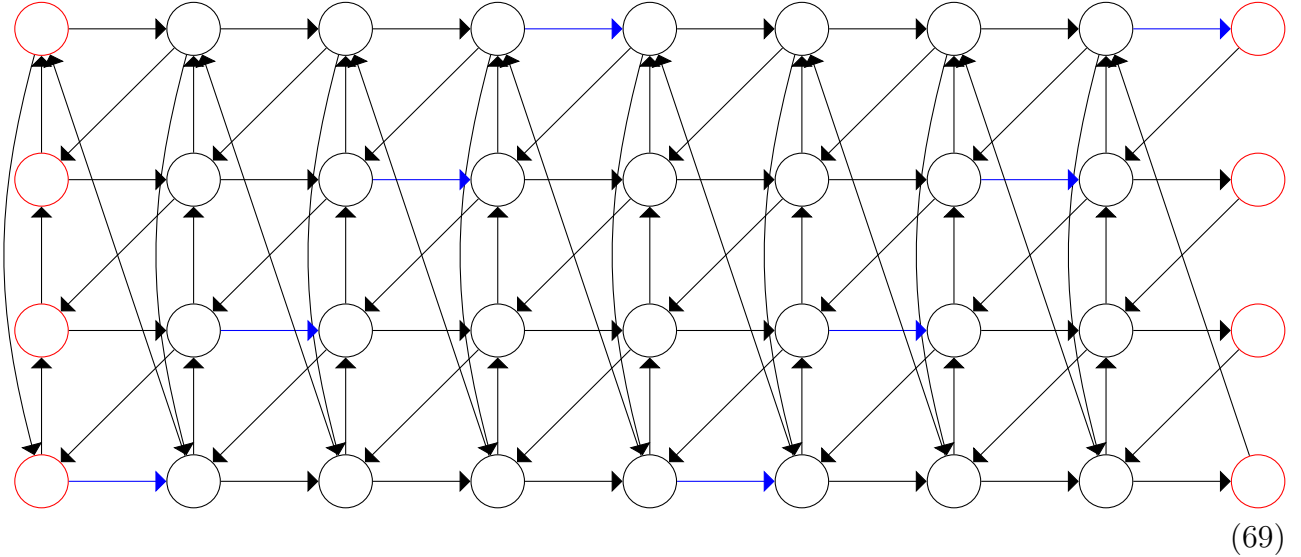
\begin{equation}\label{k4n2}
\begin{tikzpicture}
\node[draw, circle, minimum width=20pt, color=red](a1) at (0,0){};
\node[draw, circle, minimum width=20pt](b1) at (2,0){};
\node[draw, circle, minimum width=20pt](c1) at (4,0){}; 
\node[draw, circle, minimum width=20pt, color=red](a2) at (0,-2){};
\node[draw, circle, minimum width=20pt](b2) at (2,-2){};
\node[draw, circle, minimum width=20pt](c2) at (4,-2){};
\node[draw, circle, minimum width=20pt, color=red](a3) at (0,-4){};
\node[draw, circle, minimum width=20pt](b3) at (2,-4){};
\node[draw, circle, minimum width=20pt](c3) at (4,-4){};
\node[draw, circle, minimum width=20pt, color=red](a4) at (0,-6){};
\node[draw, circle, minimum width=20pt](b4) at (2,-6){};
\node[draw, circle, minimum width=20pt](c4) at (4,-6){};

\node[draw, circle, minimum width=20pt](d1) at (6,0){};
\node[draw, circle, minimum width=20pt](e1) at (8,0){};
\node[draw, circle, minimum width=20pt](f1) at (10,0){}; 
\node[draw, circle, minimum width=20pt](d2) at (6,-2){};
\node[draw, circle, minimum width=20pt](e2) at (8,-2){};
\node[draw, circle, minimum width=20pt](f2) at (10,-2){};
\node[draw, circle, minimum width=20pt](d3) at (6,-4){};
\node[draw, circle, minimum width=20pt](e3) at (8,-4){};
\node[draw, circle, minimum width=20pt](f3) at (10,-4){};
\node[draw, circle, minimum width=20pt](d4) at (6,-6){};
\node[draw, circle, minimum width=20pt](e4) at (8,-6){};
\node[draw, circle, minimum width=20pt](f4) at (10,-6){};

\node[draw, circle, minimum width=20pt](g1) at (12,0){};
\node[draw, circle, minimum width=20pt](h1) at (14,0){};
\node[draw, circle, minimum width=20pt, color=red](i1) at (16,0){};
\node[draw, circle, minimum width=20pt](g2) at (12,-2){};
\node[draw, circle, minimum width=20pt](h2) at (14,-2){};
\node[draw, circle, minimum width=20pt, color=red](i2) at (16,-2){};
\node[draw, circle, minimum width=20pt](g3) at (12,-4){};
\node[draw, circle, minimum width=20pt](h3) at (14,-4){};
\node[draw, circle, minimum width=20pt, color=red](i3) at (16,-4){};
\node[draw, circle, minimum width=20pt](g4) at (12,-6){};
\node[draw, circle, minimum width=20pt](h4) at (14,-6){};
\node[draw, circle, minimum width=20pt, color=red](i4) at (16,-6){};
\path[every node/.style={font=\sffamily\small,
  		fill=white,inner sep=1pt}]
(a1) edge [bend right=15,->]  (a4)
(b1) edge [bend right=15,->]  (b4)
(c1) edge [bend right=15,->]  (c4)
(d1) edge [bend right=15,->]  (d4)
(e1) edge [bend right=15,->]  (e4)
(f1) edge [bend right=15,->]  (f4)
(g1) edge [bend right=15,->]  (g4)
(h1) edge [bend right=15,->]  (h4);

\draw[->] (a1)--(b1); 
\draw[->] (b1)--(c1); 
\draw[->] (c1)--(d1);
\draw[->, color=blue] (d1)--(e1); 
\draw[->] (e1)--(f1);
\draw[->] (f1)--(g1); 
\draw[->] (g1)--(h1);
\draw[->, color=blue] (h1)--(i1); 
\draw[->] (b1)--(a2); 
\draw[->] (c1)--(b2); 
\draw[->] (d1)--(c2); 
\draw[->] (e1)--(d2); 
\draw[->] (f1)--(e2); 
\draw[->] (g1)--(f2);
\draw[->] (h1)--(g2); 
\draw[->] (i1)--(h2); 
\draw[->] (a2)--(b2); 
\draw[->] (b2)--(c2); 
\draw[->, color=blue] (c2)--(d2);
\draw[->] (d2)--(e2); 
\draw[->] (e2)--(f2);
\draw[->] (f2)--(g2); 
\draw[->, color=blue] (g2)--(h2);
\draw[->] (h2)--(i2); 
\draw[->] (b2)--(a3); 
\draw[->] (c2)--(b3); 
\draw[->] (d2)--(c3); 
\draw[->] (e2)--(d3); 
\draw[->] (f2)--(e3); 
\draw[->] (g2)--(f3);
\draw[->] (h2)--(g3); 
\draw[->] (i2)--(h3); 
\draw[->] (a3)--(b3); 
\draw[->, color=blue] (b3)--(c3); 
\draw[->] (c3)--(d3);
\draw[->] (d3)--(e3); 
\draw[->] (e3)--(f3);
\draw[->, color=blue] (f3)--(g3); 
\draw[->] (g3)--(h3);
\draw[->] (h3)--(i3); 
\draw[->] (b3)--(a4); 
\draw[->] (c3)--(b4); 
\draw[->] (d3)--(c4); 
\draw[->] (e3)--(d4); 
\draw[->] (f3)--(e4); 
\draw[->] (g3)--(f4);
\draw[->] (h3)--(g4); 
\draw[->] (i3)--(h4); 
\draw[->, color=blue] (a4)--(b4); 
\draw[->] (b4)--(c4); 
\draw[->] (c4)--(d4);
\draw[->] (d4)--(e4); 
\draw[->, color=blue] (e4)--(f4);
\draw[->] (f4)--(g4); 
\draw[->] (g4)--(h4);
\draw[->] (h4)--(i4); 
\draw[->] (b4)--(a1); 
\draw[->] (c4)--(b1); 
\draw[->] (d4)--(c1); 
\draw[->] (e4)--(d1); 
\draw[->] (f4)--(e1); 
\draw[->] (g4)--(f1);
\draw[->] (h4)--(g1); 
\draw[->] (i4)--(h1); 

\draw[->] (a4)--(a3); 
\draw[->] (a3)--(a2);
\draw[->] (a2)--(a1); 
\draw[->] (b4)--(b3); 
\draw[->] (b3)--(b2);
\draw[->] (b2)--(b1);  
\draw[->] (c4)--(c3); 
\draw[->] (c3)--(c2);
\draw[->] (c2)--(c1); 
\draw[->] (d4)--(d3); 
\draw[->] (d3)--(d2);
\draw[->] (d2)--(d1); 
\draw[->] (e4)--(e3); 
\draw[->] (e3)--(e2);
\draw[->] (e2)--(e1); 
\draw[->] (f4)--(f3); 
\draw[->] (f3)--(f2);
\draw[->] (f2)--(f1); 
\draw[->] (g4)--(g3); 
\draw[->] (g3)--(g2);
\draw[->] (g2)--(g1); 
\draw[->] (h4)--(h3); 
\draw[->] (h3)--(h2);
\draw[->] (h2)--(h1); 
\end{tikzpicture}
\end{equation}
The nodes in red in \eqref{k4n2} are identified so that we have 32 $SU(N)$ gauge nodes in total and there is a cubic superpotential term for each closed triangle in \eqref{k4n2}. In order to close the eight punctures and trade them for eight units of flux for $\beta_1$ we give a vev to the eight chirals denoted in blue in \eqref{k4n2}. As a result, the gauge groups connected by the blue arrows are higgsed to the diagonal $SU(N)$, leaving us with a quiver with $24$ nodes. For convenience we number them as follows: 
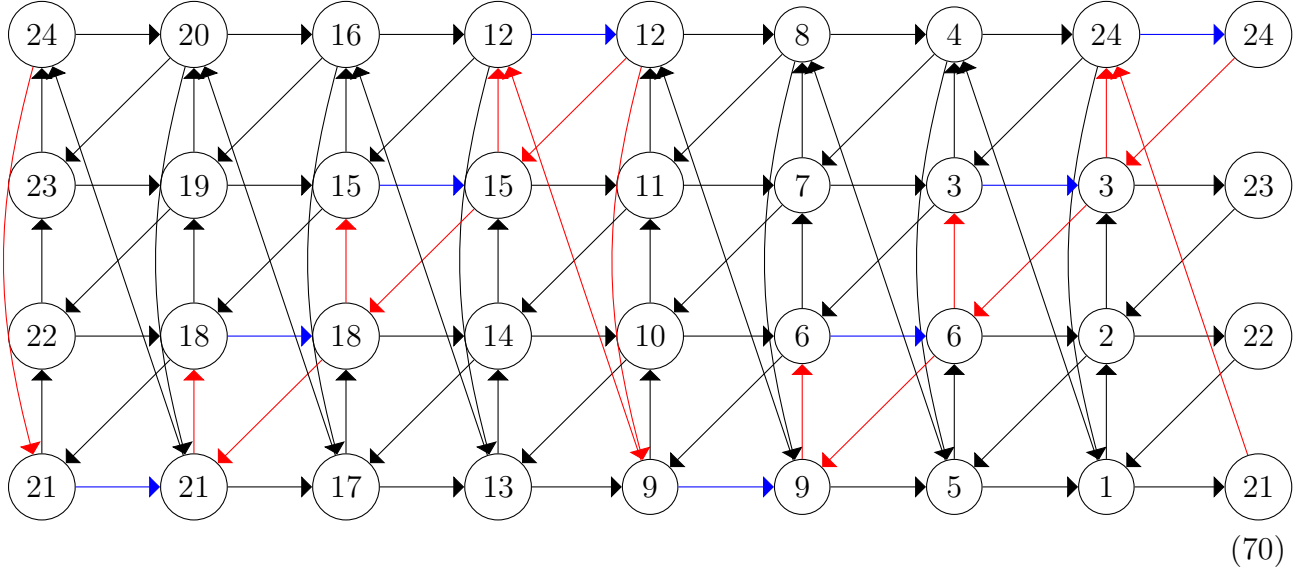
\begin{equation}\label{k4n22}
\begin{tikzpicture}
\node[draw, circle, minimum width=20pt](a1) at (0,0){24};
\node[draw, circle, minimum width=20pt](b1) at (2,0){20};
\node[draw, circle, minimum width=20pt](c1) at (4,0){16}; 
\node[draw, circle, minimum width=20pt](a2) at (0,-2){23};
\node[draw, circle, minimum width=20pt](b2) at (2,-2){19};
\node[draw, circle, minimum width=20pt](c2) at (4,-2){15};
\node[draw, circle, minimum width=20pt](a3) at (0,-4){22};
\node[draw, circle, minimum width=20pt](b3) at (2,-4){18};
\node[draw, circle, minimum width=20pt](c3) at (4,-4){18};
\node[draw, circle, minimum width=20pt](a4) at (0,-6){21};
\node[draw, circle, minimum width=20pt](b4) at (2,-6){21};
\node[draw, circle, minimum width=20pt](c4) at (4,-6){17};

\node[draw, circle, minimum width=20pt](d1) at (6,0){12};
\node[draw, circle, minimum width=20pt](e1) at (8,0){12};
\node[draw, circle, minimum width=20pt](f1) at (10,0){8}; 
\node[draw, circle, minimum width=20pt](d2) at (6,-2){15};
\node[draw, circle, minimum width=20pt](e2) at (8,-2){11};
\node[draw, circle, minimum width=20pt](f2) at (10,-2){7};
\node[draw, circle, minimum width=20pt](d3) at (6,-4){14};
\node[draw, circle, minimum width=20pt](e3) at (8,-4){10};
\node[draw, circle, minimum width=20pt](f3) at (10,-4){6};
\node[draw, circle, minimum width=20pt](d4) at (6,-6){13};
\node[draw, circle, minimum width=20pt](e4) at (8,-6){9};
\node[draw, circle, minimum width=20pt](f4) at (10,-6){9};

\node[draw, circle, minimum width=20pt](g1) at (12,0){4};
\node[draw, circle, minimum width=20pt](h1) at (14,0){24};
\node[draw, circle, minimum width=20pt](i1) at (16,0){24};
\node[draw, circle, minimum width=20pt](g2) at (12,-2){3};
\node[draw, circle, minimum width=20pt](h2) at (14,-2){3};
\node[draw, circle, minimum width=20pt](i2) at (16,-2){23};
\node[draw, circle, minimum width=20pt](g3) at (12,-4){6};
\node[draw, circle, minimum width=20pt](h3) at (14,-4){2};
\node[draw, circle, minimum width=20pt](i3) at (16,-4){22};
\node[draw, circle, minimum width=20pt](g4) at (12,-6){5};
\node[draw, circle, minimum width=20pt](h4) at (14,-6){1};
\node[draw, circle, minimum width=20pt](i4) at (16,-6){21};
\path[every node/.style={font=\sffamily\small,
  		fill=white,inner sep=1pt}]
(a1) edge [bend right=15,->, color=red]  (a4)
(b1) edge [bend right=15,->]  (b4)
(c1) edge [bend right=15,->]  (c4)
(d1) edge [bend right=15,->]  (d4)
(e1) edge [bend right=15,->, color=red]  (e4)
(f1) edge [bend right=15,->]  (f4)
(g1) edge [bend right=15,->]  (g4)
(h1) edge [bend right=15,->]  (h4);

\draw[->] (a1)--(b1); 
\draw[->] (b1)--(c1); 
\draw[->] (c1)--(d1);
\draw[->, color=blue] (d1)--(e1); 
\draw[->] (e1)--(f1);
\draw[->] (f1)--(g1); 
\draw[->] (g1)--(h1);
\draw[->, color=blue] (h1)--(i1); 
\draw[->] (b1)--(a2); 
\draw[->] (c1)--(b2); 
\draw[->] (d1)--(c2); 
\draw[->, color=red] (e1)--(d2); 
\draw[->] (f1)--(e2); 
\draw[->] (g1)--(f2);
\draw[->] (h1)--(g2); 
\draw[->, color=red] (i1)--(h2); 
\draw[->] (a2)--(b2); 
\draw[->] (b2)--(c2); 
\draw[->, color=blue] (c2)--(d2);
\draw[->] (d2)--(e2); 
\draw[->] (e2)--(f2);
\draw[->] (f2)--(g2); 
\draw[->, color=blue] (g2)--(h2);
\draw[->] (h2)--(i2); 
\draw[->] (b2)--(a3); 
\draw[->] (c2)--(b3); 
\draw[->, color=red] (d2)--(c3); 
\draw[->] (e2)--(d3); 
\draw[->] (f2)--(e3); 
\draw[->] (g2)--(f3);
\draw[->, color=red] (h2)--(g3); 
\draw[->] (i2)--(h3); 
\draw[->] (a3)--(b3); 
\draw[->, color=blue] (b3)--(c3); 
\draw[->] (c3)--(d3);
\draw[->] (d3)--(e3); 
\draw[->] (e3)--(f3);
\draw[->, color=blue] (f3)--(g3); 
\draw[->] (g3)--(h3);
\draw[->] (h3)--(i3); 
\draw[->] (b3)--(a4); 
\draw[->, color=red] (c3)--(b4); 
\draw[->] (d3)--(c4); 
\draw[->] (e3)--(d4); 
\draw[->] (f3)--(e4); 
\draw[->, color=red] (g3)--(f4);
\draw[->] (h3)--(g4); 
\draw[->] (i3)--(h4); 
\draw[->, color=blue] (a4)--(b4); 
\draw[->] (b4)--(c4); 
\draw[->] (c4)--(d4);
\draw[->] (d4)--(e4); 
\draw[->, color=blue] (e4)--(f4);
\draw[->] (f4)--(g4); 
\draw[->] (g4)--(h4);
\draw[->] (h4)--(i4); 
\draw[->] (b4)--(a1); 
\draw[->] (c4)--(b1); 
\draw[->] (d4)--(c1); 
\draw[->, color=red] (e4)--(d1); 
\draw[->] (f4)--(e1); 
\draw[->] (g4)--(f1);
\draw[->] (h4)--(g1); 
\draw[->, color=red] (i4)--(h1); 

\draw[->] (a4)--(a3); 
\draw[->] (a3)--(a2);
\draw[->] (a2)--(a1); 
\draw[->, color=red] (b4)--(b3); 
\draw[->] (b3)--(b2);
\draw[->] (b2)--(b1);  
\draw[->] (c4)--(c3); 
\draw[->, color=red] (c3)--(c2);
\draw[->] (c2)--(c1); 
\draw[->] (d4)--(d3); 
\draw[->] (d3)--(d2);
\draw[->, color=red] (d2)--(d1); 
\draw[->] (e4)--(e3); 
\draw[->] (e3)--(e2);
\draw[->] (e2)--(e1); 
\draw[->, color=red] (f4)--(f3); 
\draw[->] (f3)--(f2);
\draw[->] (f2)--(f1); 
\draw[->] (g4)--(g3); 
\draw[->, color=red] (g3)--(g2);
\draw[->] (g2)--(g1); 
\draw[->] (h4)--(h3); 
\draw[->] (h3)--(h2);
\draw[->, color=red] (h2)--(h1); 
\end{tikzpicture}
\end{equation}
Nodes with the same number in \eqref{k4n22} are always connected by a blue line. The bifundamentals to which we give a vev recombine with the generators of the broken gauge groups. Upon expanding the superpotential around the vev we generate mass terms which allow us to integrate out using the equations of motion the red arrows. We are therefore left in the IR with $24$ gauge nodes connected by the black arrows in \eqref{k4n22}. We have cubic superpotential terms for all oriented triangles in \eqref{k4n22}. We can now notice that every node has three incoming and three outgoing black arrows. Vertical black arrows connect node $i$ to node $i+1$ (identified periodically modulo $24$), diagonal black arrows connect node $i$ to node $i+3$ and horizontal black arrows connect node $i$ to node $i-4$. These are precisely the data specifying the Type IIB orbifold model 
$$ \mathbb{Z}_{24}:\; (\omega_{24},\omega_{24}^{3},\omega_{24}^{-4})\,,$$ 
in agreement with our claim \eqref{orbi1}. Notice that for $k=2$ \eqref{orbi1} correctly reduces to $Y^{n,n}$ orbifolds, in agreement with the analysis of \cite{Bah:2021iaa}, and also that  \eqref{orbi1} is consistent with the analysis for $k=3$ and $n=1$ presented in \cite{Bah:2017gph}. Finally, notice that, in closing the minimal punctures, we have not included the flipping singlets as in \cite{Gaiotto:2015usa}: The Type IIB orbifold is obtained upon removing them. 

The conclusion of the above discussion is that the torus compactification of 6d theories engineered in M-theory by probing a $\mathbb{Z}_k$ orbifold of $\mathbb{C}^2$ with $N$ $M5$ branes ($SU(k)\times SU(k)$ conformal matter) with specific choices of flux correspond to the worldvolume theory of $N$ $D3$ branes in Type IIB probing a orbifold $\mathbb{C}^3/\Gamma$. The order of $\Gamma$ encodes the information about the amount of flux and the M-theory orbifold group is a subgroup of $\Gamma$. The correspondence is that the Type IIB theory is a deformation of the compactified 6d theory, obtained by removing (or equivalently flipping) all the chiral flipping fields.  The observation of \cite{Bah:2021iaa} that for $k=2$ theories on a torus with arbitrary choices of flux correspond to $Y^{p,q}$ theories in Type IIB clearly suggests that the correspondence we found between M-theory and Type IIB for $k$ generic should extend beyond the orbifold class.

\paragraph{Flipping fields from 6d} 

As we have explained, the theory \eqref{k4n22}, associated with (a special instance of) the $\mathbb{C}^3$ orbifold 
$$ \mathbb{Z}_{nk(k-1)}:\; (\omega,\omega^{k-1},\omega^{-k})$$
is obtained from the class $\mathcal{S}_k$ theory on the torus by removing the flipping fields. We would now like to explain how to reintroduce the flipping fields once \eqref{k4n22} is given. The first observation is that in \eqref{k4n22} we have activated a vev for horizontal bifundamentals only and therefore, according to the prescription of \cite{Gaiotto:2015usa}, we should flip the baryons built from all the other horizontal arrows. These correspond to the modes arising from the adjoint chiral of $\mathcal{N}=4$ SYM associated with the third complex direction of $\mathbb{C}^3$ in the Type IIB spacetime. This can be identified with the direction which has no counterpart in the M-theory spacetime, since the $\mathbb{Z}_k$ subgroup acting in M-theory is embedded in the first two directions, as explained before.

\subsection{E-string theories on a torus with flux and permutation orientifolds}\label{E-string}

In this section we would like to come back to a family of SCFTs we encountered in Section \ref{ExampleLoworder}, namely \eqref{Z2MixedQuiver}, and explain how it is connected to the torus compactification of the E-string theory with minimal amount of flux. Revisiting this example will allow us to infer how the Type IIB dual configurations change when we pump up the flux quanta in M-theory. As is well known the higher-rank E-string theory is a 6d SCFT with $E_8\times SU(2)$ global symmetry. It can be engineered in M-theory by probing the $M9$ wall with a stack of $M5$ branes. The compactification on a torus with flux for the minimal E-string theory (a single $M5$ brane in M-theory) has been considered in detail in \cite{Kim:2017toz} and the higher-rank generalization was discussed in \cite{Pasquetti:2019hxf}. As in the class $\mathcal{S}_k$ case, we will relate the 6d theory on a torus with a special value of flux to orbi-orientifold setups in Type IIB. In the E-string case we will focus on the choice of flux preserving $E_7\times U(1)$ global symmetry. 

Before starting our analysis, we would like to point out that in Type IIB we will find symplectic quivers with fundamentals and bifundamental fields. This is the correct class of theories for the minimal E-string theory, but in the higher-rank case we should replace the bifundamental chirals, as discussed in \cite{Pasquetti:2019hxf}, with a more exotic matter system called $E[USp(2N)]$ in \cite{Pasquetti:2019hxf} (see also \cite{Bottini:2021vms, Hwang:2021ulb}). This is a strongly-coupled SCFT  with $USp(2N)\times USp(2N)\times U(1)^2$ global symmetry, defined as the IR fixed point of a lagrangian linear quiver, and it can be shown that, upon a superpotential relevant deformation, the $E[USp(2N)]$ theory flows to a ordinary $USp(2N)\times USp(2N)$ bifundamental plus a singlet flipping the baryonic operator built out of the bifundamental chiral (see Appendix E of \cite{Comi:2022aqo}). After performing this deformation and removing the flipping singlets, we land on the symplectic quivers we will discuss momentarily. As a result, we conclude again that the 6d theories on $T^2$ and the Type IIB theories are related by a relevant deformation. This just amounts to removing the flipping fields for $N=1$, while it also involves a superpotential relevant deformation in the higher-rank case.

As we have mentioned, the 4d theory corresponding to the torus compactification of the E-string theory with a minimal (integral) amount of flux has been derived in Section 3 of \cite{Kim:2017toz} and named the ``basic theory''. Such a theory differs from \eqref{Z2MixedQuiver} just by having one more term in the superpotential, namely the coupling of $\Phi^2$ to the flavors. Having a single $D7$/$O7$ stack located at $z_3=0$ clearly can never generate such a term. Instead of considering a second stack at $z_2=0$, we consider a different type of orientifold projection, which, as we will see, will turn out to be the correct thing to do to reproduce configurations with higher values of the flux. We consider an orientifold involution whose spacetime part acts by \emph{swapping} the complex coordinates $z_2$ and $z_3$ \cite{Feng:2001rh} (see also \cite{Blumenhagen:1999md}):
\be\label{permutationorientifold}
\omega\,:\qquad z_2\,\longleftrightarrow\,z_3\,.
\ee
A $D7$/$O7$ stack located at $z_3=0$ would not be invariant under such an orientifold involution. However, in the simple case of the $\mathbb{Z}_2$ orbifold, just a $D7$/$O7$ stack located at $z_2+z_3=0$ works. While the Chan-Paton representations of both orbifold and orientifold remain the same, the orientifold acts on the $D3$-brane scalars as 
\begin{eqnarray}\label{PermOrientPhi}
\Phi^{1}&=&\tilde\Omega_3\left(\Phi^1\right)^T\tilde\Omega_3\,,\nonumber\\
\Phi^{2}&=&\tilde\Omega_3\left(\Phi^3\right)^T\tilde\Omega_3\,,\nonumber\\
\Phi^{3}&=&\tilde\Omega_3\left(\Phi^2\right)^T\tilde\Omega_3\,,
\end{eqnarray}
where we put a tilde to distinguish this ``permutation orientifold'' from the previously-discussed involutions. Following the same steps as in Section \ref{ExampleLoworder}, we arrive at the quiver \eqref{Z2MixedQuiver}, which we now represent with the two bifundamentals pointing in the same direction: 
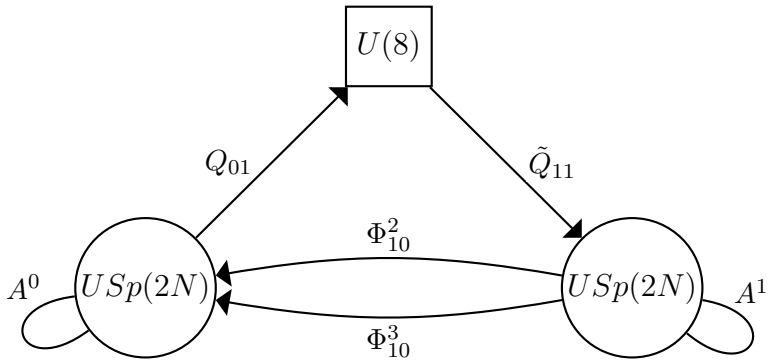
\begin{center}
\begin{tikzpicture}[thick, scale=0.8]
  \node[circle, draw, inner sep= 1pt](L1) at (6,-4){$USp(2N)$};
  \node[circle, draw, inner sep= 1pt](L2) at (14,-4){$USp(2N)$};
\node[draw, rectangle, minimum width=30pt, minimum height=30pt](L5) at (10,0){$U(8)$};
 \path[every node/.style={font=\sffamily\small,
  		fill=white,inner sep=1pt}]
(L1) edge [bend left=10, <-] node[above=0.5mm] {$\Phi^{2}_{10}$} (L2)
(L2) edge [bend left=10, ->] node[below=0.5mm] {$\Phi^{3}_{10}$} (L1)
(L5) edge [->] node[right=2.5mm] {$\tilde{Q}_{11}$} (L2)
(L1) edge [->] node[left=2.1mm] {$Q_{01}$} (L5)
(L1) edge [loop, in=-143, out=-173, looseness=6] node[above=4.5mm] {$A^0$} (L1)
(L2) edge [loop, in=-10, out=-40, looseness=6] node[above=4.8mm] {$A^1$}  (L2);
\end{tikzpicture}\vspace{-.8cm}
\end{center}\be\label{Z2MixedQuiverE}\ee
The only difference with the theory \eqref{Z2MixedQuiver} lies in the superpotential, which now schematically looks like
\be\label{WZ2MixedQuiverE}
W=(A^0-A^1)(\Phi^3_{10}\Phi^3_{10}-\Phi^2_{10}\Phi^2_{10})+\tilde{Q}_{11}(\Phi^2_{10}+\Phi^3_{10})Q_{01}\,,
\ee
where in the first four terms a contraction with the symplectic forms of the gauge groups is understood. The extra term with respect to \eqref{WZ2MixedQuiver} is due to the fact that now the coordinate transverse to the $D7$ branes is $z_2+z_3$ \cite{Aldazabal:2000sa}. 

The dimension of the conformal manifold can be determined using the Leigh-Strassler method \cite{Leigh:1995ep}, as we now explain. In the model at hand we have six matter fields with the corresponding anomalous dimensions. We will denote these as $\gamma_Q,$ $\gamma_{\widetilde{Q}}$, $\gamma_{A^0}$, $\gamma_{A^1}$, $\gamma_{\Phi^2}$, and $\gamma_{\Phi^3}$. We also have eight possible beta functions: One for each gauge coupling and six for the possible superpotential terms displayed in \eqref{WZ2MixedQuiverE}. As shown in \cite{Leigh:1995ep}, we can express the beta functions in terms of the anomalous dimensions of the various matter fields (which in turn are functions of the eight couplings), and we have to set them to zero at the fixed point, as required by conformal invariance. This leads to the system of equations 
\be\label{LSsys}
\left\{\begin{array}{l} 
(2N-2)\gamma_{A^0}+8\gamma_{Q}+2N\gamma_{\Phi^2}+2N\gamma_{\Phi^3}=0\\ 
(2N-2)\gamma_{A^1}+8\gamma_{\widetilde{Q}}+2N\gamma_{\Phi^2}+2N\gamma_{\Phi^3}=0\\
\gamma_{A^0}+2\gamma_{\Phi^2}=0\\ 
\gamma_{A^0}+2\gamma_{\Phi^3}=0 \\
\gamma_{A^1}+2\gamma_{\Phi^2}=0\\
\gamma_{A^1}+2\gamma_{\Phi^3}=0 \\
\gamma_{Q}+\gamma_{\widetilde{Q}}+\gamma_{\Phi^3}=0\\
\gamma_{Q}+\gamma_{\widetilde{Q}}+\gamma_{\Phi^2}=0\\
\end{array} \right.
\ee 
where the first two come from the gauge beta functions and the other six from the superpotential terms. It is easy to see that only five of the eight equations are independent, thus leaving one of the six functions unconstrained, and hence three combinations of the eight original variables free to vary. The solution in terms of $\gamma_{Q}$ is
\be
\gamma_{A^1}= \gamma_{A^0}=4\gamma_{Q}\,;\qquad\gamma_{\Phi^2}=\gamma_{\Phi^3}=-2\gamma_{Q}\,;\qquad\gamma_{\widetilde{Q}}=\gamma_{Q}\,.
\ee
This tells us that there is a three-dimensional conformal manifold. Had we considered the superpotential in  \eqref{WZ2MixedQuiver}, we would have ended up with only five equations (from the five couplings), indicating we do not have a conformal manifold. Therefore, in order to have an interacting conformal theory, we have to turn on at least one of the additional superpotential terms present in \eqref{WZ2MixedQuiverE}, otherwise the system flows to zero coupling.\\

Let us now double the flux in the M-theory setup: The 4d theory we should get, according to \cite{Kim:2017toz}, is encoded in the quiver
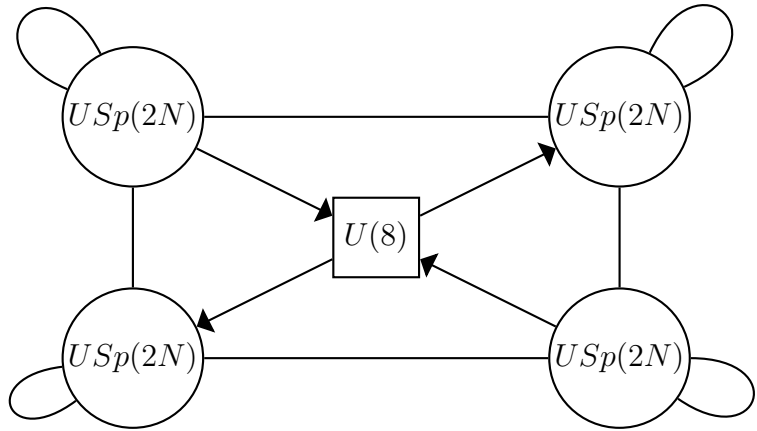
\begin{center}
\begin{tikzpicture}[thick, scale=0.8]
  \node[circle, draw, inner sep= 1pt](L3) at (6,0){$USp(2N)$};
  \node[circle, draw, inner sep= 1pt](L4) at (14,0){$USp(2N)$};
  \node[circle, draw, inner sep= 1pt](L1) at (6,-4){$USp(2N)$};
  \node[circle, draw, inner sep= 1pt](L2) at (14,-4){$USp(2N)$};
\node[draw, rectangle, minimum width=30pt, minimum height=30pt](L5) at (10,-2){$U(8)$};
 \path[every node/.style={font=\sffamily\small,
  		fill=white,inner sep=1pt}]
(L1) edge   (L2)
(L3) edge   (L4)
(L1) edge   (L3)
(L2) edge   (L4)
(L5) edge [->]  (L4)
(L3) edge [->]  (L5)
(L5) edge [<-]  (L2)
(L1) edge [<-]  (L5)
(L1) edge [loop, in=-143, out=-173, looseness=6]  (L1)
(L3) edge [loop, in=-203, out=-243, looseness=6]  (L3)
(L4) edge [loop, in=65, out=25, looseness=6]  (L4)
(L2) edge [loop, in=0, out=-35, looseness=6]   (L2);
\end{tikzpicture}\vspace{-.8cm}
\end{center}\be\label{Z4MixedQuiverE}\ee
where the loop edges are chiral fields in the (symplectic traceless) antisymmetric representation of the gauge groups, and there is a cubic term in the superpotential for each of the four triangles in the quiver. We now show that such a theory can be obtained in the Type IIB context through a $\mathbb{Z}_4$ orbifold combined with a permutation orientifold of the kind introduced above. 

Consider the orbifold action
\be
(z_1,z_2,z_3)\;\sim\;\left(z_1,{\rm i}\,z_2,-{\rm i}\,z_3\right)\,.
\ee
Instead of placing the $D7$/$O7$ stack at an orbifold-invariant location, e.g.~$z_3=0$ as we do in Appendix \ref{ExampleLoworder1}, we want to use the orientifold involution \eqref{permutationorientifold}. Here, however, things are not as easy as in the $\mathbb{Z}_2$-orbifold case, because the orbifold acts differently on the two coordinates that are exchanged by the orientifold. Calling $\gamma$ the orbifold action, this does not commute with $\omega$, but instead we have 
\be\label{OrbOrientLaw}
\omega\circ\gamma=\gamma^{-1}\circ\omega\,.
\ee
This law translates into the following relation on the $D3$-brane Chan-Paton space
\be\label{CompatPermOrient}
\tilde\Omega_3\,\Gamma_3^*\,\tilde\Omega_3^{-1}=\Gamma_3^{-1}\,,
\ee
where we have used unitarity of the representation $\Gamma_3$ of the orbifold group. Choosing for $\Gamma_3$ the regular representation of $\mathbb{Z}_4$\footnote{Here we directly implement from the start the constraint of conformality, by requiring an equal number of $D3$ branes in each fractional stack.}
\be\label{Gamma3Z4smoothPerm}
\Gamma_3={\rm diag}\,(1,{\rm i},-1,-{\rm i})\otimes I_{2N}\,,
\ee
a representation of the orientifold on the $D3$-brane Chan-Paton space that solves \eqref{CompatPermOrient} is\footnote{Contrary to the ordinary orientifold, which comes in two inequivalent versions for even-order orbifolds, the permutation orientifold is unique. This can be seen by either turning on a $\mathbb{Z}_2$ discrete torsion between the orbifold and the orientifold, or by promoting $\Gamma_3$ to a projective representation of the orbifold group.}
\be
\tilde{\Omega}_3=I_{4}\otimes {\rm i}J_{2N}\,.
\ee
This tells us that all of the four gauge nodes of the quiver are \emph{fixed} by this orientifold involution, and hence turn into symplectic gauge groups. Moreover, due to \eqref{PermOrientPhi}, the fields of type $\Phi^1$ turn into antisymmetric chirals for each gauge group, whereas the $\Phi^2$ and $\Phi^3$ bifundamentals between each pair of gauge groups are orientifold-images of one another, and thus only one combination survives the projection.

The 3-7 sector is as usual trickier to handle. In this case, placing a $D7$/$O7$ stack at $z_2+z_3=0$ is not consistent, because such a plane is not orbifold invariant. We must introduce a second $D7$/$O7$ stack at $z_2-z_3=0$. Therefore, calling $(Q_\pm,\tilde{Q}_\pm)$ the chiral/antichiral pairs originating from the $D3$-$D7$$_\pm$ open strings, they will couple to the bifundamental fields according to the superpotential
\be\label{W+-}
W\supset\tilde{Q}_+(\Phi^2+\Phi^3)Q_++\tilde{Q}_-(\Phi^2-\Phi^3)Q_-\,.
\ee
Unfortunately, however, in this basis it is hard to identify which arrows these fields correspond to in the quiver. To this end, it is much easier to work in a basis in which the orbifold action on the $D7$-brane Chan-Paton space is diagonal. Both $D7$/$O7$ stacks have components along the plane at $z_2=0$ \emph{and} the plane at $z_3=0$. Therefore there will be two pairs of chiral/antichiral fields surviving the orbifold projection, which satisfy (cf.~Eq.~\eqref{ProjQtQ})
\be
\begin{array}{rcl}e^{-\pi{\rm i}/4}Q^{(2)}&=&\Gamma_3Q^{(2)}\Gamma_{7_2}^{-1}\\ e^{-\pi{\rm i}/4}\tilde{Q}^{(2)}&=&\Gamma_{7_2}\tilde{Q}^{(2)}\Gamma_3^{-1}\,,\end{array}\qquad\qquad\begin{array}{rcl}e^{\pi{\rm i}/4}Q^{(3)}&=&\Gamma_3Q^{(3)}\Gamma_{7_3}^{-1}\\ e^{\pi{\rm i}/4}\tilde{Q}^{(3)}&=&\Gamma_{7_3}\tilde{Q}^{(3)}\Gamma_3^{-1}\,.\end{array}
\ee
This, together with Eq.~\eqref{Gamma3Z4smoothPerm} leads us to choose
\begin{eqnarray}
\Gamma_{7_2}&=&e^{\pi{\rm i}/4}\,{\rm diag}\,(I_{M_0},{\rm i}I_{M_1},-I_{M_2},-{\rm i}I_{M_3})\,,\nonumber\\
\Gamma_{7_3}&=&e^{-\pi{\rm i}/4}\,{\rm diag}\,(I_{M_3},-{\rm i}I_{M_2},-I_{M_1},{\rm i}I_{M_0})\,,
\end{eqnarray}
where we already appropriately constrained the number of the various fractional $D7$ branes so to make these representations compatible with the orientifold involution. The latter indeed requires picking an action $\tilde\Omega_7$ such that
\be\label{CompatPermOrient7}
\tilde\Omega_7\,\Gamma_{7_2}^*\,\tilde\Omega_7^{-1}=\Gamma_{7_3}^{-1}\,.
\ee
because the orientifold exchanges components along the $z_2=0$ plane with those along the $z_3=0$ plane. To make contact with the theory \eqref{Z4MixedQuiverE}, we choose $M_0=M_2=8$ and $M_1=M_3=0$, which will give us the following quiver after the orbi-orientifold projection:
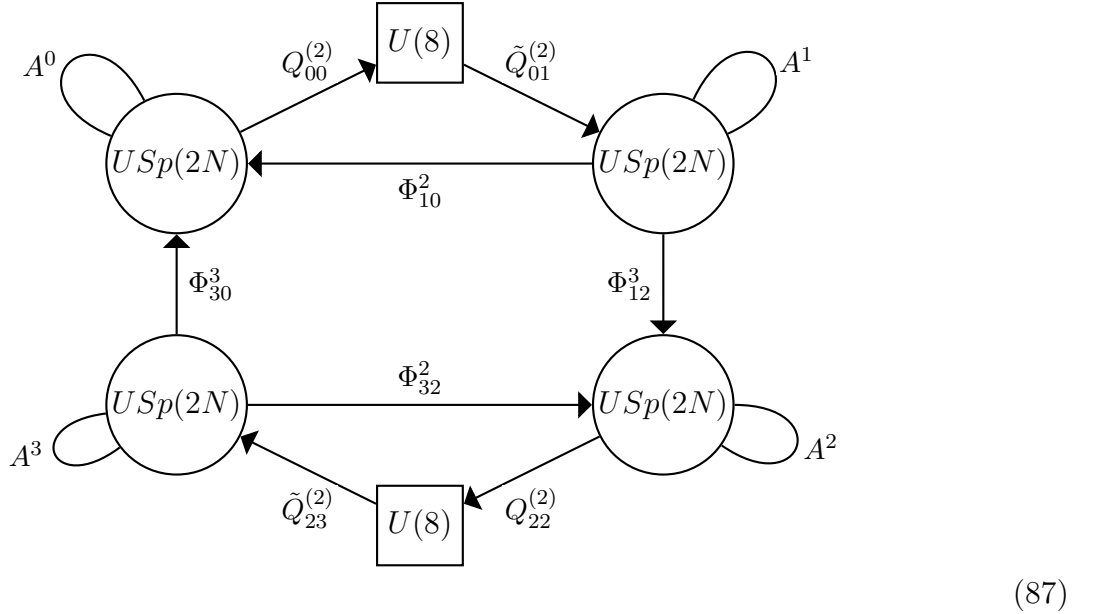
\begin{center}
\begin{tikzpicture}[thick, scale=0.8]
  \node[circle, draw, inner sep= 1pt](L1) at (6,-6){$USp(2N)$};
  \node[circle, draw, inner sep= 1pt](L2) at (14,-6){$USp(2N)$};
  \node[circle, draw, inner sep= 1pt](L3) at (6,-2){$USp(2N)$};
  \node[circle, draw, inner sep= 1pt](L4) at (14,-2){$USp(2N)$};
\node[draw, rectangle, minimum width=30pt, minimum height=30pt](L5) at (10,0){$U(8)$};
\node[draw, rectangle, minimum width=30pt, minimum height=30pt](L6) at (10,-8){$U(8)$};
 \path[every node/.style={font=\sffamily\small,
  		fill=white,inner sep=1pt}]
(L1) edge [->] node[above=1mm] {$\Phi^2_{32}$}  (L2)
(L3) edge  [<-] node[below=1mm] {$\Phi^2_{10}$} (L4)
(L1) edge [->] node[right=1mm] {$\Phi^3_{30}$}  (L3)
(L2) edge [<-] node[left=1mm] {$\Phi^3_{12}$} (L4)
(L2) edge [->] node[below=1.9mm] {$Q^{(2)}_{22}$}  (L6)
(L4) edge [<-] node[above=1.9mm] {$\tilde{Q}^{(2)}_{01}$} (L5)
(L5) edge [<-] node[above=1.9mm] {$Q^{(2)}_{00}$} (L3)
(L1) edge [<-] node[below=1.9mm] {$\tilde{Q}^{(2)}_{23}$} (L6)
(L1) edge [loop, in=-143, out=-173, looseness=6] node[left=1mm] {$A^3$}  (L1)
(L3) edge [loop, in=-203, out=-243, looseness=6] node[left=1mm] {$A^0$} (L3)
(L4) edge [loop, in=65, out=25, looseness=6] node[right=1mm] {$A^1$} (L4)
(L2) edge [loop, in=0, out=-35, looseness=6]  node[right=.4mm] {$A^2$} (L2);
\end{tikzpicture}\vspace{-.8cm}
\end{center}\be\label{Z4MixedQuiverEbis}\ee
The above theory seems to have a $U(8)^2$ flavor symmetry. However there is a superpotential that schematically looks like
\begin{eqnarray}\label{Wmixing}
W&=&A^0((\Phi^3_{30})^2-(\Phi^2_{10})^2)+A^1((\Phi^2_{10})^2-(\Phi^3_{12})^2)+A^2((\Phi^3_{12})^2-(\Phi^2_{32})^2)+A^3((\Phi^2_{32})^2-(\Phi^3_{30})^2)\nonumber\\
&&+\tilde{Q}^{(2)}_{01}\Phi^2_{10}Q^{(2)}_{00}+\tilde{Q}^{(2)}_{23}\Phi^2_{32}Q^{(2)}_{22}+\tilde{Q}^{(2)}_{01}\Phi^3_{12}Q^{(2)}_{22}+\tilde{Q}^{(2)}_{23}\Phi^3_{30}Q^{(2)}_{00}\,,
\end{eqnarray}
where in the first line contractions with the symplectic forms of the gauge groups are understood. In the second line we recognize the superpotential \eqref{W+-}, written in a field basis in which the matrix of the $\Phi$'s is non-diagonal. The mixing of the $(Q^{(2)},\tilde{Q}^{(2)})$ flavors with the $\Phi^3$ field was to be expected, because the $D7$ branes where those flavors live also have a non-trivial component along $z_3=0$. The superpotential \eqref{Wmixing} breaks $U(8)^2$ to the diagonal $U(8)$, which is the flavor group appearing in the center of the quiver \eqref{Z4MixedQuiverE}.\\

At this point it is straightforward to guess the generalization to higher values of the flux. Call $F\in\mathbb{N}$ the integral flux quanta. We propose that the low-energy theory of the smooth E-string compactified on a $T^2$ threaded by $F$ units of flux corresponds, upon a relevant superpotential deformation, to the theory on $D3$ branes probing a specific $\bZ_{2F}$ orbi-orientifold background. The M-theory setting is summarized by the following table
\begin{center}
\begin{tabular}{c|ccccccccccc}&&&&& $\Re \tilde{z}_1$&$\Im \tilde{z}_1$&$\Re \tilde{z}_2$&$\Im \tilde{z}_2$&& T&M \\
&0&1&2&3&4&5&6&7&8&9&10\\ \hline  $M5$ &$\times$&$\times$&$\times$&$\times$&&&&&&$\times$&$\times$\\ $M9$ &$\times$&$\times$&$\times$&$\times$&$\times$&$\times$&$\times$&$\times$&&$\times$&$\times$\\ \hline  
\end{tabular}
\end{center}
where T and M indicate the T-duality and the M-theory circle respectively, which together form the $T^2$, and $\Re,\Im$ stand for real and imaginary part respectively. The corresponding Type IIB background probed by $D3$ branes is
\be\label{GeneralTypeIIBSmoothEstring}
(z_1,z_2,z_3)\;\sim\;\left(z_1,e^{\pi{\rm i}/F}\,z_2,e^{-\pi{\rm i}/F}\,z_3\right)\,,
\ee
with an orientifold involution acting as in \eqref{permutationorientifold}. The two $M9$-worldvolume coordinates $\tilde{z}_1,\tilde{z}_2$ transverse to the probe $M5$ branes depend on the Type IIB CY threefold coordinates $z_1,z_2,z_3$ as follows
\be\label{CoordinatesMap}
\tilde{z}_1=z_1\,,\qquad\qquad\tilde{z}_2=z_2 z_3\,,
\ee
which are invariant combinations under both the orbifold and the orientifold transformations.

The identification \eqref{GeneralTypeIIBSmoothEstring} forces us to introduce $F$ identical stacks made of $8$ $D7$ branes and an $O7$$^-$ plane wrapping the complex surfaces $z_2+e^{2\pi {\rm i}m/F}z_3=0$ with $m=0,\ldots,F-1$. The ensuing family of conformal field theories will have $2F$ gauge groups of $USp(2N)$ type ($N$ being the number of $D3$ or $M5$ branes), each equipped with an antisymmetric chiral field, and connected by bifundamentals in a necklace shape. Matter amounts to $F$ octet pairs of fundamental chiral/antichiral fields $(q_m,\tilde{q}_m)_{m=0,\ldots,F-1}$ coupled to the bifundamentals as
\be
W\supset\sum_{m=0}^{F-1}\tilde{q}_m(\Phi^2+e^{2\pi {\rm i}m/F}\Phi^3)q_m\,.
\ee
To make contact with the fields appearing in the quiver, one can go to a non-diagonal basis $(\mathbb{Q},\mathbb{\tilde{Q}})\equiv(Q_m,\tilde{Q}_m)_{m=0,\ldots,F-1}$, where the superpontial for the matter sector becomes
\be\label{MixingFlavors}
W\supset {\mathbb{\tilde{Q}}\,{\bf \Phi} \,\mathbb{Q}}\,,\qquad\qquad{\bf \Phi}=\left(\begin{array}{cccccc}\Phi^2&\Phi^3&0&0&\cdots&0\\ 0&\Phi^2&\Phi^3&0&\cdots&0\\ 0&0&\Phi^2&\Phi^3&\cdots&0\\ \vdots&\vdots&\vdots&\vdots&\ddots&\vdots\\ \Phi^3 &0&0&0&\cdots&\Phi^2\end{array}\right)\,.
\ee
This superpotential breaks the $U(8)^F$ flavor symmetry to the diagonal $U(8)$, and the resulting quiver can be represented as follows:
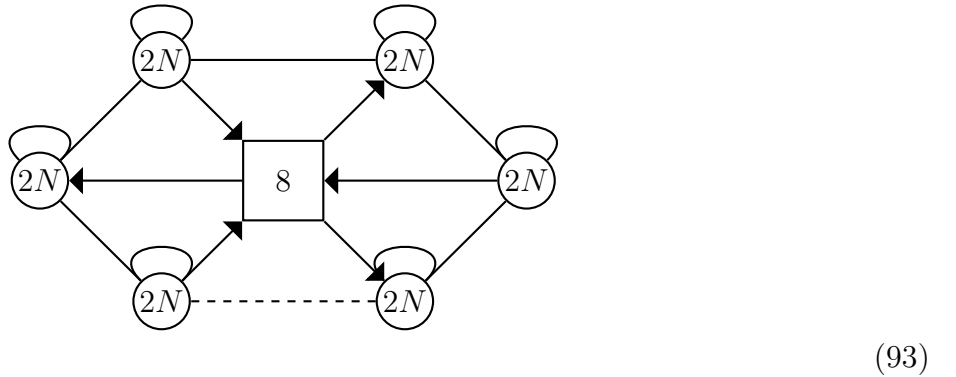
\begin{center}
\begin{tikzpicture}[thick, scale=0.8]
  \node[circle, draw, inner sep= 1pt](L1) at (2,0){$2N$};
  \node[circle, draw, inner sep= 1pt](L2) at (6,0){$2N$};
   \node[circle, draw, inner sep= 1pt](L3) at (8,-2){$2N$};
  \node[circle, draw, inner sep= 1pt](L4) at (6,-4){$2N$};
  \node[circle, draw, inner sep= 1pt](L5) at (2,-4){$2N$};
  \node[circle, draw, inner sep= 1pt](L6) at (0,-2){$2N$};
\node[draw, rectangle, minimum width=30pt, minimum height=30pt](L7) at (4,-2){$8$};
 \path[every node/.style={font=\sffamily\small,
  		fill=white,inner sep=1pt}]
(L1) edge   (L2)
(L2) edge   (L3)
(L3) edge   (L4)
(L4) edge [dashed] (L5)
(L5) edge   (L6)
(L6) edge   (L1)
(L1) edge [->]  (L7)
(L7) edge [->]  (L2)
(L3) edge [->]  (L7)
(L7) edge [->]  (L4)
(L5) edge [->]  (L7)
(L7) edge [->]  (L6)
(L1) edge [loop, looseness=4]  (L1)
(L2) edge [loop, looseness=4]  (L2)
(L3) edge [loop, looseness=4]  (L3)
(L4) edge [loop, looseness=4]  (L4)
(L5) edge [loop, looseness=4]  (L5)
(L6) edge [loop, looseness=4]  (L6);
\end{tikzpicture}\vspace{-.8cm}
\end{center}\be\label{wheel}\ee
where there is a superpotential term for each of the $2F$ triangles.

Before concluding, let us remark that the above construction works also for half-integral values of the flux. In these cases, the orbifold order is odd, and we need to introduce $2F$ $D7$/$O7$ stacks. Moreover, one of the flavor groups is fixed by the orientifold action, giving rise to an $SO(8)$ flavor symmetry. For $F=\tfrac{2p+1}2$, the resulting quiver is 
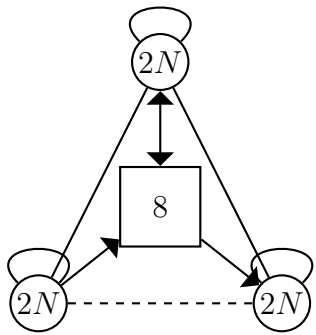
\begin{center}
\begin{tikzpicture}[thick, scale=0.8]
  \node[circle, draw, inner sep= 1pt](L1) at (2,-4){$2N$};
  \node[circle, draw, inner sep= 1pt](L2) at (6,-4){$2N$};
   \node[circle, draw, inner sep= 1pt](L3) at (4,0){$2N$};
\node[draw, rectangle, minimum width=30pt, minimum height=30pt](L7) at (4,-2.4){$8$};
 \path[every node/.style={font=\sffamily\small,
  		fill=white,inner sep=1pt}]
(L1) edge [dashed]  (L2)
(L2) edge   (L3)
(L3) edge   (L1)
(L1) edge [->]  (L7)
(L7) edge [->]  (L2)
(L3) edge [<->]  (L7)
(L1) edge [loop, looseness=4]  (L1)
(L2) edge [loop, looseness=4]  (L2)
(L3) edge [loop, looseness=4]  (L3);
\end{tikzpicture}\vspace{-.8cm}
\end{center}\be\label{Z3wheel}\ee
where there are $2p$ gauge groups in the bottom, and the doubly-oriented arrow represents $4$ chiral/antichiral pairs. The minimal case $p=0$ corresponds to the rank-$N$ $\mathcal{N}=2$ SQCD with $4$ fundamental flavors \cite{Kim:2017toz}.

\paragraph{Flipping fields from 6d} 

Let us now discuss how to characterize the flipping fields which arise in the compactification of the E-string theory on $T^2$. From the prescription of \cite{Kim:2017toz, Pasquetti:2019hxf} we conclude that,  starting from \eqref{wheel}, in order to make contact with the compactified E-string theory, we should flip the baryons built from all the $USp(2N) \times USp(2N)$ bifundamentals. These are all the fields which arise from $\Phi_2$ and $\Phi_3$, associated with the directions $z_2$ and $z_3$ in the Type IIB spacetime.

\subsection{Orbifolded E-string}

Let us now switch to the situation where the $M9$ wall wraps a singular complex surface, transverse to the $M5$ (in our parametrization its coordinates are $\tilde{z}_1,\tilde{z}_2$). We will only consider the case $\mathbb{C}^2/\bZ_2$, whose flux compactification on $T^2$ was considered in detail in \cite{Zafrir:2018hkr}. 
The 6d theory is specified by the choice of homomorphism from $\bZ_2$ to $E_8$ (see e.g. \cite{DelZotto:2014hpa, Mekareeya:2017jgc}). There are three choices preserving $E_8$, $E_7\times SU(2)$ and $SO(16)$ respectively. In what follows we will mainly concentrate on the $SO(16)$ case and comment on the $E_7\times SU(2)$ theory. It is perhaps easiest to characterize the 6d theory via its lagrangian description on a generic point of the tensor branch. This is a linear quiver with $N$ $SU(2)$ gauge groups (where $N$ is the number of $M5$ branes), with $8$ flavors on one side and $2$ on the other: 
$$\boxed{8}-SU(2)-\dots - SU(2)-\boxed{2}\,.$$ 
The $8$ flavors provide the expected $SO(16)$ global symmetry. We also have a further $SU(2)\times SU(2)$ global symmetry associated with the singularity and the isometry of the background. The effective lagrangian on the tensor branch of the $E_7\times SU(2)$ theory instead has the form 
$$\boxed{2}-SU(2)-\dots - SU(2)-\boxed{2}\,,$$
where the (say leftmost) $SU(2)$ gauge group is also coupled to the rank-$1$ E-string theory, in the sense that a $SU(2)$ subgroup of its $E_8$ global symmetry is gauged. We therefore see as a flavor symmetry only the $E_7$ commutant of the gauged $SU(2)$. The case of the $E_8$ theory is similar, the only difference being that the leftmost gauge group is now coupled to the rank-$2$ E-string theory.  

We will now focus on torus compactifications of the $SO(16)$ theory with flux breaking $SO(16)$ to $U(1)\times SU(8)$.
Let us normalize the flux $F$ such that it is (half-)integrally quantized.\footnote{The flux appearing here is twice the flux of ref.~\cite{Zafrir:2018hkr}.} As we mention in Appendix \ref{ExampleLoworder1}, for the minimal choice $F=1$, after a mass deformation for the $SU(2)$ global symmetry associated with the isometry of the M-theory background, we get a dual Type IIB model where $D3$ branes probe a $\bZ_4$ orbifold of $\mathbb{C}^3$ acting as (cf.~Eq.~\eqref{Z4actionII-1})
\be\label{Z4action-1II}
(z_1,z_2,z_3)\;\sim\;\left(-z_1,{\rm i}\,z_2,{\rm i}\,z_3\right)\,.
\ee
As was the case for \eqref{Z2storto} in the smooth situation, this minimal case has a symmetry under exchange of $z_2$ and $z_3$, so the projection under the permutation orientifold produces exactly the same quiver \eqref{ZZ4iiz1} as the ordinary projection we performed in Appendix \ref{ExampleLoworder1}. The only difference lies in the superpotential: The permutation orientifold allows for exactly the superpotential terms that were absent for the ordinary orientifold. To summarize, the theory is defined by the quiver
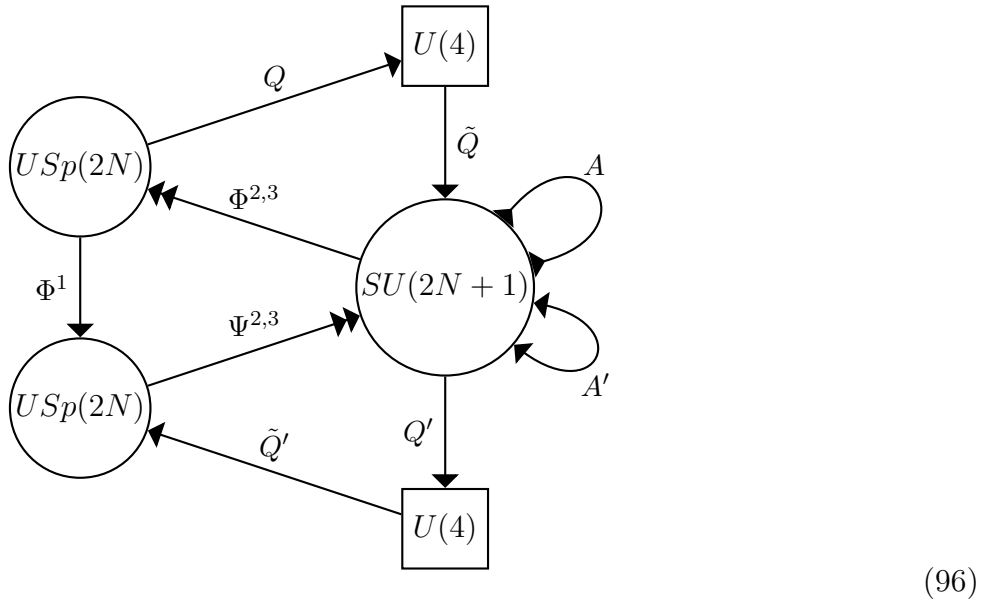
\begin{center}
\begin{tikzpicture}[thick, scale=0.8]
  \node[circle, draw, inner sep= 1pt](L1) at (6,-2){$USp(2N)$};
    \node[circle, draw, inner sep= 1pt](L7) at (6,-6){$USp(2N)$};
  \node[circle, draw, inner sep= 1pt](L2) at (12,-4){$SU(2N+1)$};
\node[draw, rectangle, minimum width=30pt, minimum height=30pt](L5) at (12,0){$U(4)$};
 \node[draw, rectangle, minimum width=30pt, minimum height=30pt](L6) at (12,-8){$U(4)$};
 \path[every node/.style={font=\sffamily\small,
  		fill=white,inner sep=1pt}]
(L1) edge [<<-] node[above=1.5mm] {$\Phi^{2,3}$} (L2)
(L1) edge [->] node[left=1mm] {$\Phi^1$} (L7)
(L7) edge [->>] node[above=1.5mm] {$\Psi^{2,3}$} (L2)
(L7) edge [<-] node[above=1mm] {$\tilde{Q}'$} (L6)
(L1) edge [->] node[above=1mm] {$Q$} (L5)
(L5) edge [->] node[right=1mm] {$\tilde{Q}$} (L2)
(L2) edge [->] node[left=1mm] {$Q'$} (L6)
(L2) edge [loop, >-<, in=15, out=50, looseness=6] node[above=2mm] {$A$}  (L2)
(L2) edge [loop,<->, in=-10, out=-40, looseness=6] node[below=1.7mm] {$A'$}  (L2);
\end{tikzpicture}\vspace{-1cm}
\end{center}\be\label{ZZ4iiz1E} \ee
with superpotential
\be\label{pot1}
W&=&\Phi^1\Psi^{[2}\Phi^{3]}+A'((\Phi^2)^2-(\Phi^3)^2)-A((\Psi^2)^2-(\Psi^3)^2)\nonumber\\
&&+\tilde{Q}(\Phi^2+\Phi^3)Q+\tilde{Q}'(\Psi^2+\Psi^3)Q'\,,
\ee
where in the first line several contractions with the symplectic forms of the $USp$ gauge groups are implicit. The second line reflects the fact that now the $D7$/$O7$ stack is located at $z_2+z_3=0$. As for \eqref{Z2MixedQuiverE}, we can perform the Leigh-Strassler analysis. The constraints on the anomalous dimensions of the chiral fields imposed by the beta functions for the three gauge couplings and the couplings for the superpotential terms appearing in \eqref{SuperpotZZ4iiz1} lead to seven independent equations and none of them are redundant. This implies we do not have any exactly marginal deformation without activating other superpotential terms. However, if we instead turn on all of the terms in \eqref{pot1}, we find thirteen equations which impose nine constraints, meaning that four of them are redundant.\footnote{We would get from \eqref{SuperpotZZ4iiz1} the two missing constraints if we added one superpotential term for each antisymmetric chiral $A$ and $A'$. The corresponding equations, together with the other seven, imply that $\Phi^2$ and $\Phi^3$ have the same anomalous dimension and analogously for $\Psi^2$ and $\Psi^3$. Using these constraints it is easy to see that the beta functions for the other four missing terms are automatically zero.} We therefore get a four (complex) dimensional manifolds of fixed points. It is therefore more natural to consider the superpotential \eqref{pot1} in order to get an interacting theory.

We note that, if we limit ourselves to imposing that the sum of the three beta functions as well as the beta function of the unitary gauge group vanish, then we get the following two-parameter family of theories
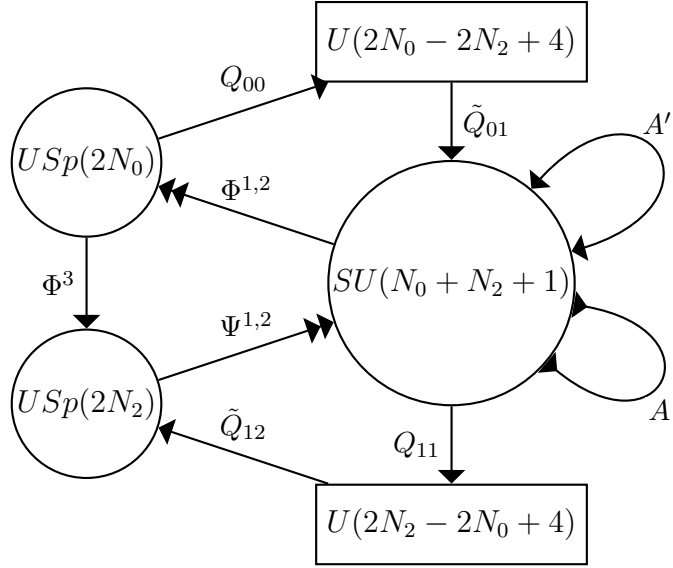
\begin{center}
\begin{tikzpicture}[thick, scale=0.8]
  \node[circle, draw, inner sep= 1pt](L1) at (6,-2){$USp(2N_0)$};
    \node[circle, draw, inner sep= 1pt](L7) at (6,-6){$USp(2N_2)$};
  \node[circle, draw, inner sep= 1pt](L2) at (12,-4){$SU(N_0+N_2+1)$};
\node[draw, rectangle, minimum width=30pt, minimum height=30pt](L5) at (12,0){$U(2N_0-2N_2+4)$};
 \node[draw, rectangle, minimum width=30pt, minimum height=30pt](L6) at (12,-8){$U(2N_2-2N_0+4)$};
 \path[every node/.style={font=\sffamily\small,
  		fill=white,inner sep=1pt}]
(L1) edge [<<-] node[above=1.5mm] {$\Phi^{1,2}$} (L2)
(L1) edge [->] node[left=1mm] {$\Phi^3$} (L7)
(L7) edge [->>] node[above=1.5mm] {$\Psi^{1,2}$} (L2)
(L7) edge [<-] node[above=1.5mm] {$\tilde{Q}_{12}$} (L6)
(L1) edge [->] node[above=2mm] {$Q_{00}$} (L5)
(L5) edge [->] node[right=1mm] {$\tilde{Q}_{01}$} (L2)
(L2) edge [->] node[left=1mm] {$Q_{11}$} (L6)
(L2) edge [loop, <->, in=15, out=50, looseness=6] node[above=2mm] {$A'$}  (L2)
(L2) edge [loop,>-<, in=-10, out=-40, looseness=6] node[below=2.3mm] {$A$}  (L2);
\end{tikzpicture}\be\label{ZZ4iiz1bis}\ee\vspace{-1cm}
\end{center}
where we recognize \eqref{ZZ4iiz1E} by putting $N_0=N_2$. The non-conformal family with $N_0=N_2+1$ was also discussed in \cite{Zafrir:2018hkr}, and corresponds to a different choice of $\mathbb{Z}_2 \to E_8$ homomorphism, i.e.~the one preserving $E_7\times SU(2)$. In this case the flux on the torus is chosen so that the global symmetry is broken to $E_6\times U(1)^2$. As in the previous case, in order to reach the Type IIB theory from the 6d SCFT, we first compactify on the torus with flux and we also need to turn on a mass deformation for the $SU(2)$ global symmetry associated with the isometry of the M-theory background.\footnote{The fact that different choices of $E_8$ embeddings of the orbifold group in M-theory map to different distributions of fractional $D3$/$D7$ branes in Type IIB was already observed in the $\mathcal{N}=2$ context in \cite{Giacomelli:2022drw}.}
\\

For general values of the flux, we argue that, after mass deformation, the dual Type IIB background is the $\mathbb{C}^3/\mathbb{Z}_{4F}$ orbifold
\be\label{GeneralTypeIIBZ2OrbEstring}
(z_1,z_2,z_3)\;\sim\;\left(-z_1,e^{\pi{\rm i}/2F}\,z_2,e^{\pi{\rm i}(2F-1)/2F}\,z_3\right)\,,
\ee
with an orientifold involution acting as in \eqref{permutationorientifold}. Again the map between M-theory and Type IIB coordinates is \eqref{CoordinatesMap}. Indeed, we have the identification
\be
(\tilde{z}_1,\tilde{z}_2)\;\sim\;\left(-\tilde{z}_1,-\tilde{z}_2\right)\,,
\ee
corresponding to the $\mathbb{Z}_2$ orbifold acting on the $M9$ wall. As was the case for the smooth E-string \eqref{GeneralTypeIIBSmoothEstring}, the $\mathcal{N}=2$ situation is recovered by setting $F=\tfrac12$.

The orbifold \eqref{GeneralTypeIIBZ2OrbEstring} forces us to introduce an invariant system of $D7$ branes and $O7$ planes, made of identical $D7$/$O7$ stacks wrapping the surfaces $z_2+e^{\pi{\rm i}(2F-1)m/2F}z_3=0$, with $m=0,\ldots,F-1$ for odd $F$ and $m=0,\ldots,2F-1$ for even $F$. Moreover, the law \eqref{OrbOrientLaw} expressing the relationship between orbifold and orientifold now becomes
\be
\omega\circ\gamma=\gamma^{2F-1}\circ\omega\,,
\ee
translating into the following relation on the $D3$-brane Chan-Paton space
\be\label{CompatPermOrientGen}
\tilde\Omega_3\,\Gamma_3^*\,\tilde\Omega_3^{-1}=\Gamma_3^{2F-1}\,.
\ee

For example, the next-to-minimal integral-flux case, $F=2$, which was also analyzed in \cite{Zafrir:2018hkr}, corresponds to a $\bZ_8$ orbifold. Choosing for $\Gamma_3$ as usual the regular representation, we immediately realize that an action of the permutation orientifold on the $D3$-brane Chan-Paton space solving \eqref{CompatPermOrientGen} is
\be
\tilde\Omega_3=\left(\begin{array}{cccccccc}{\rm i}J_{2N}&0&0&0&0&0&0&0\\0&0&0&0&0&{\rm i}I_{2N+1}&0&0\\ 0&0& {\rm i}J_{2N}&0&0&0&0&0\\ 0&0&0&0&0&0&0&{\rm i} I_{2N+1}\\ 0&0& 0&0&{\rm i}J_{2N}&0&0&0\\ 0&-{\rm i}I_{2N+1}&0&0&0&0&0&0\\ 0&0& 0&0&0&0&{\rm i}J_{2N}&0\\  0&0&0&-{\rm i}I_{2N+1}&0&0&0&0\end{array}\right)\,,
\ee
where we have already constrained the number of the various fractional $D3$ branes in order to meet the requirements of conformal invariance. After the projection, this creates four $USp(2N)$ and two $SU(2N+1)$ gauge groups. Scalars in the 3-3 sector behave as \eqref{PermOrientPhi} under orientifold involution. As for the 3-7 sector, the fields surviving the orbifold projection satisfy
\be
\begin{array}{rcl}e^{-\pi{\rm i}/8}Q^{(2)}&=&\Gamma_3Q^{(2)}\Gamma_{7_2}^{-1}\\ e^{-\pi{\rm i}/8}\tilde{Q}^{(2)}&=&\Gamma_{7_2}\tilde{Q}^{(2)}\Gamma_3^{-1}\,,\end{array}\qquad\qquad\begin{array}{rcl}e^{-3\pi{\rm i}/8}Q^{(3)}&=&\Gamma_3Q^{(3)}\Gamma_{7_3}^{-1}\\ e^{-3\pi{\rm i}/8}\tilde{Q}^{(3)}&=&\Gamma_{7_3}\tilde{Q}^{(3)}\Gamma_3^{-1}\,.\end{array}
\ee
Hence we can choose
\begin{eqnarray}
\Gamma_{7_2}={\rm diag}(e^{\pi{\rm i}/8}I_{M_0},e^{3\pi{\rm i}/8}I_{M_1},e^{5\pi{\rm i}/8}I_{M_2},e^{7\pi{\rm i}/8}I_{M_3},e^{-7\pi{\rm i}/8}I_{M_4},e^{-5\pi{\rm i}/8}I_{M_5},e^{-3\pi{\rm i}/8}I_{M_6},e^{-\pi{\rm i}/8}I_{M_7})\,,\nonumber\\
\Gamma_{7_3}={\rm diag}\,(e^{3\pi{\rm i}/8}I_{M_7},e^{-7\pi{\rm i}/8}I_{M_6},e^{-\pi{\rm i}/8}I_{M_5},e^{5\pi{\rm i}/8}I_{M_4},e^{-5\pi{\rm i}/8}I_{M_3},e^{\pi{\rm i}/8}I_{M_2},e^{7\pi{\rm i}/8}I_{M_1},e^{-3\pi{\rm i}/8}I_{M_0})\,.\nonumber
\end{eqnarray}
which yields a spectrum invariant under the orientifold involution
\be
\tilde\Omega_7\,\Gamma_{7_2}^*\,\tilde\Omega_7^{-1}=-\Gamma_{7_3}^{3}\,.
\ee
To make contact with the corresponding theory discussed in \cite{Zafrir:2018hkr}, we set $M_0=M_2=M_4=M_6=0$, and choose $M_1=M_3=M_5=M_7=4$ to meet conformal invariance. Performing the orbi-orientifold projection according to the usual rules, we arrive at the following quiver:\footnote{Note that the fields $\Phi^2_{72}$ and $\Phi^3_{67}$ satisfy $\Gamma_3\Phi^2_{72}\Gamma_3^\dagger=e^{-3\pi{\rm i}/4}\Phi^2_{72}$ and $\Gamma_3\Phi^3_{67}\Gamma_3^\dagger=e^{-\pi{\rm i}/4}\Phi^3_{67}$ respectively. Therefore they can be identified with $(\Phi^3_{27})^\dagger$ and $(\Phi^2_{76})^\dagger$ respectively.}
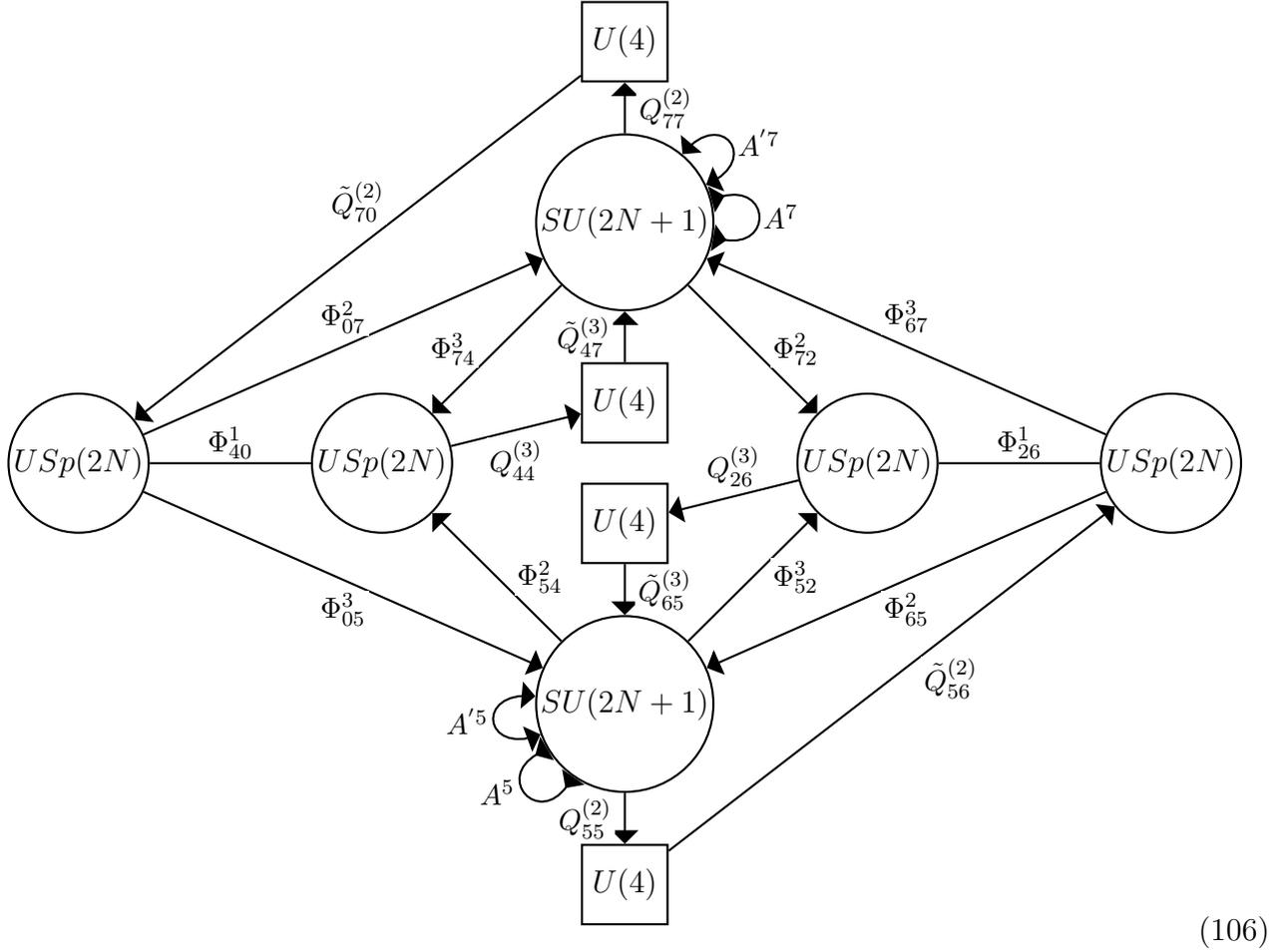
\begin{center}
\begin{tikzpicture}[thick, scale=0.8]
  \node[circle, draw, inner sep= 1pt](L0) at (-5,0){$USp(2N)$};
  \node[circle, draw, inner sep= 1pt](L4) at (0,0){$USp(2N)$};
   \node[circle, draw, inner sep= 1pt](L2) at (8,0){$USp(2N)$};
  \node[circle, draw, inner sep= 1pt](L6) at (13,0){$USp(2N)$};
  \node[circle, draw, inner sep= 1pt](L7) at (4,4){$SU(2N+1)$};
  \node[circle, draw, inner sep= 1pt](L5) at (4,-4){$SU(2N+1)$};
\node[draw, rectangle, minimum width=30pt, minimum height=30pt](L72) at (4,7){$U(4)$};
\node[draw, rectangle, minimum width=30pt, minimum height=30pt](L43) at (4,1){$U(4)$};
\node[draw, rectangle, minimum width=30pt, minimum height=30pt](L63) at (4,-1){$U(4)$};
\node[draw, rectangle, minimum width=30pt, minimum height=30pt](L52) at (4,-7){$U(4)$};
 \path[every node/.style={font=\sffamily\small, fill=white,inner sep=1pt}]
 (L0) edge node[above=.1mm] {$\Phi^1_{40}$} (L4)
  (L2) edge node[above=.1mm] {$\Phi^1_{26}$} (L6)
   (L72) edge [->] node[above=3mm] {$\tilde{Q}^{(2)}_{70}$} (L0)
     (L0) edge [->] node[above=1.3mm] {$\Phi^2_{07}$} (L7)
     (L7) edge [->] node[left=2.3mm] {$\Phi^3_{74}$} (L4)
        (L4) edge [->] node[below=.9mm] {$Q^{(3)}_{44}$} (L43)
             (L5) edge [->] node[right=2.2mm] {$\Phi^2_{54}$} (L4)
                  (L0) edge [->] node[below=1.3mm] {$\Phi^3_{05}$} (L5)
                          (L7) edge [->] node[right=1.4mm] {$Q^{(2)}_{77}$} (L72)
  (L5) edge [->] node[left=1.2mm] {$Q^{(2)}_{55}$} (L52)
    (L43) edge [->] node[left=1.4mm] {$\tilde{Q}^{(3)}_{47}$} (L7)
    (L63) edge [->] node[right=1.3mm] {$\tilde{Q}^{(3)}_{65}$} (L5)
    (L52) edge [->] node[right=3.8mm] {$\tilde{Q}^{(2)}_{56}$} (L6)
                 (L6) edge [->] node[below=1.3mm] {$\Phi^2_{65}$} (L5)
                 (L5) edge [->] node[right=2.2mm] {$\Phi^3_{52}$} (L2)
                  (L2) edge [->] node[above=0.9mm] {$Q^{(3)}_{26}$} (L63)
                  (L7) edge [->] node[right=2.2mm] {$\Phi^2_{72}$} (L2)
                  (L6) edge [->] node[above=1.4mm] {$\Phi^3_{67}$} (L7)
                  (L7) edge [loop, >-<, in=-10, out=15, looseness=4] node[right=.4mm] {$A^7$}  (L7)
                  (L7) edge [loop, <->, in=25, out=50, looseness=4] node[right=.6mm] {$A^{'7}$}  (L7)
                  (L5) edge [loop, >-<, in=-150, out=-125, looseness=4] node[left=1.2mm] {$A^5$}  (L5)
                  (L5) edge [loop, <->, in=-185, out=-160, looseness=4] node[left=.3mm] {$A^{'5}$}  (L5);
\end{tikzpicture}\vspace{-1cm}
\end{center}\be\label{ZafrirFig7} \ee
There is a cubic superpotential term for every triangle in the above figure. The four antisymmetric fields (the loop edges) have couplings of the schematic form $(A^i-A^{'i})(\Phi^3)^2$ and $(A^{'i}-A^i)(\Phi^2)^2$, $i=5,7$, where contractions with the symplectic form are understood. Finally, in complete analogy with the smooth E-string, there are mixing terms for the flavors induced by superpotential terms of the form \eqref{MixingFlavors}. In particular, here these are
\be
\tilde{Q}^{(3)}_{65}\,\Phi^2_{54}\,Q^{(3)}_{44}\qquad{\rm and}\qquad  \tilde{Q}^{(3)}_{47}\,\Phi^2_{72}\,Q^{(3)}_{26}\,,
\ee
which identify the two flavor nodes in the center, and
\be
\tilde{Q}^{(2)}_{70}\,\Phi^3_{05}\,Q^{(2)}_{55}\qquad{\rm and}\qquad  \tilde{Q}^{(2)}_{56}\,\Phi^3_{67}\,Q^{(2)}_{77}\,,
\ee
which identify the top with the bottom flavor node. Therefore the $U(4)^4$ flavor symmetry, which would naively be present, is actually broken to $U(4)\times U(4)$. The resulting theory is exactly the one described in Fig.~7 of \cite{Zafrir:2018hkr}.\\

To conclude, it is tempting to speculate that, in general, the $\bZ_k$-orbifolded E-string, compactified on a two-torus with $F$ units of flux, and suitably mass-deformed, gives rise at low-energy to a family of 4d SCFTs obtainable by probing with $D3$ branes a Type IIB $\mathbb{C}^3/\bZ_{2kF}$ orbifold background of the form
\be\label{GenOrbiInst}
(z_1,z_2,z_3)\;\sim\;\left(e^{2\pi{\rm i}/k}\,z_1,e^{-\pi{\rm i}/kF}\,z_2,e^{-\pi{\rm i}(2F-1)/kF}\,z_3\right)\,,
\ee
with an orientifold involution that exchanges the coordinates $z_2$ and $z_3$. Again the map between M-theory and Type IIB coordinates is \eqref{CoordinatesMap}, which gives us
\be
(\tilde{z}_1,\tilde{z}_2)\;\sim\;\left(e^{2\pi{\rm i}/k}\,\tilde{z}_1,e^{-2\pi{\rm i}/k}\,\tilde{z}_2\right)\,,
\ee
corresponding to the $\mathbb{Z}_k$ orbifold acting on the $M9$ wall. As before, the $\mathcal{N}=2$ situation is recovered by setting $F=\tfrac12$.

Such theories have not been derived yet, and thus we leave the check of this claim to future studies.

\section{Conclusions}\label{sec5}

In this paper we have shown that there is a deep connection between the torus compactification of 6d theories defined on the worldvolume of $M5$ branes in M-theory and 4d theories on $D3$ branes probing an F-theory background. The two theories are connected by an RG flow, which involve in general a mass deformation and the removal of gauge singlets (flipping fields). The latter can be easily reintroduced once the Type IIB theory is given, since they flip fields whose vev parametrize the motion of the $D3$ branes along spacetime directions which have no counterpart in M-theory. This is true at least in all the cases we have discussed explicitly. We have determined the detailed correspondence for a specific choice of flux in each 6d theory, namely the choice which maps to orbifold data in Type IIB. We checked our claim for 6d theories defined on a M-theory background involving either generic Abelian orbifolds, or $M9$ walls, or both of them (albeit in this case for the $\mathbb{Z}_2$ orbifold only). 

We expect the correspondence to hold much more generally: Our conjecture for the Type IIB dual of generic orbi-instanton theories is contained in formula \eqref{GenOrbiInst}, with $F$ specifying the units of flux turned on in the M-theory setup. Moreover, as suggested by \cite{Bah:2021iaa} which considered in detail the special case of class $\mathcal{S}_2$ theories with arbitrary flux on the torus, we expect such a correspondence to hold for generic flux configurations too, although this requires going beyond the orbi-orientifold Calabi-Yau setups of Type IIB we considered here. It would be very interesting to explore this idea further, possibly without restricting to perturbative Type IIB backgrounds, and understand how general this correspondence is. In particular, it is not clear to us if it applies to torus compactifications without punctures only, or it can be extended to more general choices of Riemann surface. We hope this provides a useful new perspective on both the study of 6d compactifications and of generic $\mathcal{N}=1$ $\mathcal{S}$-folds. Based on the analogy with the more supersymmetric case, we expect non-trivial $\mathcal{S}$-fold configurations in F-theory preserving minimal supersymmetry to correspond to 6d compactifications involving almost commuting holonomies on the torus. So far these have been studied only in the absence of flux.\footnote{See \cite{Bah:2017gph} for a general discussion about the case with flux in the context of class $\mathcal{S}_k$ theories.}

There are several other interesting directions for future research. First of all it is worth pointing out that we have not touched upon the case of Abelian orbifolds with two generators, $\mathcal{A}\simeq\mathbb{Z}_{n_1}\times\mathbb{Z}_{n_2}$. When $n_1$ and $n_2$ are \emph{not} coprime, a new ingredient needs to be specified to define the string vacuum, and hence to derive the low-energy field theory on the probe. This data is called ``discrete torsion'', and it amounts to introducing a phase $\epsilon(\gamma_1,\gamma_2)$ in the action of $\gamma_1\in\mathbb{Z}_{n_1}$ on states in the closed-string sector twisted by $\gamma_2\in\mathbb{Z}_{n_2}$. This phase must be a GCD$(n_1,n_2)$-root of unity. At the perturbartive level, the effect of this phase on the open-string sector is to turn the embedding of the orbifold twist in the Chan-Paton bundle into a projective representation of the orbifold group, i.e.~the matrices \eqref{CPembOrb} obey the group law of $\mathcal{A}$ up to the aforementioned phase:
\be
\Gamma(\gamma_1)\Gamma(\gamma_2)=\epsilon(\gamma_1,\gamma_2) \Gamma(\gamma_2)\Gamma(\gamma_1)\,.
\ee
This in turn affects the low-energy spectrum on the $D3$ branes. It would be interesting to understand in general what is the feature of the M-theory setup which the discrete torsion originates from under the correspondence we discussed in this paper. It would also be nice to extend our analysis to non-Abelian orbifold groups. Finally, it could be interesting to consider the inclusion of $T$-branes in the Type IIB side\footnote{See \cite{Kimura:2020hgw} for a discussion about $T$-branes in the context of $\mathcal{N}=2$ $\mathcal{S}$-fold theories.} and elucidate their M-theory counterpart.

\subsection*{Acknowledgements}

We would like to thank Massimo Bianchi, Federico Bonetti, Sara Pasquetti, and Angel Uranga for helpful discussions. RS would like to thank the Department of Physics of the University of Milan Bicocca for kind hospitality in the final stage of this project. The work of SG is supported by the INFN grant ``Per attivit\`a di formazione per sostenere progetti di ricerca'' (GRANT 73/STRONGQFT).

\appendix 

\section{Conformal theories at low orbifold order}\label{ExampleLoworder1}  

In this section we will go in detail through \emph{all} the perturbative anomaly-free and conformal configurations which appear for Abelian orbifolds of orders three and four.

\subsubsection*{$\mathbb{C}^3/\ZZ_3$} 

We start by taking the $\ZZ_3$ to only act on two of the three internal coordinates:
\be\label{NnChiZ3}
(z_1,z_2,z_3)\;\sim\;\left(z_1,e^{2\pi{\rm i}/3}z_2,e^{-2\pi{\rm i}/3}z_3\right)\,.
\ee
First consider a $D7$/$O7$ stack placed as usual at $z_3=0$.\footnote{The case in which both coordinates affected by the orbifold action are longitudinal to the stack leads to $\mathcal{N}=2$ theories, and was analyzed in depth in \cite{Giacomelli:2022drw}.} The rules of Section \ref{Pert:Sec} now tell us that there is a one-parameter family of superconformal gauge theories, encoded in the following quiver
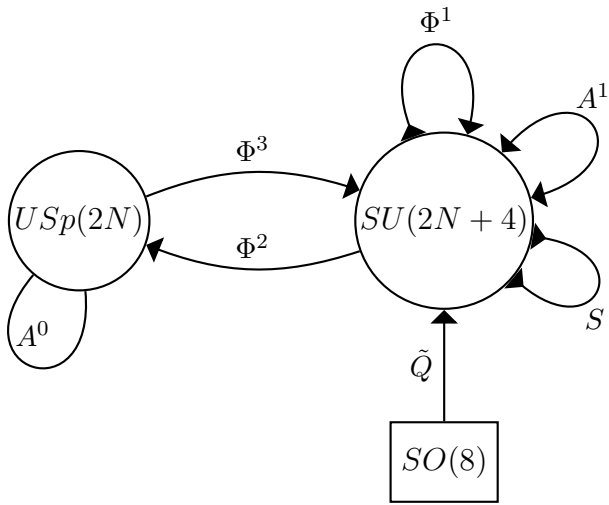
\begin{center}
\begin{tikzpicture}[thick, scale=0.8]
  \node[circle, draw, inner sep= 1pt](L1) at (6,-4){$USp(2N)$};
  \node[circle, draw, inner sep= 1pt](L2) at (12,-4){$SU(2N+4)$};
 \node[draw, rectangle, minimum width=30pt, minimum height=30pt](L6) at (12,-8){$SO(8)$};
 \path[every node/.style={font=\sffamily\small,
  		fill=white,inner sep=1pt}]
(L1) edge [bend right=20,<-] node[above=1mm] {$\Phi^{2}$} (L2)
(L1) edge [bend left=20,->] node[above=1mm] {$\Phi^{3}$} (L2)
(L1) edge [loop, in=-85, out=-130, looseness=6] node[above=2.3mm] {$A^0$}  (L1)
(L2) edge [<-] node[left=1mm] {$\tilde{Q}$} (L6)
(L2) edge [loop, >->, in=75, out=110, looseness=6] node[above=1.4mm] {$\Phi^{1}$}  (L2)
(L2) edge [loop, <->, in=15, out=50, looseness=6] node[above=2mm] {$A^{1}$}  (L2)
(L2) edge [loop,>-<, in=-10, out=-40, looseness=6] node[below=2mm] {$S$}  (L2);
\end{tikzpicture}\vspace{-.5cm}
\end{center}\be\label{ZZ31} \ee
Here the original flavor symmetry is left unbroken. Besides the two bifundamental chiral fields, this theory features a chiral field $A^0$ in the antisymmetric representation of the symplectic gauge group and an antichiral $A^1$ in the antisymmetric of the unitary gauge group; the latter also has an adjoint $\Phi^1$, a symmetric $S$, and $8$ antifundamentals. The superpotential of this family of theories looks schematically like
\be
W=J^{\alpha\beta}A^0_{\beta\gamma}\Phi^{[2}_\alpha\Phi^{3]\gamma} +\Phi^1A^1S+\Tr_{SO(8)}\left(\tilde{Q}S\tilde{Q}\right)\,,
\ee
where $J$ is the symplectic form of $USp(2N)$, and indices of the unitary gauge group have been suppressed. It is immediate to see that no mass deformations for the $SO(8)$ flavor symmetry are allowed.\\

Two more families of SCFTs can be found for the orbifold action \eqref{NnChiZ3}, now involving no $D7$ branes but just pure $O3$$^\pm$ projections \cite{Bianchi:2020fuk}. They are the following $\mathcal{N}=2$ models
\begin{center}
\begin{tikzpicture}[thick, scale=0.8]
  \node[circle, draw, inner sep= 1pt](L1) at (6,-4){$USp(2N)$};
  \node[circle, draw, inner sep= 1pt](L2) at (12,-4){$SU(2N+2)$};
 \path[every node/.style={font=\sffamily\small,
  		fill=white,inner sep=1pt}]
(L2) edge [loop,-, in=15, out=50, looseness=6]node[above=2mm] {$S$}  (L2)
(L1) edge  (L2);
\end{tikzpicture}\vspace{-1cm}
\end{center}\be\label{ZZ3O3+} \ee\\
corresponding to the $O3$$^+$ projection, with the loop indicating a hypermultiplet transforming in the symmetric representation of the unitary gauge group, and
\begin{center}
\begin{tikzpicture}[thick, scale=0.8]
  \node[circle, draw, inner sep= 1pt](L1) at (6,-4){$SO(2N+2)$};
  \node[circle, draw, inner sep= 1pt](L2) at (12,-4){$SU(2N)$};
 \path[every node/.style={font=\sffamily\small,
  		fill=white,inner sep=1pt}]
(L2) edge [loop,-, in=10, out=50, looseness=6]node[above=2mm] {$A$}  (L2)
(L1) edge  (L2);
\end{tikzpicture}\vspace{-1cm}
\end{center}\be\label{ZZ3O3-} \ee\\
corresponding to the $O3$$^-$ projection, with the loop indicating a hypermultiplet transforming in the antisymmetric representation of the unitary gauge group. Note that \eqref{ZZ3O3-} and \eqref{ZZ2O3} coincide for $N=1$.\\

Let us now switch to the following $\ZZ_3$ orbifold action on the three internal complex coordinates
\be
(z_1,z_2,z_3)\;\sim\;e^{2\pi{\rm i}/3}\left(z_1,z_2,z_3\right)\,.
\ee
In this case, it is easy to see that with only $O3$ planes and no $D7$ branes it is not possible to construct anomaly-free configurations with all beta functions simultaneously vanishing \cite{Bianchi:2020fuk}. Hence the only option we have is to introduce a tadpole-free $D7$/$O7$ stack. Following the rules reviewed in Section \ref{Pert:Sec}, we conclude that there exists a one-parameter family of superconformal gauge theories, codified by the following quiver
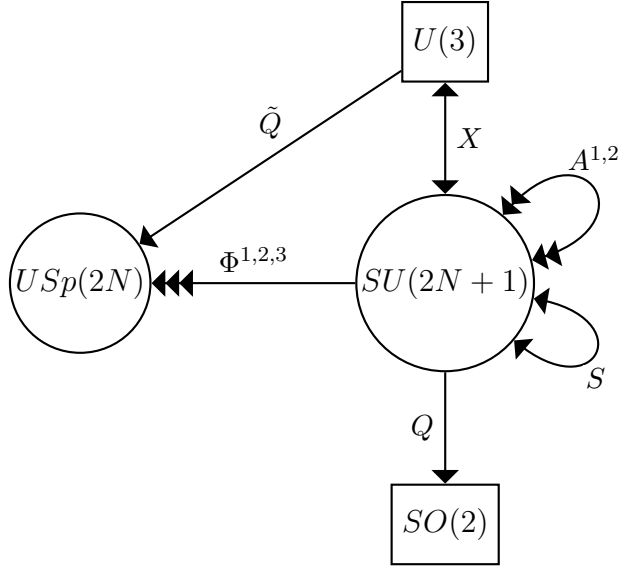
\begin{center}
\begin{tikzpicture}[thick, scale=0.8]
  \node[circle, draw, inner sep= 1pt](L1) at (6,-4){$USp(2N)$};
  \node[circle, draw, inner sep= 1pt](L2) at (12,-4){$SU(2N+1)$};
\node[draw, rectangle, minimum width=30pt, minimum height=30pt](L5) at (12,0){$U(3)$};
 \node[draw, rectangle, minimum width=30pt, minimum height=30pt](L6) at (12,-8){$SO(2)$};
 \path[every node/.style={font=\sffamily\small,
  		fill=white,inner sep=1pt}]
(L1) edge [<<<-] node[above=1mm] {$\Phi^{1,2,3}$} (L2)
(L1) edge [<-] node[above=2mm] {$\tilde{Q}$} (L5)
(L5) edge [<->] node[right=1mm] {$X$} (L2)
(L2) edge [->] node[left=1mm] {$Q$} (L6)
(L2) edge [loop, <<->>, in=15, out=50, looseness=6] node[above=2mm] {$A^{1,2}$}  (L2)
(L2) edge [loop,<->, in=-10, out=-40, looseness=6] node[below=1.5mm] {$S$}  (L2);
\end{tikzpicture}\vspace{-1cm}
\end{center}\be\label{ZZ3} \ee
The fields $X,A^{1,2},S$ all originate from non-orientable open strings, and are antichiral transforming in the fundamental, antisymmetric, and symmetric representation of the unitary gauge group respectively. 

The superpotential of this family of theories looks schematically like
\be
W=J^{\alpha\beta}\Phi^{[1}_\alpha\Phi^{2]}_\beta\, S+\Tr_{SO(2)}\left(QSQ\right)+J^{\alpha\beta}\Tr_{U(3)}\left(\tilde{Q}_\alpha (X\Phi^3)_\beta\right)\,,
\ee
where $J$ is the symplectic form of $USp(2N)$, and indices of the unitary gauge group have been suppressed. The couplings to both the antisymmetric fields can be easily seen to vanish identically.

As far as mass deformations are concerned, Eq.~\eqref{OrbifoldMass}, together with the second Eq.~in \eqref{PhiOmega}, tells us that, of the original mass matrix in the adjoint of $SO(8)$, the following two pieces survive the orientifold projection
\be\label{massesZ3}
\mathcal{M}_{[IJ]}\quad{\rm and}\quad \tilde{\mathcal{M}}^I_A\,,\qquad\qquad I=1,2,3\,,\;\;\;A=1,2\,,
\ee
where the first one transforms in the antisymmetric representation of $U(3)$ and the second one is a bifundamental of the two flavor groups. The first matrix in \eqref{massesZ3} corresponds to a mass term for the fields $\tilde{Q}$, whereas the second matrix couples $Q$ to $X$, i.e.~we have the following mass terms
\be
\mathcal{M}_{[IJ]}\tilde{Q}^I_\alpha\tilde{Q}^J_\beta J^{\alpha\beta}\qquad{\rm and}\qquad \tilde{\mathcal{M}}^I_AQ_BX_I\delta^{AB}\,.
\ee

\subsubsection*{$\mathbb{C}^3/\ZZ_4$}

We start from the following $\ZZ_4$ action on the three internal complex coordinates
\be\label{Z4actionII-1}
(z_1,z_2,z_3)\;\sim\;\left({\rm i}\,z_1,{\rm i}\,z_2,-z_3\right)\,.
\ee
At this point, we need to distinguish the type of orientifold projection. Consider first the projection with fixed nodes: In this case only the $D7$/$O7$ background allows for SCFTs on the probe.\footnote{Without $D7$ branes, and using mixed $O3$$^\pm$ projections, one can construct quiver gauge theories with vanishing sum of beta functions, but none of them can have all beta functions simultaneously vanishing \cite{Bianchi:2020fuk}.} The resulting family of field theories, however, depends on which is the transverse direction of the $D7$/$O7$ stack. In particular, if we place the stack at $z_3=0$, we find three families of them with different flavor structure:
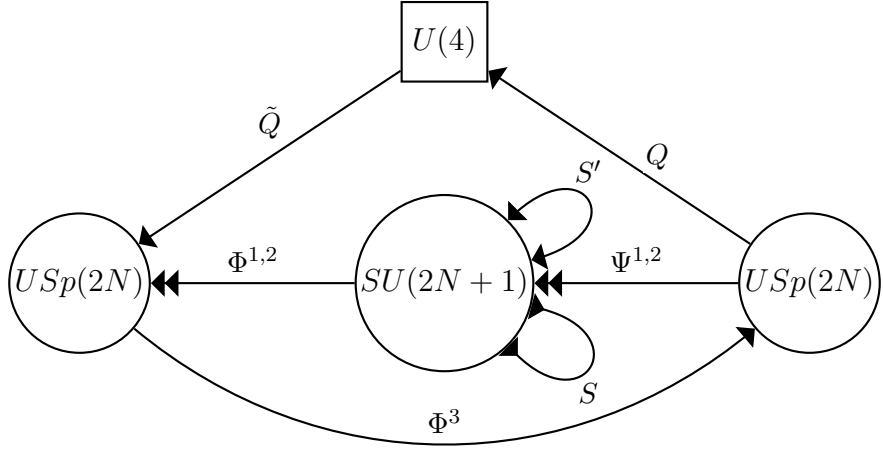
\begin{center}
\begin{tikzpicture}[thick, scale=0.8]
  \node[circle, draw, inner sep= 1pt](L1) at (6,-4){$USp(2N)$};
  \node[circle, draw, inner sep= 1pt](L2) at (12,-4){$SU(2N+1)$};
\node[draw, rectangle, minimum width=30pt, minimum height=30pt](L5) at (12,0){$U(4)$};
 \node[circle, draw, inner sep= 1pt](L4) at (18,-4){$USp(2N)$};
 \path[every node/.style={font=\sffamily\small,
  		fill=white,inner sep=1pt}]
(L1) edge [<<-] node[above=1mm] {$\Phi^{1,2}$} (L2)
(L1) edge [<-] node[above=2mm] {$\tilde{Q}$} (L5)
(L5) edge [<-] node[right=3mm] {$Q$} (L4)
(L2) edge [loop, <->, in=15, out=45, looseness=6] node[above=2mm] {$S'$}  (L2)
(L2) edge [loop,>-<, in=-15, out=-45, looseness=6] node[below=1.8mm] {$S$}  (L2)
(L4) edge [->>] node[above=1mm] {$\Psi^{1,2}$} (L2)
(L1) edge [bend right=40, ->] node[above=1mm] {$\Phi^{3}$} (L4);
\end{tikzpicture}\vspace{-1.3cm}
\end{center}\be\label{ZZ4v1} \ee
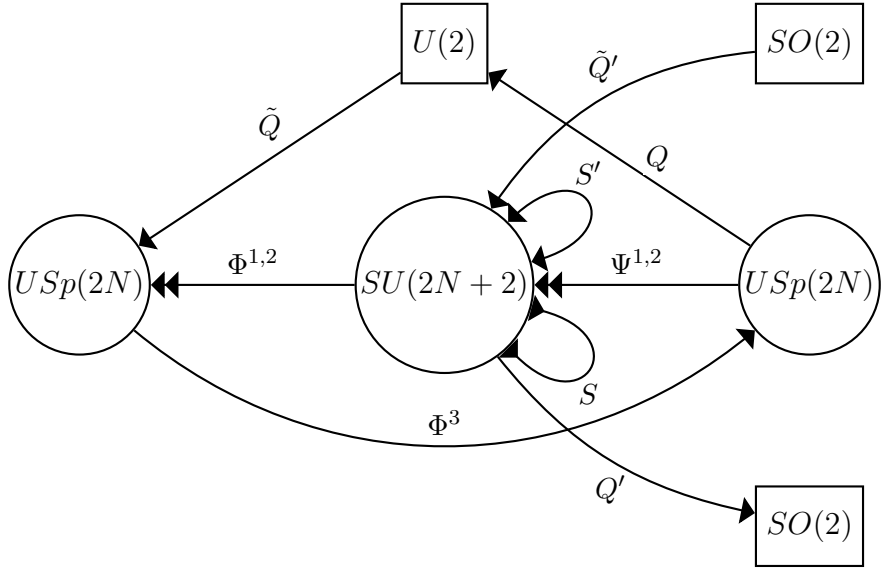
\begin{center}
\begin{tikzpicture}[thick, scale=0.8]
  \node[circle, draw, inner sep= 1pt](L1) at (6,-4){$USp(2N)$};
  \node[circle, draw, inner sep= 1pt](L2) at (12,-4){$SU(2N+2)$};
\node[draw, rectangle, minimum width=30pt, minimum height=30pt](L5) at (12,0){$U(2)$};
 \node[draw, rectangle, minimum width=30pt, minimum height=30pt](L6) at (18,0){$SO(2)$};
 \node[circle, draw, inner sep= 1pt](L4) at (18,-4){$USp(2N)$};
 \node[draw, rectangle, minimum width=30pt, minimum height=30pt](L7) at (18,-8){$SO(2)$};
 \path[every node/.style={font=\sffamily\small,
  		fill=white,inner sep=1pt}]
(L1) edge [<<-] node[above=1mm] {$\Phi^{1,2}$} (L2)
(L1) edge [<-] node[above=2mm] {$\tilde{Q}$} (L5)
(L5) edge [<-] node[right=3mm] {$Q$} (L4)
(L2) edge [loop, <->, in=15, out=45, looseness=6] node[above=2mm] {$S'$}  (L2)
(L2) edge [loop,>-<, in=-15, out=-45, looseness=6] node[below=1.8mm] {$S$}  (L2)
(L4) edge [->>] node[above=1mm] {$\Psi^{1,2}$} (L2)
(L1) edge [bend right=40, ->] node[above=1mm] {$\Phi^{3}$} (L4)
(L2) edge [bend right=20, ->] node[below=1.5mm] {$Q'$} (L7)
(L2) edge [bend left=25, <-] node[above=1.5mm] {$\tilde{Q}'$} (L6);
\end{tikzpicture}\vspace{-1.3cm}
\end{center}\be\label{ZZ4v2} \ee
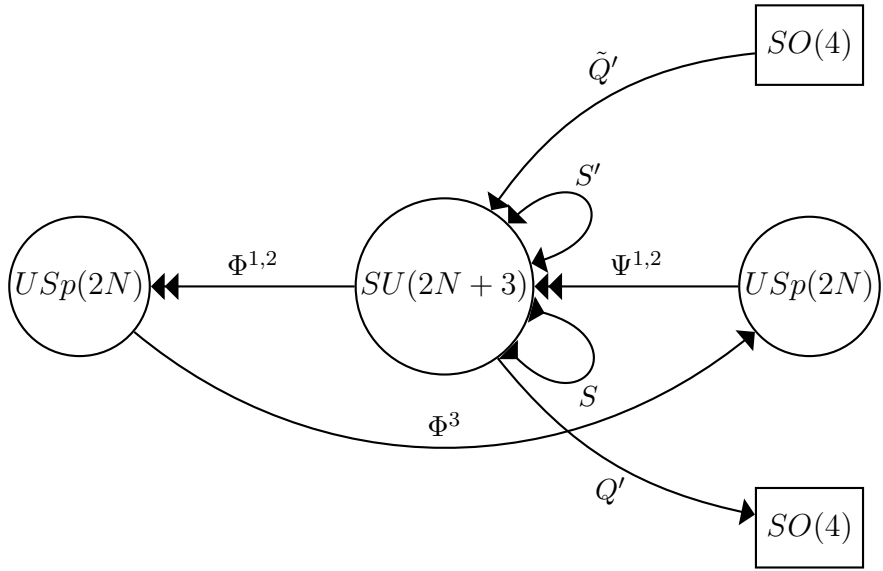
\begin{center}
\begin{tikzpicture}[thick, scale=0.8]
  \node[circle, draw, inner sep= 1pt](L1) at (6,-4){$USp(2N)$};
  \node[circle, draw, inner sep= 1pt](L2) at (12,-4){$SU(2N+3)$};
 \node[draw, rectangle, minimum width=30pt, minimum height=30pt](L6) at (18,0){$SO(4)$};
 \node[circle, draw, inner sep= 1pt](L4) at (18,-4){$USp(2N)$};
 \node[draw, rectangle, minimum width=30pt, minimum height=30pt](L7) at (18,-8){$SO(4)$};
 \path[every node/.style={font=\sffamily\small,
  		fill=white,inner sep=1pt}]
(L1) edge [<<-] node[above=1mm] {$\Phi^{1,2}$} (L2)
(L2) edge [loop, <->, in=15, out=45, looseness=6] node[above=2mm] {$S'$}  (L2)
(L2) edge [loop,>-<, in=-15, out=-45, looseness=6] node[below=1.8mm] {$S$}  (L2)
(L4) edge [->>] node[above=1mm] {$\Psi^{1,2}$} (L2)
(L1) edge [bend right=40, ->] node[above=1mm] {$\Phi^{3}$} (L4)
(L2) edge [bend right=20, ->] node[below=1.5mm] {$Q'$} (L7)
(L2) edge [bend left=25, <-] node[above=1.5mm] {$\tilde{Q}'$} (L6);
\end{tikzpicture}\vspace{-1.3cm}
\end{center}\be\label{ZZ4v3} \ee
The fields $S,S'$ are chiral and antichiral respectively, transforming in the symmetric representation of the unitary gauge group. The three above quivers are interconnected by higgsing processes, whereby one gives vevs to fundamental flavors only. Starting from the bottom quiver \eqref{ZZ4v3}, we give a non-trivial vev to one fundamental and one antifundamental chiral field of $SU(2N+3)$, i.e.~$Q',\tilde{Q}'$ respectively. This breaks the flavor symmetry $SO(4)^2$ to $SO(2)^2$ and, at the same time, the gauge symmetry $SU(2N+3)$ to $SU(2N+2)$. As a consequence, the two neighboring $USp(2N)$ gauge groups gain two favors each, transforming in the fundamental/antifundamental of the right/left $USp(2N)$, thereby making a new $U(2)$ flavor symmetry emerge. This is now the theory represented by the middle quiver \eqref{ZZ4v2}. By the same token, we can give vev to the remaining chiral fields $Q',\tilde{Q}'$, thus breaking completely the orthogonal flavor symmetry, reducing the gauge group to $SU(2N+1)$, and enhancing the unitary flavor symmetry to $U(4)$. We now landed on the theory associated to the top quiver \eqref{ZZ4v1}. At this point, we can give non-trivial vevs to the pair $Q,Q'$ such that we completely break the $U(4)$ flavor symmetry. This operation partially breaks both the $USp(2N)$ gauge groups to a pair of $USp(2N-2)$ gauge groups. Consequently, the $SU(2N+1)$ gauge group gains four fundamental and four antifundamental chiral fields, thus making an $SO(4)^2$ flavor symmetry appear. Therefore we see that, by this chain of higgsing, we got back to the bottom quiver \eqref{ZZ4v3}, but with $N\to N-1$. This process continues until all gauge groups disappear (first the symplectic ones and last the unitary one), when $N$ hits zero, yielding a bunch of free chiral fields.

The superpotential of these families of theories looks schematically like
\begin{eqnarray}\label{SuperpotZZ4v}
W= \Phi_{\;\;\rho}^{3\alpha}\Psi^{[1\rho}\Phi^{2]}_\alpha+J^{\alpha\beta}\Phi^{1}_\alpha\Phi^{2}_\beta S'+J_{\rho\sigma}\Psi^{1\rho}\Psi^{2\sigma} S+\Tr_{SO}\left(Q'S'Q'+\tilde{Q}'S\tilde{Q}'\right)+\Tr_{U}\left(\tilde{Q}_\alpha\Phi_{\;\;\rho}^{3\alpha}Q^\rho\right)\hspace{-.12cm}\,,\nonumber
\end{eqnarray}
where $J^{\alpha\beta},J_{\rho\sigma}$ are the symplectic forms of the left-most and right-most gauge groups respectively, and indices of the unitary gauge group have been suppressed.

As far as mass deformations are concerned, Eq.~\eqref{OrbifoldMass}, together with the second Eq.~in \eqref{PhiOmega}, tells us that, of the original mass matrix in the adjoint of $SO(8)$, the following three pieces survive the orientifold projection
\be\label{massesZ4}
\mathcal{M}_{[IJ]}\,,\qquad\bar{\mathcal{M}}^{[IJ]}\,,\qquad \tilde{\mathcal{M}}_A^P\,,
\ee
where the first two matrices transform in the antisymmetric representation (and in its conjugate) of the unitary flavor group, while the third matrix is a bifundamental of the two orthogonal flavor groups. The matrices \eqref{massesZ4} lead respectively to the following mass terms
\be
\mathcal{M}_{[IJ]}\tilde{Q}^I_\alpha\tilde{Q}^I_\beta J^{\alpha\beta}\,,\qquad\bar{\mathcal{M}}^{[IJ]}Q_I^\rho Q_J^\sigma J_{\rho\sigma}\,,\qquad \tilde{\mathcal{M}}^P_A\tilde{Q}'^AQ'_P\,,
\ee
where gauge indices have been suppressed in the last term.\\

Consider now placing the $D7$/$O7$ stack at e.g.~$z_1=0$.\footnote{Placing it at $z_2=0$ is of course equivalent.} Recalling the general discussion of Section \ref{orbi-orient}, in order to deal with the 3-7 sector, we need to choose the orbifold action on the $D3$ and $D7$ Chan-Paton spaces respectively as $\Gamma_3={\rm diag}(I_{N_0},{\rm i}I_{N_1},-I_{N_2},-{\rm i}I_{N_3})$ and $\Gamma_7=e^{\pi{\rm i}/4}{\rm diag}(I_{M_0},{\rm i}I_{M_1},-I_{M_2},-{\rm i}I_{M_3})$.\footnote{This choice is compatible with the orientifold involution that fixes two gauge nodes and no flavor nodes. For the other orientifold involution, that fixes two flavor nodes and no gauge nodes, one needs to choose $\Gamma_3=e^{\pi{\rm i}/4}{\rm diag}(I_{N_0},{\rm i}I_{N_1},-I_{N_2},-{\rm i}I_{N_3})$ and $\Gamma_7={\rm diag}(I_{M_0},{\rm i}I_{M_1},-I_{M_2},-{\rm i}I_{M_3})$.} This leads to the following 3-7 and 7-3 fields surviving the orbifold projection \eqref{ProjQtQ}
\be
Q=\left(\begin{array}{cccc}Q_{00}&0&0&0\\ 0&Q_{11}&0&0\\ 0&0&Q_{22}&0\\ 0&0&0&Q_{33} \end{array}\right)\,,\qquad \tilde{Q}=\left(\begin{array}{cccc}0&\tilde{Q}_{01}&0&0\\ 0&0&\tilde{Q}_{12}&0\\ 0&0&0&\tilde{Q}_{23}\\ \tilde{Q}_{30}&0&0&0 \end{array}\right)\,.
\ee
Similarly to what we have seen for the $\mathbb{Z}_2$ orbifold action \eqref{Z2storto}, in order for the orbifold-projected spectrum to be orientifold invariant, we are forced to choose a \emph{mixed} orientifold involution, which treats differently gauge and flavor nodes. Again, insisting on conformal invariance, the two inequivalent choices lead to either twice the expected number of $D7$ branes or to no $D7$ branes. While the second case can be equivalently formulated in terms of a mixed $O3$$^\pm$ projection and will be discussed shortly, the first case corresponds to a projection with two fixed gauge nodes and no fixed flavor nodes, i.e.
\be\label{2G0F}
\Omega_3=\left(\begin{array}{cccc}{\rm i}J_{2N_0}&0&0&0\\ 0&0&0&{\rm i} I_{N_1}\\ 0&0&{\rm i}J_{2N_2}&0\\ 0& -{\rm i} I_{N_1}&0&0\end{array}\right)\,,\qquad \hat{\Omega}_7=\left(\begin{array}{cccc}0&0&0&I_{M_0}\\ 0&0&I_{M_1}&0\\ 0&I_{M_1}&0&0\\  I_{M_0}&0&0&0\end{array}\right)\,.
\ee
Using Eq.~\eqref{OmegaProjAPhiQ}, we deduce that the gauge group is broken to $USp(2N_0)\times SU(N_1)\times USp(2N_2)$, while the flavor group to $U(M_0)\times U(M_1)$; moreover the surviving spectrum comprises a chiral/antichiral-multiplet pair $A,A'$ transforming in the antisymmetric representation of $SU(N_1)$, a pair of bifundamental chiral multiplets between each symplectic gauge group and the unitary gauge group $\Phi^{1,2},\Psi^{1,2}$, a bifundamental chiral multiplet between the symplectic gauge nodes $\Phi^3$, and a number of fundamental chiral multiplets connecting gauge and flavor nodes $Q_{00},Q_{11},\tilde{Q}_{01},\tilde{Q}_{12}$. Anomaly cancelation for the unitary gauge group imposes $M_0-M_1=4(N_0-N_2)$. Using \eqref{SCTcond}, it is immediate to see that, as anticipated, the vanishing of the sum of the three beta functions imposes $M_0+M_1=8$. If we also force all beta functions to simultaneously vanish, we land on the one-parameter family of $\mathcal{N}=1$ superconformal field theories, described by the following quiver
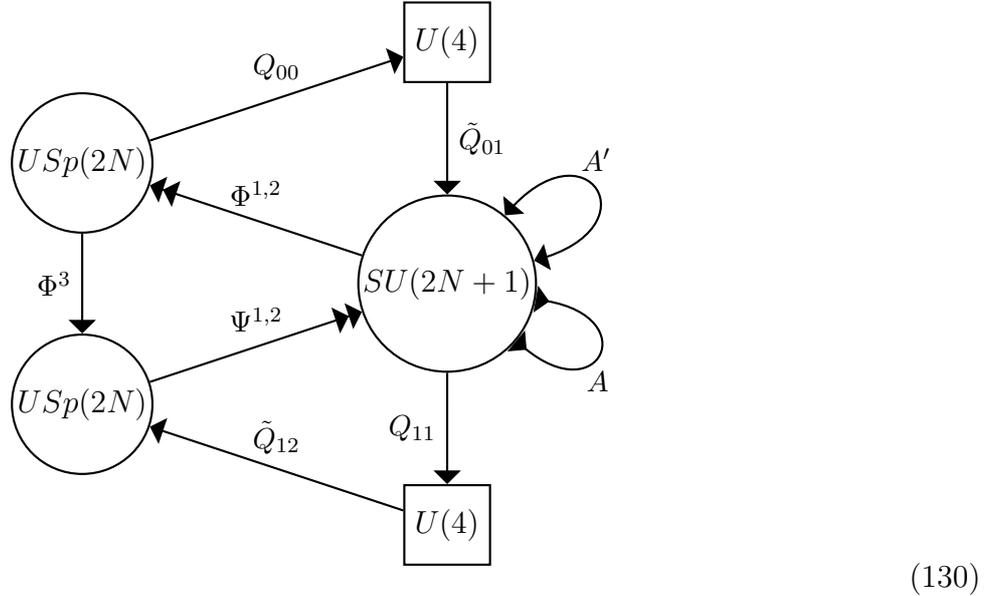
\begin{center}
\begin{tikzpicture}[thick, scale=0.8]
  \node[circle, draw, inner sep= 1pt](L1) at (6,-2){$USp(2N)$};
    \node[circle, draw, inner sep= 1pt](L7) at (6,-6){$USp(2N)$};
  \node[circle, draw, inner sep= 1pt](L2) at (12,-4){$SU(2N+1)$};
\node[draw, rectangle, minimum width=30pt, minimum height=30pt](L5) at (12,0){$U(4)$};
 \node[draw, rectangle, minimum width=30pt, minimum height=30pt](L6) at (12,-8){$U(4)$};
 \path[every node/.style={font=\sffamily\small,
  		fill=white,inner sep=1pt}]
(L1) edge [<<-] node[above=1.5mm] {$\Phi^{1,2}$} (L2)
(L1) edge [->] node[left=1mm] {$\Phi^3$} (L7)
(L7) edge [->>] node[above=1.5mm] {$\Psi^{1,2}$} (L2)
(L7) edge [<-] node[above=1.5mm] {$\tilde{Q}_{12}$} (L6)
(L1) edge [->] node[above=2mm] {$Q_{00}$} (L5)
(L5) edge [->] node[right=1mm] {$\tilde{Q}_{01}$} (L2)
(L2) edge [->] node[left=1mm] {$Q_{11}$} (L6)
(L2) edge [loop, <->, in=15, out=50, looseness=6] node[above=2mm] {$A'$}  (L2)
(L2) edge [loop,>-<, in=-10, out=-40, looseness=6] node[below=2mm] {$A$}  (L2);
\end{tikzpicture}\vspace{-1cm}
\end{center}\be\label{ZZ4iiz1} \ee
The superpotential for the above family of theories looks schematically like
\begin{eqnarray}\label{SuperpotZZ4iiz1}
W= \Phi_{\;\;\rho}^{3\alpha}\Psi^{[1\rho}\Phi^{2]}_\alpha+\Tr_{F}\left(\tilde{Q}_{01}\Phi^1_\alpha Q_{00}^\alpha+\tilde{Q}_{12}\Psi^1_\rho Q_{11}^\rho\right)\,,
\end{eqnarray}
where we have suppressed flavor as well as unitary-gauge indices, and $\Tr_{F}$ denotes the trace over the flavor groups.

Such a family has recently been obtained in \cite{Zafrir:2018hkr} as a mass-deformed compactification with minimal flux (half-integral) of the six-dimensional theory obtained by probing with $M5$ branes an $M9$ wall wrapped on $\mathbb{C}^2/\mathbb{Z}_2$ (i.e.~the orbifolded E-string). Differently from \cite{Zafrir:2018hkr}, however, we have no superpotential couplings for the two antisymmetric fields, and moreover $\Phi^2$ and $\Psi^2$ are not coupled to the fundamental fields. The reason why we miss these terms is essentially the same as for the quiver \eqref{Z2MixedQuiver}, when compared to the analogous one in ref.~\cite{Kim:2017toz}.  We analyze this in Section \ref{E-string}.

As far as mass deformations of the theory \eqref{ZZ4iiz1} are concerned, Eq.~\eqref{OrbifoldMass}, together with the second Eq.~in \eqref{PhiOmega}, tells us that the following three matrices survive the orientifold projection
\be\label{massesZZ4iiz1}
\mathcal{M}_{[MN]}\,,\qquad\bar{\mathcal{M}}^{[IJ]}\,,\qquad \tilde{\mathcal{M}}^M_I\,,
\ee
where $I,J$ and $M,N$ are indices of the top and the bottom flavor group respectively. The matrices \eqref{massesZZ4iiz1} lead respectively to the following mass terms
\be\label{massTermsZZ4iiz1}
\mathcal{M}_{[MN]}(\tilde{Q}_{12})^M_\rho(\tilde{Q}_{12})^N_\sigma J^{\rho\sigma}\,,\qquad\bar{\mathcal{M}}^{[IJ]}(Q_{00})_I^\alpha (Q_{00})_J^\beta J_{\alpha\beta}\,,\qquad \tilde{\mathcal{M}}^M_I(\tilde{Q}_{01})^I(Q_{11})_M\,,
\ee
where gauge indices have been suppressed in the last term.
\\

Let us now switch to the orientifold projection with no fixed nodes. Contrary to the previous case, here there is only one family of SCFTs, that arises from probing a background with a \emph{mixed} $O3$$^\pm$ projection and no $D7$ branes.\footnote{This holds regardless of where the $D7$/$O7$ stack is placed.} This has been described in \cite{Bianchi:2020fuk} and is given by the following quiver
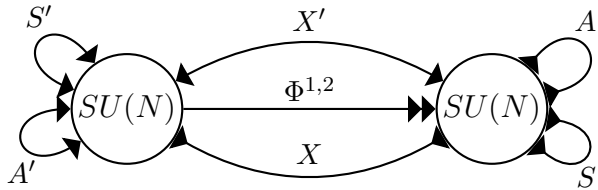
\begin{center}
\begin{tikzpicture}[thick, scale=0.8]
  \node[circle, draw, inner sep= 1pt](L1) at (6,0){$SU(N)$};
  \node[circle, draw, inner sep= 1pt](L2) at (12,0){$SU(N)$};
  \path[every node/.style={font=\sffamily\small,
  		fill=white,inner sep=1pt}]
(L1) edge [->>] node[above=.5mm] {$\Phi^{1,2}$} (L2)
(L1) edge [bend right=30, >-<] node[above=1mm] {$X$} (L2)
(L1) edge [bend left=30, <->] node[above=1mm] {$X'$} (L2)
(L2) edge [loop, >-<, in=15, out=50, looseness=6] node[above=2mm] {$A$}  (L2)
(L2) edge [loop,>-<, in=-10, out=-40, looseness=6] node[below=1.5mm] {$S$}  (L2)
(L1) edge [loop, <->, in=-145, out=-180, looseness=6] node[below=2mm] {$A'$}  (L1)
(L1) edge [loop,<->, in=125, out=160, looseness=6] node[above=1.5mm] {$S'$}  (L1);
\end{tikzpicture}\be\label{ZZ4e} \ee\vspace{-.5cm}
\end{center}
The chiral fields $A,S$ transform in the antisymmetric and symmetric representation respectively, whereas $A',S'$ transform in the corresponding conjugate representations. The superpotential of this family of theories looks schematically like\footnote{There is a completely analogous family of theories obtained by exchanging $1\leftrightarrow2$.}
\begin{eqnarray}
W= \Tr\left(\Phi^1A'X-\Phi^2AX'+\Phi^1SX'-\Phi^2S'X\right)\,,
\end{eqnarray}
where all indices have been suppressed.
\\

What is left to discuss, to exhaust the $\mathbb{C}^3/\ZZ_4$ orbifold, is an action that only involves two of the three internal complex coordinates, e.g.
\be\label{Z4actionz2z3}
(z_1,z_2,z_3)\;\sim\;\left(z_1,-{\rm i}\,z_2,{\rm i}\,z_3\right)\,.
\ee
First, let us discuss the cases without $D7$/$O7$. Like for the $\mathbb{C}^3/\ZZ_2$ orbifold,, there appear two families of SCFTs with $\mathcal{N}=2$ supersymmetry by taking a \emph{mixed} $O3$$^\pm$ projection: They are distinguished according to whether two or no nodes are fixed by the involution. In the first case, the mixed projection manifests itself at the level of the gauge groups, and the ensuing family of field theories is described by the following quiver
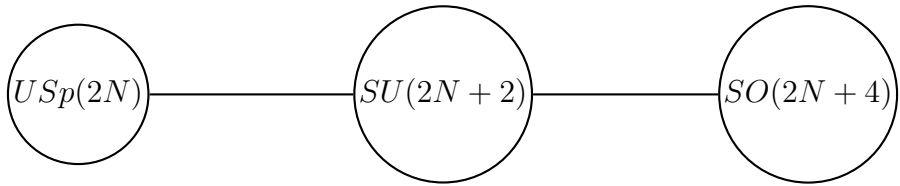
\begin{center}
\begin{tikzpicture}[thick, scale=0.8]
  \node[circle, draw, inner sep= 1pt](L1) at (6,-4){$USp(2N)$};
  \node[circle, draw, inner sep= 1pt](L2) at (12,-4){$SU(2N+2)$};
   \node[circle, draw, inner sep= 1pt](L3) at (18,-4){$SO(2N+4)$};
 \path[every node/.style={font=\sffamily\small,
  		fill=white,inner sep=1pt}]
(L2) edge  (L3)
(L1) edge  (L2);
\end{tikzpicture}\vspace{-.9cm}
\end{center}\be\label{ZZ2O3} \ee\\
where the links represent hypermultiplets in the bifundamental representation. In the second case, the mixed projection manifests itself at the level of the matter fields, and we get the following quiver
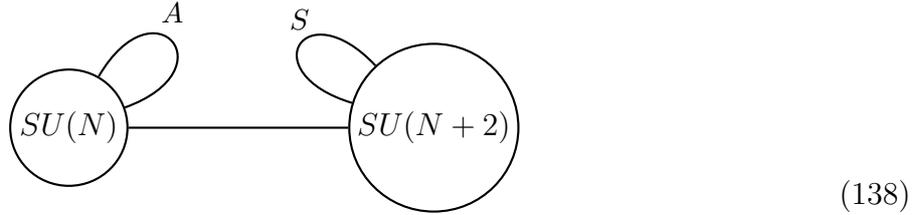
\begin{center}
\begin{tikzpicture}[thick, scale=0.8]
  \node[circle, draw, inner sep= 2pt](L1) at (4,-2){$SU(N)$};
  \node[circle, draw, inner sep= 2pt](L2) at (10,-2){$SU(N+2)$};
   \path[every node/.style={font=\sffamily\small,
  		fill=white,inner sep=1pt}]
(L1) edge  (L2)
(L1) edge [loop, in=20, out=60, looseness=7] node[above=2mm] {$A$}  (L1)
(L2) edge [loop, in=133, out=163, looseness=6] node[above=1.5mm] {$S$}  (L2);
\end{tikzpicture}\vspace{-1cm}\be\label{Z4MixedQuiver2}\ee
\end{center}
where the two loops represent two hypermultiplets, the left one in the antisymmetric and the right one in the symmetric representation of the corresponding gauge group.
\\

Now let us introduce a $D7$/$O7$ stack. The cases which we will consider here are those where the stack is transverse to one of the coordinates involved by the orbifold action, say $z_3$,\footnote{One can check that choosing $z_2$ leads to the same conclusions.} because the $\mathcal{N}=2$ theories one gets by placing the $D7$/$O7$ stack transverse to $z_1$ have already been fully covered in \cite{Giacomelli:2022drw}. By a simple inspection, we can see that the only possibility compatible with conformality is when the $O7$$^-$ involution fixes two gauge nodes and no flavor nodes, i.e.~Eq.~\eqref{2G0F}. Following the usual rules, we arrive at the one-parameter family of SCFTs encoded in the quiver
 
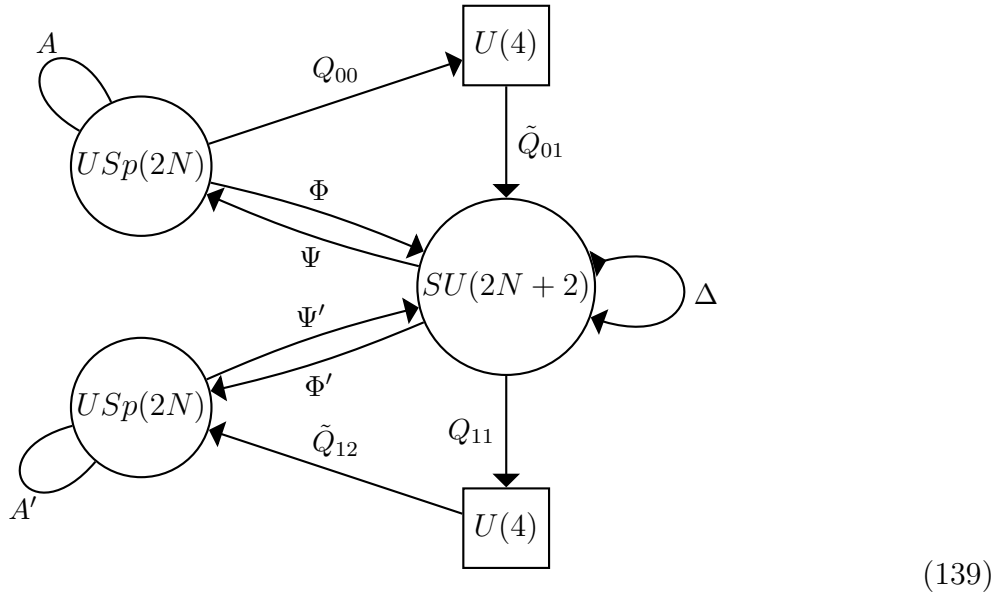
\begin{center}
\begin{tikzpicture}[thick, scale=0.8]
  \node[circle, draw, inner sep= 1pt](L1) at (6,-2){$USp(2N)$};
    \node[circle, draw, inner sep= 1pt](L7) at (6,-6){$USp(2N)$};
  \node[circle, draw, inner sep= 1pt](L2) at (12,-4){$SU(2N+2)$};
\node[draw, rectangle, minimum width=30pt, minimum height=30pt](L5) at (12,0){$U(4)$};
 \node[draw, rectangle, minimum width=30pt, minimum height=30pt](L6) at (12,-8){$U(4)$};
 \path[every node/.style={font=\sffamily\small,
  		fill=white,inner sep=1pt}]
(L1) edge [loop, in=-245, out=-210, looseness=6] node[above=1mm] {$A$}  (L1)
(L7) edge [loop, in=-165, out=-130, looseness=6] node[below=1.4mm] {$A'$}  (L7)
(L1) edge [bend right=5,<-] node[below=1mm] {$\Psi$} (L2)
(L1) edge [bend left=5,->] node[above=1mm] {$\Phi$} (L2)
(L7) edge [bend right=5,<-] node[below=1mm] {$\Phi'$} (L2)
(L7) edge [bend left=5,->] node[above=1mm] {$\Psi'$} (L2)
(L7) edge [<-] node[above=1.5mm] {$\tilde{Q}_{12}$} (L6)
(L1) edge [->] node[above=2mm] {$Q_{00}$} (L5)
(L5) edge [->] node[right=1mm] {$\tilde{Q}_{01}$} (L2)
(L2) edge [->] node[left=1mm] {$Q_{11}$} (L6)
(L2) edge [loop, >->, in=-20, out=15, looseness=6] node[right=1mm] {$\Delta$}  (L2);
\end{tikzpicture}\vspace{-1cm}
\end{center}\be\label{ZZ4iiz2} \ee
where $A,A'$ transform in the antisymmetric representation of the symplectic gauge groups, whereas $\Delta$ is an adjoint chiral field of the unitary gauge group. The superpotential for the above family of theories looks schematically like
\begin{eqnarray}\label{SuperpotZZ4iiz2}
W=J_{\alpha\beta}\Phi^\beta\Psi_\gamma A^{\alpha\gamma}-J^{\rho\sigma}\Phi'_\sigma \Psi'^\tau A'_{\rho\tau} -\Delta\Phi^\alpha\Psi_\alpha+\Delta\Phi'_\rho\Psi^\rho+\Tr_{F}\left(\tilde{Q}_{01}\Psi_\alpha Q_{00}^\alpha+\tilde{Q}{_{12}}_\rho\Psi'^\rho Q_{11}\right)\,,\nonumber
\end{eqnarray}
where we have suppressed flavor as well as unitary-gauge indices, and $\Tr_{F}$ denotes the trace over the flavor groups. $J_{\alpha\beta},J^{\rho\sigma}$ are the symplectic forms of the upper and lower symplectic gauge group respectively.

As far as mass deformations are concerned, the discussion goes exactly as for the quiver \eqref{ZZ4iiz1}, cf.~Eqs.~\eqref{massesZZ4iiz1} and \eqref{massTermsZZ4iiz1}.\\

Let us end with the following observation. It can be easily seen that \emph{no unoriented} quiver SCFTs exist for the orbifold $\mathbb{C}^3/\ZZ_5$ with action $(1,1,3)$, either with $O3$$^\pm$ planes or with a $D7$/$O7$ stack at $z_3=0$, hinting at the fact that such theories become more and more sparse as we increase the orbifold order.

\bibliographystyle{JHEP}
\bibliography{N1Refs.bib}
\end{document}